\pdfoutput=1

\documentclass[11pt,twoside,a4paper,cmspaper,final,collab]{cms-tdr}
\def\svnVersion{292397}\def\cmsCernNo{2013-037}\def\cmsCernDate{\today}\def\cmsMessage{Published in the European Physical Journal C as \href{http://dx.doi.org/10.1140/epjc/s10052-015-3351-7}{\doi{10.1140/epjc/s10052-015-3351-7.}}}

\begin{document}\cmsNoteHeader{HIG-14-009}

\hyphenation{had-ron-i-za-tion}
\hyphenation{cal-or-i-me-ter}
\hyphenation{de-vices}

\RCS$Revision: 292365 $
\RCS$HeadURL: svn+ssh://svn.cern.ch/reps/tdr2/papers/HIG-14-009/trunk/HIG-14-009.tex $
\RCS$Id: HIG-14-009.tex 292365 2015-06-15 07:19:14Z mschen $
\newlength\cmsFigWidth
\ifthenelse{\boolean{cms@external}}{\setlength\cmsFigWidth{0.85\columnwidth}}{\setlength\cmsFigWidth{0.4\textwidth}}
\ifthenelse{\boolean{cms@external}}{\providecommand{\cmsLeft}{top}}{\providecommand{\cmsLeft}{left}}
\ifthenelse{\boolean{cms@external}}{\providecommand{\cmsRight}{bottom}}{\providecommand{\cmsRight}{right}}
\ifthenelse{\boolean{cms@external}}{\providecommand{\cmsBold}{\relax}}{\providecommand{\cmsBold}{\boldmath}}
\hyphenation{ATLAS}
\newcommand{\bb}{\ensuremath{\PQb\PQb}\xspace}
\newcommand{\bbh}{\ensuremath{\cPqb\cPqb\PH}\xspace}
\newcommand{\BRBSM}{\ensuremath{\mathrm{BR_{BSM}}}\xspace}
\newcommand{\BRinv}{\ensuremath{\mathrm{BR_{inv}}}\xspace}
\newcommand{\BRundet}{\ensuremath{\mathrm{BR}_\text{undet}}\xspace}
\newcommand{\calM}{\ensuremath{\mathcal{M}}\xspace}
\newcommand{\chisq}{\ensuremath{\chi^{2}}\xspace}
\newcommand{\chisqdof}{\ensuremath{\chisq/\dof}\xspace}
\newcommand{\dof}{\text{dof}\xspace}
\newcommand{\ee}{\ensuremath{\Pe\Pe}\xspace}
\newcommand{\eee}{\ensuremath{\ee\Pe}\xspace}
\newcommand{\eeee}{\ensuremath{4\Pe}\xspace}
\newcommand{\eem}{\ensuremath{\ee\PGm}\xspace}
\newcommand{\eemm}{\ensuremath{2\Pe2\Pgm}\xspace}
\newcommand{\emu}{\ensuremath{\Pe\PGm}\xspace}
\newcommand{\ggh}{\ensuremath{\Pg\Pg\PH}\xspace}
\newcommand{\ggzh}{\ensuremath{\Pg\Pg\to\cPZ\PH}\xspace}
\newcommand{\gluglu}{\ensuremath{\Pg\Pg}\xspace}
\newcommand{\GSM}{\ensuremath{\Gamma_\mathrm{SM}}\xspace}
\newcommand{\hbb}{\ensuremath{\PH\to\bb}\xspace}
\newcommand{\hee}{\ensuremath{\PH\to\ee}\xspace}
\newcommand{\hgg}{\ensuremath{\PH\to\gg}\xspace}
\newcommand{\hhad}{\ensuremath{\PH\to\text{hadrons}}\xspace}
\newcommand{\hinv}{\ensuremath{\PH(\mathrm{inv})}\xspace}
\newcommand{\hinvisible}{\ensuremath{\PH\to\mathrm{invisible}}\xspace}
\newcommand{\hlep}{\ensuremath{\PH\to\text{leptons}}\xspace}
\newcommand{\hmm}{\ensuremath{\PH\to\mumu}\xspace}
\newcommand{\hthth}{\ensuremath{\PH\to\thth}\xspace}
\newcommand{\htt}{\ensuremath{\PH\to\PGt\PGt}\xspace}
\newcommand{\hww}{\ensuremath{\PH\to\ww}\xspace}
\newcommand{\hwwlnln}{\ensuremath{\hww\to\ell\cPgn\ell\cPgn}\xspace}
\newcommand{\hzz}{\ensuremath{\PH\to\zz}\xspace}
\newcommand{\hzzD}{\ensuremath{\mathcal{D}^{\text{kin}}_\text{bkg}}\xspace}
\newcommand{\hzzllll}{\ensuremath{\hzz\to 4\ell}\xspace}
\newcommand{\JP}{\ensuremath{J^{P}}\xspace}
\newcommand{\JPC}{\ensuremath{J^{PC}}\xspace}
\newcommand{\kb}{\ensuremath{\kappa_{\cPqb}}\xspace}
\newcommand{\kc}{\ensuremath{\kappa_{\PQc}}\xspace}
\newcommand{\kd}{\ensuremath{\kappa_{\PQd}}\xspace}
\newcommand{\kf}{\ensuremath{\kappa_{\mathrm{f}}}\xspace}
\newcommand{\kgam}{\ensuremath{\kappa_{\PGg}}\xspace}
\newcommand{\kglu}{\ensuremath{\kappa_{\mathrm{\Pg}}}\xspace}
\newcommand{\kgluZ}{\ensuremath{\kappa_{\Pg\cPZ}}\xspace}
\newcommand{\kH}{\ensuremath{\kappa_{\PH}}\xspace}
\newcommand{\kl}{\ensuremath{\kappa_{\ell}}\xspace}
\newcommand{\kmu}{\ensuremath{\kappa_{\Pgm}}\xspace}
\newcommand{\kq}{\ensuremath{\kappa_{\PQq}}\xspace}
\newcommand{\ks}{\ensuremath{\kappa_{\PQs}}\xspace}
\newcommand{\ktau}{\ensuremath{\kappa_{\PGt}}\xspace}
\newcommand{\ktop}{\ensuremath{\kappa_{\cPqt}}\xspace}
\newcommand{\ku}{\ensuremath{\kappa_{\PQu}}\xspace}
\newcommand{\kV}{\ensuremath{\kappa_{\mathrm{V}}}\xspace}
\newcommand{\kW}{\ensuremath{\kappa_{\mathrm{\PW}}}\xspace}
\newcommand{\kZ}{\ensuremath{\kappa_{\mathrm{\cPZ}}}\xspace}
\newcommand{\lbbgg}{\ensuremath{\lambda_{\cPqb\cPqb,\PGg\PGg}}\xspace}
\newcommand{\lbZ}{\ensuremath{\lambda_{\cPqb\cPZ}}\xspace}
\newcommand{\ldu}{\ensuremath{\lambda_{\mathrm{\PQd\PQu}}}\xspace}
\newcommand{\lgamZ}{\ensuremath{\lambda_{\PGg\cPZ}}\xspace}
\newcommand{\llq}{\ensuremath{\lambda_{\mathrm{\ell\PQq}}}\xspace}
\newcommand{\ltauZ}{\ensuremath{\lambda_{\PGt\cPZ}}\xspace}
\newcommand{\ltopglu}{\ensuremath{\lambda_{\cPqt\cPg}}\xspace}
\newcommand{\lttgg}{\ensuremath{\lambda_{\PGt\PGt,\PGg\PGg}}\xspace}
\newcommand{\ltth}{\ensuremath{\lambda_{\tth}}\xspace}
\newcommand{\ltthbb}{\ensuremath{\lambda_{\tth}^{\bb}}\xspace}
\newcommand{\ltthgg}{\ensuremath{\lambda_{\tth}^{\gg}}\xspace}
\newcommand{\ltthtt}{\ensuremath{\lambda_{\tth}^{\tt}}\xspace}
\newcommand{\ltthww}{\ensuremath{\lambda_{\tth}^{\ww}}\xspace}
\newcommand{\ltthzz}{\ensuremath{\lambda_{\tth}^{\zz}}\xspace}
\newcommand{\lvbf}{\ensuremath{\lambda_{\vbf}}\xspace}
\newcommand{\lvbfbb}{\ensuremath{\lambda_{\vbf}^{\bb}}\xspace}
\newcommand{\lvbfgg}{\ensuremath{\lambda_{\vbf}^{\gg}}\xspace}
\newcommand{\lvbftt}{\ensuremath{\lambda_{\vbf}^{\tt}}\xspace}
\newcommand{\lvbfww}{\ensuremath{\lambda_{\vbf}^{\ww}}\xspace}
\newcommand{\lvbfzz}{\ensuremath{\lambda_{\vbf}^{\zz}}\xspace}
\newcommand{\lvh }{\ensuremath{\lambda_{\vh }}\xspace}
\newcommand{\lvhbb }{\ensuremath{\lambda_{\vh }^{\bb}}\xspace}
\newcommand{\lvhgg }{\ensuremath{\lambda_{\vh }^{\gg}}\xspace}
\newcommand{\lvhtt }{\ensuremath{\lambda_{\vh }^{\tt}}\xspace}
\newcommand{\lvhww }{\ensuremath{\lambda_{\vh }^{\ww}}\xspace}
\newcommand{\lvhzz }{\ensuremath{\lambda_{\vh }^{\zz}}\xspace}
\newcommand{\lwwgg}{\ensuremath{\lambda_{\PW\PW,\PGg\PGg}}\xspace}
\newcommand{\lWZ}{\ensuremath{\lambda_{\mathrm{\PW\cPZ}}}\xspace}
\newcommand{\lZglu}{\ensuremath{\lambda_{\cPZ\Pg}}\xspace}
\newcommand{\lzzgg}{\ensuremath{\lambda_{\cPZ\cPZ,\PGg\PGg}}\xspace}
\newcommand{\mh}{\ensuremath{\mathrm{m}_{\PH}}\xspace}
\newcommand{\mH}{\ensuremath{m_{\PH}}\xspace}
\newcommand{\mjj}{\ensuremath{m_{\text{jj}}}}
\newcommand{\mme}{\ensuremath{\mumu\Pe}\xspace}
\newcommand{\mmee}{\ensuremath{2\Pgm2\Pe}\xspace}
\newcommand{\mmm}{\ensuremath{\mumu\PGm}\xspace}
\newcommand{\mmmm}{\ensuremath{4\Pgm}\xspace}
\newcommand{\mt}{\ensuremath{m_{\text{T}}}\xspace}
\newcommand{\mubb}{\ensuremath{\mu_{\bb}}\xspace}
\newcommand{\muf}{\ensuremath{\mu_{\ggh,\tth}}\xspace}
\newcommand{\mufbb}{\ensuremath{\mu_{\ggh,\tth}^{\bb}}\xspace}
\newcommand{\mufgg}{\ensuremath{\mu_{\ggh,\tth}^{\gg}}\xspace}
\newcommand{\muftt}{\ensuremath{\mu_{\ggh,\tth}^{\tt}}\xspace}
\newcommand{\mufww}{\ensuremath{\mu_{\ggh,\tth}^{\ww}}\xspace}
\newcommand{\mufzz}{\ensuremath{\mu_{\ggh,\tth}^{\zz}}\xspace}
\newcommand{\mugg}{\ensuremath{\mu_{\gg}}\xspace}
\newcommand{\muggh}{\ensuremath{\mu_{\ggh}}\xspace}
\newcommand{\mumu}{\ensuremath{\PGm\PGm}\xspace}
\newcommand{\mutt}{\ensuremath{\mu_{\tt}}\xspace}
\newcommand{\mutth}{\ensuremath{\mu_{\tth}}\xspace}
\newcommand{\muv}{\ensuremath{\mu_{\vbf,\vh}}\xspace}
\newcommand{\muvbf}{\ensuremath{\mu_{\vbf}}\xspace}
\newcommand{\muvh }{\ensuremath{\mu_{\vh}}\xspace}
\newcommand{\muww}{\ensuremath{\mu_{\ww}}\xspace}
\newcommand{\muzz}{\ensuremath{\mu_{\zz}}\xspace}
\newcommand{\MX}{\ensuremath{m_\mathrm{\PH}}\xspace}
\newcommand{\njet}[1][n]{\mbox{#1-jet}\xspace}
\newcommand{\pT}{\pt}
\newcommand{\pTV}{\ensuremath{\pT(\mathrm{V})}\xspace}
\newcommand{\pTZ}{\ensuremath{\pT(\cPZ)}\xspace}
\newcommand{\pval}{\ensuremath{p\text{-value}}\xspace}
\newcommand{\pvals}{\ensuremath{p\text{-values}}\xspace}
\newcommand{\rankM}{\ensuremath{\operatorname{rank}(\calM)}\xspace}
\newcommand{\rvf}{\ensuremath{\muv/\muf}\xspace}
\newcommand{\tauh}{\ensuremath{\Pgt_{\mathrm{h}}}\xspace}
\newcommand{\tete}{\ensuremath{\Pe\Pe}\xspace}
\newcommand{\teth}{\ensuremath{\Pe\tauh}\xspace}
\newcommand{\tetm}{\ensuremath{\Pe\Pgm}\xspace}
\newcommand{\thth}{\ensuremath{\tauh\tauh}\xspace}
\newcommand{\tlth}{\ensuremath{\ell\tauh}\xspace}
\newcommand{\tmth}{\ensuremath{\Pgm\tauh}\xspace}
\newcommand{\tmtm}{\ensuremath{\Pgm\Pgm}\xspace}
\newcommand{\tth}{\ensuremath{\cPqt\cPqt\PH}\xspace}
\newcommand{\vbf}{\ensuremath{\mathrm{VBF}}\xspace}
\newcommand{\vh}{\ensuremath{\mathrm{V}\PH}\xspace}
\newcommand{\wh}{\ensuremath{\PW\PH}\xspace}
\newcommand{\wlnbb}{\ensuremath{\PW(\ell\cPgn)\PH(\bb)}\xspace}
\newcommand{\wtnbb}{\ensuremath{\PW(\tauh\cPgn)\PH(\bb)}\xspace}
\newcommand{\ww}{\ensuremath{\PW\PW}\xspace}
\newcommand{\wz}{\ensuremath{\PW\cPZ}\xspace}
\newcommand{\zh}{\ensuremath{\cPZ\PH}\xspace}
\newcommand{\zllbb}{\ensuremath{\cPZ(\ell\ell)\PH(\bb)}\xspace}
\newcommand{\znnbb}{\ensuremath{\cPZ(\cPgn\cPgn)\PH(\bb)}\xspace}
\newcommand{\zz}{\ensuremath{\cPZ\cPZ}\xspace}
\renewcommand{\gg}{\ensuremath{\PGg\PGg}\xspace}
\renewcommand{\tt}{\ensuremath{\PGt\PGt}\xspace}

\newcommand{\BRBSMCSeven}{\ensuremath{[0.00,0.57]}\xspace}           
\newcommand{\BRBSMCSevenEXPOneSig}{\ensuremath{[0.00,0.27]}\xspace}           
\newcommand{\BRBSMCSevenEXPTwoSig}{\ensuremath{[0.00,0.52]}\xspace}           %
\newcommand{\BRBSMCSevenOneSig}{\ensuremath{\leq0.34}\xspace}
\newcommand{\BRBSMHinvBestFit}   {\ensuremath{0.03}\xspace}
\newcommand{\BRBSMHinvKSMBestFit}   {\ensuremath{0.06}\xspace}
\newcommand{\BRBSMHinvKSMOneSig}{\ensuremath{\BRBSMHinvKSMBestFit~^{+0.11}_{-0.06}}\xspace}
\newcommand{\BRBSMHinvKSMOneSigEXP}{\ensuremath{[0.00,0.12]}\xspace}
\newcommand{\BRBSMHinvKSMTwoSig}{\ensuremath{[0.00,0.27]}\xspace}
\newcommand{\BRBSMHinvKSMTwoSigEXP}{\ensuremath{[0.00,0.21]}\xspace}
\newcommand{\BRBSMHinvOneSig}   {\ensuremath{\BRBSMHinvBestFit~^{+0.15}_{-0.03}}\xspace}
\newcommand{\BRBSMHinvOneSigEXP}   {\ensuremath{[0.00,0.16]}\xspace}
\newcommand{\BRBSMHinvTwoSig}   {\ensuremath{[0.00,0.32]}\xspace}
\newcommand{\BRBSMHinvTwoSigEXP}   {\ensuremath{[0.00,0.29]}\xspace}
\newcommand{\BRBSMOneSig}      {\ensuremath{\leq0.14}\xspace}
\newcommand{\BRBSMOneSigEXP}{\ensuremath{[0.00,0.22]}\xspace}           
\newcommand{\BRBSMTwoSig}{\ensuremath{[0.00,0.32]}\xspace}           
\newcommand{\BRBSMTwoSigEXP}{\ensuremath{[0.00,0.42]}\xspace}           
\newcommand{\BRinvHinvCEight}  {\ensuremath{[0.00,0.49]}\xspace}           
\newcommand{\BRinvHinvCEightEXPOneSig}  {\ensuremath{[0.00,0.18]}\xspace}           
\newcommand{\BRinvHinvCEightEXPTwoSig}  {\ensuremath{[0.00,0.32]}\xspace}           
\newcommand{\BRinvHinvCEightOneSig}{\ensuremath{0.17\pm0.17}\xspace}
\newcommand{\BRinvHinvCSeven}  {\ensuremath{[0.00,0.49]}\xspace}           
\newcommand{\BRinvHinvCSevenEXPOneSig}  {\ensuremath{[0.00,0.18]}\xspace}           
\newcommand{\BRinvHinvCSevenEXPTwoSig}  {\ensuremath{[0.00,0.32]}\xspace}           
\newcommand{\BRinvHinvCSevenOneSig}{\ensuremath{0.17\pm0.17}\xspace}
\newcommand{\BRundetHinvCEight}{\ensuremath{[0.00,0.52]}\xspace}           
\newcommand{\BRundetHinvCEightEXPOneSig}{\ensuremath{[0.00,0.27]}\xspace}           
\newcommand{\BRundetHinvCEightEXPTwoSig}{\ensuremath{[0.00,0.51]}\xspace}           
\newcommand{\BRundetHinvCEightOneSig}{\ensuremath{\leq0.23}\xspace}
\newcommand{\CFIVEkb}   {\ensuremath{[0.09,1.44]}\xspace}
\newcommand{\CFIVEkbOneSig}  {\ensuremath{0.74~^{+0.33}_{-0.29}}\xspace}
\newcommand{\CFIVEkmu}  {\ensuremath{[0.00,2.77]}\xspace}
\newcommand{\CFIVEkmuOneSig} {\ensuremath{0.49~^{+1.38}_{-0.49}}\xspace}
\newcommand{\CFIVEktau} {\ensuremath{[0.50,1.24]}\xspace}
\newcommand{\CFIVEktauOneSig}{\ensuremath{0.84~^{+0.19}_{-0.18}}\xspace}
\newcommand{\CFIVEktop} {\ensuremath{[0.53,1.20]}\xspace}
\newcommand{\CFIVEktopOneSig}{\ensuremath{0.81~^{+0.19}_{-0.15}}\xspace}
\newcommand{\CFIVEkW}   {\ensuremath{[0.68,1.23]}\xspace}
\newcommand{\CFIVEkWOneSig}  {\ensuremath{0.95~^{+0.14}_{-0.13}}\xspace}
\newcommand{\CFIVEkZ}   {\ensuremath{[0.72,1.35]}\xspace}
\newcommand{\CFIVEkZOneSig}  {\ensuremath{1.05~^{+0.16}_{-0.16}}\xspace}
\newcommand{\ChiCouplingRatios}    {\ensuremath{9.6 / 7}\xspace}
\newcommand{\ChiCSix}    {\ensuremath{7.9 / 6}\xspace}
\newcommand{\ChiFiveDec} {\ensuremath{1.0 / 5}\xspace}
\newcommand{\ChiFourProd}{\ensuremath{5.5 / 4}\xspace}
\newcommand{\ChiResolvedCSix}    {\ensuremath{1.6 / 5}\xspace}
\newcommand{\ChiSixteen} {\ensuremath{10.5 / 16}\xspace}
\newcommand{\CSIXkb}     {\ensuremath{[0.00,1.23]}\xspace}        
\newcommand{\CSIXkbOneSig}  {\ensuremath{0.64 ^{+0.28}_{-0.29}}\xspace}
\newcommand{\CSIXkgam}   {\ensuremath{[0.67,1.33]}\xspace}      
\newcommand{\CSIXkgamOneSig}{\ensuremath{0.98 ^{+0.17}_{-0.16}}\xspace}
\newcommand{\CSIXkglu}   {\ensuremath{[0.52,1.07]}\xspace}      
\newcommand{\CSIXkgluOneSig}{\ensuremath{0.75 ^{+0.15}_{-0.13}}\xspace}
\newcommand{\CSIXktau}   {\ensuremath{[0.48,1.20]}\xspace}      
\newcommand{\CSIXktauOneSig}{\ensuremath{0.82 ^{+0.18}_{-0.18}}\xspace}
\newcommand{\CSIXktop}   {\ensuremath{[0.97,2.28]}\xspace}      
\newcommand{\CSIXktopOneSig}{\ensuremath{1.60 ^{+0.34}_{-0.32}}\xspace}
\newcommand{\CSIXkV}     {\ensuremath{[0.66,1.23]}\xspace}        
\newcommand{\CSIXkVOneSig}  {\ensuremath{0.96 ^{+0.14}_{-0.15}}\xspace}
\newcommand{\CTWOkF}     {\ensuremath{[0.63,1.15]}\xspace}        
\newcommand{\CTWOkFOneSig}  {\ensuremath{0.87 ^{+0.14}_{-0.13}}\xspace}
\newcommand{\CTWOkgam}   {\ensuremath{[0.89,1.40]}\xspace}      
\newcommand{\CTWOkgamBEST}{\ensuremath{1.14}\xspace}   
\newcommand{\CTWOkgamOneSig}{\ensuremath{1.14 ^{+0.12}_{-0.13}}\xspace}
\newcommand{\CTWOkglu}   {\ensuremath{[0.69,1.11]}\xspace}     
\newcommand{\CTWOkgluBEST}{\ensuremath{0.89}\xspace}   
\newcommand{\CTWOkgluOneSig}{\ensuremath{0.89 ^{+0.11}_{-0.10}}\xspace}
\newcommand{\CTWOkV}     {\ensuremath{[0.87,1.14]}\xspace}        
\newcommand{\CTWOkVOneSig}  {\ensuremath{1.01 ^{+0.07}_{-0.07}}\xspace}
\newcommand{\DeltaMMeas}{\ensuremath{-0.89^{+0.56}_{-0.57}\GeV}\xspace}
\newcommand{\DRbbgg}{\ensuremath{0.63^{+0.44}_{-0.35}}\xspace}
\newcommand{\DRbbggTwoSig}{\ensuremath{[0.00,1.76]}\xspace}
\newcommand{\DRbbtt}{\ensuremath{0.87^{+0.69}_{-0.49}}\xspace}
\newcommand{\DRbbttTwoSig}{\ensuremath{[0.00,2.84]}\xspace}
\newcommand{\DRbbww}{\ensuremath{0.74^{+0.61}_{-0.41}}\xspace}
\newcommand{\DRbbwwTwoSig}{\ensuremath{[0,00,2.30]}\xspace}
\newcommand{\DRbbzz}{\ensuremath{0.65^{+0.59}_{-0.37}}\xspace}
\newcommand{\DRbbzzTwoSig}{\ensuremath{[0.00,2.16]}\xspace}
\newcommand{\DRggbb}{\ensuremath{1.60^{+1.86}_{-0.70}}\xspace}
\newcommand{\DRggbbTwoSig}{\ensuremath{[0.54,-]}\xspace}
\newcommand{\DRggtt}{\ensuremath{1.41^{+0.75}_{-0.45}}\xspace}
\newcommand{\DRggttTwoSig}{\ensuremath{[0.67,3.76]}\xspace}
\newcommand{\DRggww}{\ensuremath{1.21^{+0.41}_{-0.31}}\xspace}
\newcommand{\DRggwwTwoSig}{\ensuremath{[0.67,2.22]}\xspace}
\newcommand{\DRggzz}{\ensuremath{1.06^{+0.44}_{-0.31}}\xspace}
\newcommand{\DRggzzTwoSig}{\ensuremath{[0.54,2.16]}\xspace}
\newcommand{\DRttbb}{\ensuremath{1.14^{+1.34}_{-0.52}}\xspace}
\newcommand{\DRttbbTwoSig}{\ensuremath{[0.33,-]}\xspace}
\newcommand{\DRttgg}{\ensuremath{0.71^{+0.43}_{-0.25}}\xspace}
\newcommand{\DRttggTwoSig}{\ensuremath{[0.26,1.45]}\xspace}
\newcommand{\DRttww}{\ensuremath{0.86^{+0.42}_{-0.32}}\xspace}
\newcommand{\DRttwwTwoSig}{\ensuremath{[0.30,1.90]}\xspace}
\newcommand{\DRttzz}{\ensuremath{0.76^{+0.43}_{-0.30}}\xspace}
\newcommand{\DRttzzTwoSig}{\ensuremath{[0.26,1.80]}\xspace}
\newcommand{\DRwwbb}{\ensuremath{1.32^{+1.57}_{-0.59}}\xspace}
\newcommand{\DRwwbbTwoSig}{\ensuremath{[0.41,-]}\xspace}
\newcommand{\DRwwgg}{\ensuremath{0.83^{+0.27}_{-0.22}}\xspace}
\newcommand{\DRwwggTwoSig}{\ensuremath{[0.44,1.48]}\xspace}
\newcommand{\DRwwtt}{\ensuremath{1.15^{+0.68}_{-0.44}}\xspace}
\newcommand{\DRwwttTwoSig}{\ensuremath{[0.50,3.22]}\xspace}
\newcommand{\DRwwzz}{\ensuremath{0.88^{+0.38}_{-0.26}}\xspace}
\newcommand{\DRwwzzTwoSig}{\ensuremath{[0.44,1.78]}\xspace}
\newcommand{\DRzzbb}{\ensuremath{1.48^{+1.85}_{-0.70}}\xspace}
\newcommand{\DRzzbbTwoSig}{\ensuremath{[0.44,-]}\xspace}
\newcommand{\DRzzgg}{\ensuremath{0.92^{+0.38}_{-0.27}}\xspace}
\newcommand{\DRzzggTwoSig}{\ensuremath{[0.44,1.78]}\xspace}
\newcommand{\DRzztt}{\ensuremath{1.31^{+0.81}_{-0.48}}\xspace}
\newcommand{\DRzzttTwoSig}{\ensuremath{[0.54,3.79]}\xspace}
\newcommand{\DRzzww}{\ensuremath{1.10^{+0.44}_{-0.33}}\xspace}
\newcommand{\DRzzwwTwoSig}{\ensuremath{[0.54,2.18]}\xspace}
\newcommand{\GammaExpComb}{2.3\xspace}
\newcommand{\GammaExpHgg} {3.1\xspace}
\newcommand{\GammaExpHzz} {2.8\xspace}
\newcommand{\GammaObsComb}{1.7\xspace}
\newcommand{\GammaObsHgg} {2.4\xspace}
\newcommand{\GammaObsHzz} {3.4\xspace}
\newcommand{\ggHNonOneSign}{\ensuremath{-0.8}\xspace}
\newcommand{\ggHNonZeroExpSign}{\ensuremath{7.4}\xspace}
\newcommand{\ggHNonZeroSign}{\ensuremath{6.6}\xspace}
\newcommand{\ggHobsEightOneSig}{\ensuremath{0.79^{+0.19}_{-0.17}}\xspace}
\newcommand{\ggHobsOneSig}{\ensuremath{0.85 ^{+0.19}_{-0.16}}\xspace}
\newcommand{\ggHobsSevenOneSig}{\ensuremath{1.03^{+0.37}_{-0.33}}\xspace}
\newcommand{\kFBEST}{\ensuremath{[0.63,1.15]}\xspace}        
\newcommand{\kVBEST}{\ensuremath{[0.87,1.14]}\xspace}        
\newcommand{\lduOneSig}     {\ensuremath{0.99 ^{+0.19}_{-0.18}}\xspace}
\newcommand{\lduTwoSig}  {\ensuremath{[0.65,1.39]}\xspace}             
\newcommand{\llqOneSig}     {\ensuremath{1.03 ^{+0.23}_{-0.21}}\xspace}
\newcommand{\llqTwoSig}  {\ensuremath{[0.62,1.50]}\xspace}             
\newcommand{\LumiEight}{19.7}   
\newcommand{\LumiSeven}{5.1}    
\newcommand{\lwzONE}{\ensuremath{[0.61,1.45]}\xspace}        
\newcommand{\lwzONEOneSig}  {\ensuremath{0.94 ^{+0.22}_{-0.18}}\xspace}
\newcommand{\lwzTWO}{\ensuremath{[0.71,1.24]}\xspace}        
\newcommand{\lwzTWOOneSig}  {\ensuremath{0.92 ^{+0.14}_{-0.12}}\xspace}
\newcommand{\MASSH}{\ensuremath{125.02 \MASSHtot\GeV}\xspace}
\newcommand{\MASSHdetail}{\ensuremath{125.02\,\MASSHstat\stat\,\MASSHsyst\syst\GeV}\xspace}
\newcommand{\MASSHExpectedFullPostFit}{\ensuremath{\,^{+0.26}_{-0.25}\stat\,\pm 0.14\syst\GeV}\xspace}
\newcommand{\MASSHExpectedHybridSM}{\ensuremath{\,\pm 0.28\stat\,\pm 0.13\syst\GeV}\xspace}
\newcommand{\MASSHstat}{\ensuremath{^{+0.26}_{-0.27}}\xspace}  
\newcommand{\MASSHsyst}{\ensuremath{^{+0.14}_{-0.15}}\xspace}  
\newcommand{\MASSHtot} {\ensuremath{^{+0.29}_{-0.31}}\xspace}  
\newcommand{\mbAtmH}{\ensuremath{2.76\GeV}\xspace}
\newcommand{\mepsEOneSig}{\ensuremath{0.014 ^{+0.041}_{-0.036}}\xspace}
\newcommand{\mepsETwoSig}{\ensuremath{[-0.054,0.100]}\xspace}
\newcommand{\mepsMOneSig}{\ensuremath{245 \pm 15}\xspace}
\newcommand{\mepsMTwoSig}{\ensuremath{[217,279]}\xspace}
\newcommand{\mtopTheo}{\ensuremath{172.5\GeV}\xspace}
\newcommand{\MUggH}{\ensuremath{0.85 ^{+0.19}_{-0.16}}\xspace}
\newcommand{\MUggHdetail}{\ensuremath{0.85\,^{+0.11}_{-0.09}\stat\,^{+0.11}_{-0.08}\thy\,^{+0.10}_{-0.09}\syst}\xspace}
\newcommand{\MUHAT}{\ensuremath{1.00 ^{+0.14}_{-0.13}}\xspace}  
\newcommand{\MUHATdetail}{\ensuremath{1.00\,\pm0.09\stat\,^{+0.08}_{-0.07}\thy\,\pm0.07\syst}\xspace}
\newcommand{\mX}{\ensuremath{125.0\GeV}\xspace}
\newcommand{\Nbins}{207\xspace}
\newcommand{\NHggBgNuisances}{219\xspace}
\newcommand{\Nnuisances}{2519\xspace}
\newcommand{\pvalCouplingRatios}    {\ensuremath{0.21}\xspace}
\newcommand{\pvalCSix}    {\ensuremath{0.24}\xspace}
\newcommand{\pvalFiveDec} {\ensuremath{0.96}\xspace}
\newcommand{\pvalFourProd}{\ensuremath{0.24}\xspace}
\newcommand{\pvalResolvedCSix}    {\ensuremath{0.90}\xspace}
\newcommand{\pvalSixteen} {\ensuremath{0.84}\xspace}
\newcommand{\qqHNonOneSign}{\ensuremath{+0.4}\xspace}
\newcommand{\qqHNonZeroExpSign}{\ensuremath{3.3}\xspace}
\newcommand{\qqHNonZeroSign}{\ensuremath{3.7}\xspace}
\newcommand{\qqHobsEightOneSig}{\ensuremath{1.02^{+0.39}_{-0.36}}\xspace}
\newcommand{\qqHobsOneSig}{\ensuremath{1.16 ^{+0.37}_{-0.34}}\xspace}
\newcommand{\qqHobsSevenOneSig}{\ensuremath{1.77^{+0.99}_{-0.90}}\xspace}
\newcommand{\rankQobs}{\ensuremath{12.2}\xspace}
\newcommand{\rvfExp}{\ensuremath{1.00 ^{+0.49}_{-0.35}}\xspace}
\newcommand{\rvfObs}{\ensuremath{1.25 ^{+0.62}_{-0.44}}\xspace}
\newcommand{\rvfObsNonOneSign}{\ensuremath{+0.5}\xspace}
\newcommand{\rvfObsNonZeroExpSign}{\ensuremath{4.3}\xspace}
\newcommand{\rvfObsNonZeroSign}{\ensuremath{4.3}\xspace}
\newcommand{\SMVeV}{\ensuremath{246.22\GeV}\xspace}
\newcommand{\ttHNonOneSign}{\ensuremath{+\ttHNonOneSignVal}\xspace}
\newcommand{\ttHNonOneSignVal}{2.2}
\newcommand{\ttHNonZeroExpSign}{\ensuremath{1.2}\xspace}
\newcommand{\ttHNonZeroSign}{\ensuremath{3.5}\xspace}
\newcommand{\ttHobsEightOneSig}{\ensuremath{3.27^{+1.20}_{-1.04}}\xspace}
\newcommand{\ttHobsOneSig}{\ensuremath{2.90 ^{+1.08}_{-0.94}}\xspace}
\newcommand{\ttHobsSevenOneSig}{\ensuremath{<2.19}\xspace}
\newcommand{\ttHtagNonOneSignVal}{2.0}
\newcommand{\VHxNonOneSign}{\ensuremath{-0.2}\xspace}
\newcommand{\VHxNonZeroExpSign}{\ensuremath{2.9}\xspace}
\newcommand{\VHxNonZeroSign}{\ensuremath{2.7}\xspace}
\newcommand{\VHxobsEightOneSig}{\ensuremath{0.96^{+0.41}_{-0.39}}\xspace}
\newcommand{\VHxobsOneSig}{\ensuremath{0.92 ^{+0.38}_{-0.36}}\xspace}
\newcommand{\VHxobsSevenOneSig}{\ensuremath{<0.99}\xspace}

\title{Precise determination of the mass of the Higgs boson and tests of
compatibility of its couplings with the standard model predictions
using proton collisions at 7 and 8\TeV}

\titlerunning{Higgs mass and couplings using 7 and 8\TeV proton collisions}

\date{\today}

\abstract{
Properties of the Higgs boson with mass near 125\GeV are measured
in proton-proton collisions with the CMS experiment at the LHC.
Comprehensive sets of production and decay measurements are combined.
The decay channels include \gg, \zz, \ww, \tt, \bb, and \mumu pairs.
The data samples were collected in 2011 and 2012 and correspond to
integrated luminosities of up to 5.1\fbinv at 7\TeV and up to 19.7\fbinv
at 8\TeV.
From the high-resolution \gg and \zz channels, the mass of the Higgs boson is
determined to be \MASSHdetail.
For this mass value, the event yields obtained in the different analyses
tagging specific decay channels and production mechanisms are consistent with
those expected for the standard model Higgs boson.
The combined best-fit signal relative to the standard model
expectation is
\MUHATdetail
at the measured mass.
The couplings of the
Higgs boson are probed for deviations in magnitude from the standard model
predictions in multiple ways, including searches for invisible and undetected
decays.
No significant deviations are found.\\[3ex]
\textit{This paper is dedicated to the memory of Robert Brout
and Gerald Guralnik, whose seminal contributions helped elucidate the
mechanism for spontaneous breaking of the electroweak symmetry.}
}

\hypersetup{%
pdfauthor={CMS Collaboration},
pdftitle={Precise determination of the mass of the Higgs boson and tests of
compatibility of its couplings with the standard model predictions
using proton collisions at 7 and 8 TeV},%
pdfsubject={CMS},%
pdfkeywords={CMS, physics, Higgs, properties, mass, couplings}}

\maketitle

\section{Introduction}
\label{sec:introduction}
One of the most important objectives of
the physics programme at the CERN LHC
is to understand the mechanism behind electroweak symmetry breaking (EWSB).
In the standard model
(SM)~\cite{Glashow1961579,Weinberg19671264,Salam1968367} EWSB is
achieved by a complex scalar doublet field that leads to the prediction of one physical Higgs boson
(\PH)~\cite{Englert:1964et,Higgs:1964ia,Higgs:1964pj,Guralnik:1964eu,Higgs:1966ev,Kibble:1967sv}.
Through Yukawa interactions, the Higgs scalar field can also
account for fermion masses~\cite{Nambu:1961fr,NambuNobel,GellMann:1960np}.

In 2012 the ATLAS and CMS Collaborations at the LHC reported
the observation of a new boson with mass near
125\GeV~\cite{ATLASObservation2012,CMSObservation2012,CMSLong2013}, a value
confirmed in later
measurements~\cite{CMSHzzLegacyRun1,ATLASMassLegacyRun1,CMSHggLegacyRun1}.
Subsequent studies of the production and decay
rates~\cite{CMSttHbb7TeV,ATLASDiboson2013,CMSHbbLegacyRun1,CMSHwwLegacyRun1,
CMSHzzLegacyRun1,CMSHttLegacyRun1,CMSHggLegacyRun1,CMSHzg2013,ATLASHzg2014,
ATLASInvisible2014,CMSOffShellWidth2014,CMSHinvLegacyRun1,CMSttHLegacyRun1,
CMSHmmLegacyRun1,ATLASHmm2014,ATLASHggFiducial2014,ATLASHzzFiducial2014,
ATLASHzzCouplings2014,ATLASHggCouplings2014,ATLASHggttH2014,ATLASVHbbLegacyRun1,
ATLASHwwLegacyRun1}
and of the spin-parity quantum
numbers~\cite{CMSMassParity2012,ATLASSpin2013,CMSHwwLegacyRun1,CMSHzzLegacyRun1,
CMSAnomalousHVV2014}
of the new boson show that its properties are compatible with
those expected for the SM Higgs boson.
The CDF and D0 experiments have also reported an excess of events consistent
with the LHC observations~\cite{TevatronJul2012,TevatronLegacy}.

Standard model predictions have improved with
time, and the results presented in this paper
make use of a large number of theory tools and
calculations~\cite{Georgi:1977gs,Djouadi:1991tka,Dawson:1990zj,
Spira:1995rr,Anastasiou:2002yz,Ravindran:2003um,Catani:2003zt,Aglietti:2004nj,
Degrassi:2004mx,Actis:2008ug,Anastasiou:2008tj,deFlorian:2009hc,Baglio:2010ae,
HqT2,deFlorian:2011xf,Passarino:2010qk,Anastasiou:2011pi,deFlorian_2012yg,
Spira:1995mt,FeHiPro2,deFlorian:2012mx,Grazzini:2013mca,Stewart:2011cf,Kauer:2012hd,
Kauer:2012ma,ggww,Goria:2011wa,Passarino:2012ri,Cahn:1983ip,Altarelli:1987ue,Han:1992hr,
Ciccolini:2007jr,Ciccolini:2007ec,Denner:2011rn,Figy:2003nv,Rainwater:1999sd,
Rainwater:1997dg,Rainwater:1998kj,Jager:2006zc,Bolzoni:2010xr,Glashow:1978ab,
Harlander:2002wh,Harlander:2002vv,Harlander:2003ai,HAWK3,Ciccolini:2003jy,
Ferrera:2011bk,Han:1991ia,Brein:2003wg,Hamberg:1990np,Bredenstein:2010rs,
Maltoni:2002jr,Raitio:1978pt,Ng:1983jm,Kunszt:1984ri,Beenakker:2001rj,
Beenakker:2002nc,Dawson:2002tg,Dawson:2003zu,Garzelli:2011vp,Czakon:2013goa,
Campbell:2012dh,Garzelli:2012bn,Kidonakis:2012rm,Melnikov:2011ta,Bredenstein:2006rh,
Bredenstein:2006ha,Denner_2011mq,HDECAY,Actis:2008ts,cite:higgsDalitz,Cahn:1978nz,
Gainer:2011aaa,massRecoCollinearApprox,MCFM,MCFMHiggsProduction,MCFMVVProduction,
Campbell:2011cu,Maltoni:2002qb,MG4,MG5,MCatNLO,GEANT4-1,GEANT4-2,herwigpp,herwig,
pythia64,Alioli:2008gx,powhegVBF,powhegGF,
powhegZZ,Alioli:2010xa,Hamilton:2012np,Bagnaschi:2011tu,Re:2010bp,Alioli:2009je,
powheg1,powheg2,powheg3,powheg-Zjj,Gleisberg:2008ta,ALPGEN,tauola,phantom,
Arnold:2008rz,Guzzi:2011sv,CTEQ10,Ball:2010de,mstw,Botje:2011sn,Alekhin:2011sk,
WgammaXsec,Gavin:2010az,Li:2012wna,Gavin:2012sy,Cacciari:2008zb,Kidonakis:2012db,
Martin:2012xc,Dixon:2013haa,HqT1,Dixon:2003yb,tune_P0,tune_proq20,tune_z2,antikt},
summarized in Refs.~\cite{LHCHXSWG1,LHCHXSWG2,LHCHXSWG3}.
In proton-proton (pp) collisions at $\sqrt{s}=\text{7--8}\TeV$,
the gluon-gluon fusion Higgs boson production mode
(\ggh) has the largest cross section.
It is followed by vector boson fusion (\vbf),
associated \wh and \zh production (\vh),
and production in association with a top quark pair (\tth).
The cross section values for the Higgs boson production modes
and the values for the decay branching fractions,
together with their uncertainties, are tabulated in
Ref.~\cite{LHCHXSWG3} and regular online updates.
For a Higgs boson mass of 125\GeV, the total production cross section is
expected to be 17.5\unit{pb} at $\sqrt{s}=7\TeV$ and 22.3\unit{pb} at 8\TeV,
and varies with the mass at a rate of about $-1.6\%$ per \GeVns.

This paper presents results from a comprehensive analysis combining the CMS
measurements of the properties of the Higgs boson targeting its decay to
\bb~\cite{CMSHbbLegacyRun1},
\ww~\cite{CMSHwwLegacyRun1},
\zz~\cite{CMSHzzLegacyRun1},
\tt~\cite{CMSHttLegacyRun1},
\gg~\cite{CMSHggLegacyRun1},
and
\mumu~\cite{CMSHmmLegacyRun1}
as well as measurements of the \tth production mode~\cite{CMSttHLegacyRun1}
and
searches for invisible decays of the Higgs boson~\cite{CMSHinvLegacyRun1}.
For simplicity, \bb is used to denote \bbbar,
\tt to denote $\Pgt^+\Pgt^-$, etc.
Similarly, \zz is used to denote
$\zz^{(\ast)}$ and \ww to denote $\ww^{(\ast)}$.
The broad complementarity of measurements targeting different production
and decay modes enables a variety of
studies of the couplings of the new boson to be performed.

The different analyses have different sensitivities to the presence of the SM
Higgs boson.
The \hgg and \hzzllll (where $\ell =
\Pe,\PGm$) channels play a special role because of their high sensitivity
and excellent mass resolution of the reconstructed diphoton and four-lepton
final states, respectively.
The \hwwlnln measurement has a high sensitivity due to large expected yields
but relatively poor mass resolution because of the presence of neutrinos in the
final state.
The \bb and \tt decay modes are beset by large
background contributions and have relatively poor mass resolution,
resulting in lower sensitivity compared to the other channels;
combining the results from \bb and \tt, the CMS Collaboration has published
evidence for the decay of the Higgs boson to
fermions~\cite{CMSFermionCombo2013}.
In the SM the \ggh process is dominated by a virtual top quark loop.
However, the direct coupling of top quarks to the Higgs boson can be probed
through the study of events tagged as having been produced via the \tth
process.

The mass of the Higgs boson is determined by combining the
measurements performed in the \hgg and \hzzllll
channels~\cite{CMSHggLegacyRun1,CMSHzzLegacyRun1}.
The SM Higgs boson is predicted to have even
parity, zero electric charge, and zero spin.
All its other properties can be derived if the boson's mass is specified.
To investigate the couplings of the Higgs boson to SM particles,
we perform a combined analysis of all measurements to extract
ratios between the observed coupling strengths and those predicted by the SM.

The couplings of the
Higgs boson are probed for deviations in magnitude using the formalism
recommended by the LHC Higgs Cross Section Working Group in
Ref.~\cite{LHCHXSWG3}.
This formalism assumes, among other things, that the
observed state has quantum numbers $\JPC=0^{++}$ and that the narrow-width
approximation holds, leading to a factorization of the couplings in the production and decay
of the boson.

The data sets were processed with updated alignment and calibrations of the
CMS detector and correspond to
integrated luminosities of up to \LumiSeven\fbinv at $\sqrt{s}=7\TeV$ and
\LumiEight\fbinv at 8\TeV
for pp collisions collected in 2011 and 2012.
The central feature of the CMS detector is a
13\unit{m} long superconducting solenoid of 6\unit{m} internal diameter that
generates a uniform 3.8\unit{T} magnetic field parallel to the direction of the LHC beams.
Within the solenoid volume are a silicon pixel and strip
tracker, a lead tungstate crystal electromagnetic calorimeter, and a
brass and scintillator hadron calorimeter.
Muons are identified and measured in gas-ionization detectors embedded in the steel
magnetic flux-return yoke of the solenoid. The detector is subdivided into a cylindrical barrel
and two endcap disks.
Calorimeters on either side of the detector complement the
coverage provided by the barrel and endcap detectors.
A more detailed description of the CMS detector, together with a definition of
the coordinate system used and the relevant kinematic variables, can be found
in Ref.~\cite{CMSJinst2008}.

This paper is structured as follows: Section~\ref{sec:analyses} summarizes the
analyses contributing to the combined measurements.
Section~\ref{sec:method} describes the statistical method used to extract the
properties of the boson; some expected differences between the results of the
 combined analysis and those of the individual analyses are also explained.
The results of the combined analysis are reported in the following four
sections.
A precise determination of the mass of the boson and direct limits on its width
are presented in Section~\ref{sec:masswidth}.
We then discuss the significance of the observed excesses of events in
Section~\ref{sec:significance}.
Finally, Sections~\ref{sec:deviations} and \ref{sec:kappas} present multiple
evaluations of the compatibility of the data with the SM expectations for the
magnitude of the Higgs boson's couplings.

\section{Inputs to the combined analysis}
\label{sec:analyses}

Table~\ref{tab:channels} provides an overview of all inputs used in this
combined analysis, including the following information:
the final states selected, the production and decay modes targeted in the
analyses, the integrated luminosity used, the expected mass resolution,
and the number of event categories in each channel.

Both Table~\ref{tab:channels} and the descriptions of the different inputs
make use of the following notation.
The expected relative mass resolution, $\sigma_{\mH}/\mH$, is estimated using
different $\sigma_{\mH}$ calculations: the \hgg, \hzzllll, \hwwlnln, and \hmm
analyses quote $\sigma_{\mH}$ as half of the width of the shortest interval
containing $68.3\%$ of the signal events,
the \htt analysis quotes the RMS of the signal distribution, and
the analysis of \vh with \hbb quotes the standard deviation of the
Gaussian core of a function that also describes non-Gaussian tails.
Regarding leptons, $\ell$ denotes an electron or a muon,
\tauh denotes a $\Pgt$ lepton identified via its decay into hadrons,
and $L$ denotes any charged lepton.
Regarding lepton pairs,
SF (DF) denotes same-flavour (different-flavour) pairs
and
SS (OS) denotes same-sign (opposite-sign) pairs.
Concerning reconstructed jets,
CJV denotes a central jet veto,
\pT is the magnitude of the transverse momentum vector,
\MET refers to the magnitude of the missing transverse momentum vector,
$\mathrm{j}$ stands for a reconstructed jet,
and $\PQb$ denotes a jet tagged as originating from the hadronization of a
bottom quark.

\begin{table*}[tp]
\centering
\topcaption{
\small
Summary of the channels in the analyses included in this
combination.
The first and second columns indicate which decay mode and production
mechanism is targeted by an analysis.
Notes on the expected composition of the signal are given in the third column.
Where available, the fourth column specifies the expected relative mass
resolution for the SM Higgs boson.
Finally, the last columns provide the number of event categories and
the integrated luminosity for the 7 and 8\TeV data sets.
The notation is explained in the text.
}
\label{tab:channels}
\resizebox{0.95\textwidth}{!}{
\begin{tabular}{cl>{\small}l>{\small}r>{\small}c>{\small}c}
\hline
\multicolumn{2}{c}{\multirow{2}{*}{Decay tag and production tag}} &
\multirow{2}{*}{Expected signal composition} &
\multirow{2}{*}{$\sigma_{\mH}/\mH$} &
\multicolumn{2}{c}{\begin{tabular}[c]{@{}c@{}}\textit{Luminosity (\fbinv)}\\No. of categories\end{tabular}} \\ \cline{5-6}
\multicolumn{2}{c}{} &  &  & 7\TeV & 8\TeV \\ \hline

\rowcolor[HTML]{EFEFEF}
\multicolumn{1}{l}{\cellcolor[HTML]{EFEFEF}\hgg \cite{CMSHggLegacyRun1}, Section~\ref{sec:hgg}} & \multicolumn{3}{l}{\cellcolor[HTML]{EFEFEF}} & \textit{5.1} & \textit{19.7} \\
 & Untagged & 76--93\% \ggh & 0.8--2.1\% & 4 & 5 \\
 & \njet[2] \vbf & 50--80\% \vbf & 1.0--1.3\% & 2 & 3 \\
 & Leptonic \vh & ${\approx}95\%$ \vh ($\wh/\zh\approx5$) & 1.3\% & 2 & 2 \\
 & \MET \vh & 70--80\% \vh ($\wh/\zh\approx1$) & 1.3\% & 1 & 1 \\
  & \njet[2] \vh & ${\approx}65\%$ \vh ($\wh/\zh\approx5$) & 1.0--1.3\% & 1 & 1 \\
 & Leptonic \tth & ${\approx}95\%$ \tth & 1.1\% & \multirow{2}{*}{$1^{\dagger}$} & 1 \\
\multirow{-7}{*}{\gg} & Multijet \tth & ${>}90\%$ \tth & 1.1\% &  & 1\\  \hline

\rowcolor[HTML]{EFEFEF}
\multicolumn{1}{l}{\cellcolor[HTML]{EFEFEF}\hzzllll \cite{CMSHzzLegacyRun1}, Section~\ref{sec:hzz}} & \multicolumn{3}{l}{\cellcolor[HTML]{EFEFEF}} & \textit{5.1} & \textit{19.7} \\
 & \njet[0/1] & ${\approx}90\%$ \ggh &  & 3 & 3 \\
\multirow{-2}{*}{$\mmmm$, $\eemm/\mmee$, $\eeee$} & \njet[2] & $42\%$ $(\vbf+\vh)$ & \multirow{-2}{*}{1.3, 1.8, 2.2\%$^{\ddagger}$} & 3 & 3 \\ \hline
\rowcolor[HTML]{EFEFEF}

\multicolumn{2}{l}{\cellcolor[HTML]{EFEFEF}\hwwlnln \cite{CMSHwwLegacyRun1}, Section~\ref{sec:hww}} & \multicolumn{2}{l}{\cellcolor[HTML]{EFEFEF}} & \textit{4.9} & \textit{19.4} \\
 & \njet[0] & 96--98\% \ggh & 16\%$^{\ddagger}$ & 2 & 2 \\
 & \njet[1] & 82--84\% \ggh & 17\%$^{\ddagger}$ & 2 & 2 \\
 & \njet[2] \vbf & 78--86\% \vbf &  & 2 & 2 \\
\multirow{-4}{*}{$\ee+\mumu$, \emu} & \njet[2] \vh & 31--40\% \vh &  & 2 & 2 \\ \cline{2-6}
$3\ell3\cPgn$ (\wh) & SF-SS, SF-OS & \multicolumn{2}{l}{${\approx}100\%$ \wh, up to 20\% \tt} & 2 & 2 \\ \cline{2-6}
$\ell\ell+\ell^{\prime}\cPgn\mathrm{jj}$ (\zh) & \eee, \eem, \mmm, \mme & \multicolumn{2}{l}{${\approx}100\%$ \zh} & 4 & 4 \\ \hline

\rowcolor[HTML]{EFEFEF}
\multicolumn{1}{l}{\cellcolor[HTML]{EFEFEF}\htt \cite{CMSHttLegacyRun1}, Section~\ref{sec:htt}} & \multicolumn{3}{l}{\cellcolor[HTML]{EFEFEF}} & \textit{4.9} & \textit{19.7} \\
 & \njet[0] & ${\approx}98\%$ \ggh & 11--14\% & 4 & 4 \\
 & \njet[1] & 70--80\% \ggh & 12--16\% & 5 & 5 \\
\multirow{-3}{*}{\teth, \tmth} & \njet[2] \vbf & 75--83\% \vbf & 13--16\% & 2 & 4
\\ \cline{2-6} & \njet[1] & 67--70\% \ggh & 10--12\% & --- & 2 \\
\multirow{-2}{*}{\thth} & \njet[2] \vbf & 80\% \vbf & 11\% & --- & 1 \\ \cline{2-6}
 & \njet[0] & ${\approx}$98\% \ggh, 23--30\% \ww & 16--20\% & 2 & 2 \\
 & \njet[1] & 75--80\% \ggh, 31--38\% \ww & 18--19\% & 2 & 2 \\
\multirow{-3}{*}{\tetm} & \njet[2] \vbf & 79--94\% \vbf, 37--45\% \ww & 14--19\% & 1 & 2 \\ \cline{2-6}
 & \njet[0] & \multicolumn{2}{l}{88--98\% \ggh }& 4 & 4 \\
 & \njet[1] & \multicolumn{2}{l}{74--78\% \ggh, ${\approx}17\%$ \ww$^{\star}$ }& 4 & 4 \\
\multirow{-3}{*}{\tete, \tmtm} & \njet[2] CJV & \multicolumn{2}{l}{${\approx}50\%$ \vbf, ${\approx}45\%$ \ggh, 17--24\% \ww$^{\star}$ }& 2 & 2 \\ \cline{2-6}
$\ell\ell+LL^{\prime}$ (\zh) & $LL^{\prime}=\thth,\tlth,\tetm$ & \multicolumn{2}{l}{${\approx}15\%$ (70\%) \ww for $LL^{\prime}=\tlth$ $(\tetm)$} & 8 & 8 \\
$\ell+\thth$ (\wh) &  & \multicolumn{2}{l}{${\approx}96\%$ \vh, $\zh/\wh\approx0.1$} & 2 & 2 \\
$\ell+\ell^{\prime}\tauh$ (\wh) &  & \multicolumn{2}{l}{$\zh/\wh\approx5\%$, 9--11\% \ww} & 2 & 4 \\ \hline

\rowcolor[HTML]{EFEFEF}
\multicolumn{2}{l}{\cellcolor[HTML]{EFEFEF}\vh production with \hbb
\cite{CMSHbbLegacyRun1}, Section~\ref{sec:vhbb}} & \multicolumn{2}{l}{\cellcolor[HTML]{EFEFEF}} & \textit{5.1} & \textit{18.9} \\
\wlnbb & $\pTV$ bins & ${\approx}$100\% \vh, 96--98\% \wh &  & 4 & 6 \\
\wtnbb & --- & 93\% \wh &  & --- & 1 \\
\zllbb & $\pTV$ bins & ${\approx}$100\% \zh &  & 4 & 4 \\
\znnbb & $\pTV$ bins & ${\approx}$100\% \vh, 62--76\% \zh & \multirow{-4}{*}{${\approx}10\%$} & 2 & 3 \\ \hline

\rowcolor[HTML]{EFEFEF}
\multicolumn{3}{l}{\cellcolor[HTML]{EFEFEF}\tth production with \hhad or \hlep \cite{CMSttHLegacyRun1}, Section~\ref{sec:tth}} & \multicolumn{1}{l}{\cellcolor[HTML]{EFEFEF}} & \textit{5.0} & \textit{$\leq$19.6} \\
 & $\ttbar$ lepton+jets & \multicolumn{2}{l}{${\approx}90\%$ \bb but ${\approx}24\%$ \ww in ${\ge}6\mathrm{j}+2\mathrm{b}$} & 7 & 7 \\
\multirow{-2}{*}{\hbb} & $\ttbar$ dilepton & \multicolumn{2}{l}{45--85\% \bb, 8--35\% \ww, 4--14\% \tt} & 2 & 3 \\ \cline{2-6}
\hthth & $\ttbar$ lepton+jets & \multicolumn{2}{l}{68--80\% \tt, 13--22\% \ww, 5--13\% \bb} & --- & 6 \\

\cline{2-6}
$2\ell$ SS &  & \multicolumn{2}{l}{$\ww/\tt\approx3$} & --- & 6 \\
$3\ell$ &  & \multicolumn{2}{l}{$\ww/\tt\approx3$} & --- & 2 \\
$4\ell$ & \multirow{-3}{*}{$\geq2$ jets, $\geq1$ b jet} & \multicolumn{2}{l}{$\ww:\tt:\zz\approx3:2:1$} & --- & 1 \\ \hline

\rowcolor[HTML]{EFEFEF}
\multicolumn{1}{l}{\cellcolor[HTML]{EFEFEF}\hinvisible \cite{CMSHinvLegacyRun1}, Section~\ref{sec:hinv}} & \multicolumn{3}{l}{\cellcolor[HTML]{EFEFEF}\textbf{}} & \textit{4.9} & \textit{$\leq$19.7} \\
\hinv & \njet[2] VBF & \multicolumn{2}{l}{${\approx}94\%$ \vbf, ${\approx}6\%$ \ggh} & --- & 1 \\
\cline{2-6}
 & \njet[0] & & & 2 & 2 \\
\multirow{-2}{*}{$\zh\to\cPZ(\ee,\mumu)\hinv$}  & \njet[1] & \multicolumn{2}{l}{\multirow{-2}{*}{${\approx}100\%$ \zh}}& 2 & 2 \\
\hline

\rowcolor[HTML]{EFEFEF}
\multicolumn{1}{l}{\cellcolor[HTML]{EFEFEF}\hmm \cite{CMSHmmLegacyRun1}, Section~\ref{sec:hmm}} & \multicolumn{3}{l}{\cellcolor[HTML]{EFEFEF}\textbf{}} & \textit{5.0} & \textit{19.7} \\
 & Untagged & 88--99\% \ggh & 1.3--2.4\% & 12 & 12 \\
 & \njet[2] \vbf & ${\approx}80\%$ \vbf & 1.9\% & 1 & 1 \\
 & \njet[2] boosted & ${\approx}50\%$ \ggh, ${\approx}50\%$ \vbf & 1.8\% & 1 & 1 \\
\multirow{-4}{*}{\mumu} & \njet[2] other & ${\approx}68\%$ \ggh, ${\approx}17\%$ \vh, ${\approx}15\%$ \vbf & 1.9\% & 1 & 1 \\
\hline

\multicolumn{6}{l}{\footnotesize$^{\dagger}$ Events fulfilling the requirements of either selection are combined into one category.}\\
\multicolumn{6}{l}{\footnotesize$^{\ddagger}$ Values for analyses dedicated to the measurement of the mass that do not use the same categories and/or observables.}\\
\multicolumn{6}{l}{\footnotesize$^{\star}$ Composition in the regions for which the ratio of signal and background $s/(s+b)>0.05$.}\\
\end{tabular}
}\end{table*}

\subsection{\texorpdfstring{\hgg}{Higgs boson decay to diphoton}}
\label{sec:hgg}

The \hgg analysis~\cite{CMSHggLegacyRun1,CMSPhotonLegacyRun1} measures a narrow
signal mass peak situated on a smoothly falling background due to events originating from prompt
nonresonant diphoton production or due to events with at least one jet
misidentified as an isolated photon.

The sample of selected events containing a photon pair
is split into mutually exclusive event categories targeting
the different Higgs boson production processes, as listed in
Table~\ref{tab:channels}.
Requiring the presence of two jets with a large rapidity gap favours events
produced by the \vbf mechanism,
while event categories designed to preferentially select \vh or \tth production
require the presence of muons, electrons, \MET, a pair of jets compatible with the
decay of a vector boson, or jets arising from the hadronization
of bottom quarks.
For 7\TeV data, only one \tth-tagged event category is used, combining the
events selected by the leptonic \tth and multijet \tth selections.
The \njet[2] \vbf-tagged categories are further split according to
a multivariate (MVA) classifier that is trained to discriminate \vbf events
from both background and \ggh events.

Fewer than 1\% of the selected events are tagged according to production mode.
The remaining ``untagged'' events are subdivided into different categories based
on the output of an MVA classifier that assigns a high
score to signal-like events and to events with a good mass resolution,
based on a combination of
i)~an event-by-event estimate of the diphoton mass resolution,
ii)~a photon identification score for each photon, and
iii)~kinematic information about the photons and the diphoton system.
The photon identification score is obtained from a separate MVA
classifier that uses shower shape information and
variables characterizing how isolated the photon candidate is
to discriminate prompt photons from those arising in jets.

The same event categories and observables are used for the mass measurement and
to search for deviations in the magnitudes of the scalar couplings of the Higgs
boson.

In each event category, the background in the signal region is estimated from a
fit to the observed diphoton mass distribution in data.
The uncertainty due to the choice of function used to describe the background is
incorporated into the statistical procedure: the likelihood maximization is
also performed for a discrete variable that selects which of the functional
forms is evaluated.
This procedure is found to have correct coverage probability and negligible bias
in extensive tests using pseudo-data extracted from fits of multiple
families of functional forms to the data.
By construction, this ``discrete profiling'' of the background functional form
leads to confidence intervals for any estimated parameter that are at least as
large as those obtained when considering any single functional form.
Uncertainty in the parameters of the background functional forms contributes
to the statistical uncertainty of the measurements.

\subsection{\texorpdfstring{\hzz}{Higgs boson decay to ZZ}}
\label{sec:hzz}

In the \hzzllll analysis~\cite{CMSHzzLegacyRun1,CMSElectronLegacyRun1},
we measure a four-lepton mass peak over a small continuum background.
To further separate signal and background, we build a discriminant, \hzzD,
using the leading-order matrix elements
for signal and background.
The value of \hzzD is calculated from the observed kinematic variables,
namely the masses of the two dilepton pairs and five angles,
which uniquely define a four-lepton configuration
in its centre-of-mass frame.

Given the different mass resolutions and
different background rates arising from jets misidentified as leptons,
the $\mmmm$, $\eemm/\mmee$, and $\eeee$ event categories are analysed separately.
A stricter dilepton mass selection is performed for the
lepton pair with invariant mass closest to the nominal $\cPZ$ boson mass.

The dominant irreducible background in this channel is due to nonresonant \zz
production with both $\cPZ$ bosons decaying to a pair of charged leptons
and is estimated from simulation.
The smaller reducible backgrounds with misidentified leptons, mainly
from the production of $\cPZ+\text{jets}$, top quark pairs, and
$\PW\cPZ+\text{jets}$, are estimated from data.

For the mass measurement an event-by-event estimator of the mass resolution is built from the
single-lepton momentum resolutions evaluated from the study of a large number of
$\JPsi\to\PGm\PGm$ and $\cPZ\to\ell\ell$ data events.
The relative mass resolution, $\sigma_{m_{4\ell}}/m_{4\ell}$, is then used
together with $m_{4\ell}$ and \hzzD to measure the mass of the boson.

To increase the sensitivity to the different production mechanisms,
the event sample is split into two categories based on jet multiplicity:
i) events with fewer than two jets and
ii) events with at least two jets.
In the first category, the four-lepton transverse momentum is used to
discriminate \vbf and \vh production from \ggh production.
In the second category, a linear discriminant, built from
the values of the invariant mass of the two leading jets and their
pseudorapidity difference,
is used to separate the \vbf and \ggh processes.

\subsection{\texorpdfstring{\hww}{Higgs boson decay to WW}}
\label{sec:hww}

In the \hww analysis~\cite{CMSHwwLegacyRun1},
we measure an excess of events with two OS leptons or three
charged leptons with a total charge of $\pm1$, moderate \MET, and up to two
jets.

{\tolerance=500
The two-lepton events are divided into eight categories, with different
background compositions and signal-to-background ratios.
The events are split
into SF and DF dilepton event categories, since
the background from Drell--Yan production ($\PQq\PQq \to \gamma^{\ast}/\cPZ^{(\ast)} \to \ell\ell$)
is much larger for SF dilepton events.
For events with no jets, the main background is due to nonresonant \ww
production.
For events with one jet, the dominant backgrounds are nonresonant \ww
production and top quark production.
The \njet[2] \vbf tag is optimized to take advantage of the \vbf
production signature and the main background is due to top quark production.
The \njet[2] \vh tag targets the decay of the vector boson into two jets,
$\mathrm{V}\to\text{jj}$.
The selection requires two centrally-produced jets with invariant mass in the range $65<\mjj<105\GeV$.
To reduce the top quark, Drell--Yan, and \ww backgrounds in all previous
categories, a selection is performed on the dilepton mass and on the angular
separation between the leptons.
All background rates, except for very small contributions from \wz, \zz,
and $\PW\Pgg$ production, are evaluated from data.
The two-dimensional distribution of events in the $(m_{\ell\ell},\mt)$ plane
is used for the measurements in the DF dilepton categories with
zero and one jets; $m_{\ell\ell}$ is the invariant mass of the dilepton and
$\mt$ is the transverse mass reconstructed from the dilepton transverse
momentum and the \MET vector.
For the DF \njet[2] \vbf tag the binned distribution of $m_{\ell\ell}$ is
used.
For the SF dilepton categories and for the \njet[2] \vh tag
channel, only the total event counts are used.

In the $3\ell3\cPgn$ channel targeting the $\wh \to \PW\PW\PW$ process, we
search for an excess of events with three leptons, electrons or muons, large
\MET, and low hadronic activity.
The dominant background is due to $\wz \to 3\ell\cPgn$ production,
which is largely reduced by requiring that all SF and OS lepton pairs have
invariant masses away from the $\cPZ$ boson mass.
The smallest angular distance between OS reconstructed
lepton tracks is the observable chosen to perform the measurement.
The background processes with jets misidentified as leptons,
\eg $\cPZ+\text{jets}$ and top quark production,
as well as the $\wz \to 3\ell\cPgn$ background, are estimated from data.
The small contribution from the $\zz \to 4\ell$ process with one of the leptons
escaping detection is estimated using simulated samples.
In the $3\ell3\cPgn$ channel, up to 20\% of the signal
events are expected to be due to \htt decays.
\par}

In the $3\ell\cPgn\text{jj}$ channel, targeting the
$\zh\to\cPZ+\PW\PW\to\ell\ell+\ell^{\prime}\cPgn\text{jj}$ process, we first
identify the leptonic decay of the $\cPZ$ boson and then require the dijet system to satisfy $|\mjj-m_{\PW}|\leq60\GeV$.
The transverse mass of the $\ell\cPgn\text{jj}$ system is the observable chosen
to perform the measurement.
The main backgrounds are due to the production of \wz, \zz, and tribosons, as
well as processes involving nonprompt leptons.
The first three are estimated from simulated samples, while the last one is evaluated from data.

{\tolerance=500
Finally, a dedicated analysis for the measurement of the boson mass
 is performed in the \njet[0] and \njet[1] categories in the
\emu channel, employing observables that are extensively used in searches for
supersymmetric particles.
A resolution of 16--17\% for $\mH=125\GeV$ has been achieved.
\par}

\subsection{\texorpdfstring{\htt}{Higgs boson decay to tau leptons}}
\label{sec:htt}

The \htt analysis~\cite{CMSHttLegacyRun1} measures an excess of events over the
SM background expectation using multiple final-state signatures.
For the \tetm, \teth, \tmth, and \thth final states, where electrons and muons
arise from leptonic $\Pgt$ decays,
the event samples are further divided into categories based
on the number of reconstructed jets in the event: 0 jets, 1 jet, or 2 jets.
The \njet[0] and \njet[1] categories are further subdivided according to the
reconstructed $\pT$ of the leptons.
The \njet[2] categories require a \vbf-like topology and are subdivided according to
selection criteria applied to the dijet kinematic properties.
In each of these categories, we search for a broad excess in the reconstructed \tt mass distribution.
The \njet[0] category is used to constrain background normalizations, identification efficiencies, and energy scales.
Various control samples in data are used to evaluate
the main irreducible background from $\cPZ \to \tt$ production
and the largest reducible backgrounds from $\PW+\text{jets}$ and multijet production.
The \tete and \tmtm final states are similarly subdivided into jet categories as
above, but the search is
performed on the combination of two MVA discriminants.
The first is trained to distinguish $\cPZ \to \ell\ell$ events from $\cPZ \to \tt$ events
while the second is trained to separate $\cPZ \to \tt$ events from \htt events.
The expected SM Higgs boson signal in the \tetm, \tete, and \tmtm categories
has a sizeable contribution from \hww decays: 17--24\% in the \tete and \tmtm
event categories, and 23--45\% in the \tetm categories, as shown in
Table~\ref{tab:channels}.

The search for \tt decays of Higgs bosons produced in association with a $\PW$
or $\cPZ$ boson is conducted in events where the vector bosons
are identified through the $\PW\to\ell\Pgn$ or $\cPZ\to\ell\ell$ decay modes.
The analysis targeting \wh production selects events that have electrons or muons
and one or two hadronically decaying
tau leptons: $\Pgm+\tmth$, $\Pe+\tmth$ or $\Pgm+\teth$, $\Pgm+\thth$, and $\Pe+\thth$.
The 
analysis targeting \zh production selects events
with an identified $\cPZ \to \ell\ell$ decay
and a Higgs boson candidate decaying to \tetm, \teth, \tmth, or \thth.
The main irreducible backgrounds to the \wh and \zh searches are \wz and \zz diboson events, respectively.
The irreducible backgrounds are estimated using simulated event samples
corrected by measurements from control samples in data.
The reducible backgrounds in both analyses are due to the production of $\PW$
bosons, $\cPZ$ bosons, or top quark pairs with at least one jet misidentified as
an isolated $\Pe$, $\Pgm$, or $\tauh$.
These backgrounds are estimated exclusively from data by measuring the probability for jets
to be misidentified as isolated leptons in background-enriched control regions,
and weighting the selected events that fail the lepton requirements with the
misidentification probability.
For the SM Higgs boson, the expected fraction of \hww events in the \zh 
analysis is 10--15\% for the $\zh\to\cPZ+\tlth$ channel and 70\% for the
$\zh\to\cPZ+\tetm$ channel, as shown in Table~\ref{tab:channels}.

\subsection{\texorpdfstring{\vh with \hbb}{Associated VH production with Higgs to bb} }
\label{sec:vhbb}

Exploiting the large expected \hbb branching fraction,
the analysis of \vh production and \hbb decay
examines the \wlnbb, \wtnbb, \zllbb, and \znnbb topologies~\cite{CMSHbbLegacyRun1}.

The Higgs boson candidate is reconstructed by requiring
two b-tagged jets.
The event sample is divided into
 categories defined by the transverse momentum of the vector boson,
 \pTV.
An MVA regression is used to estimate the true energy of the bottom quark
after being trained on reconstructed b jets in simulated \hbb events.
This regression algorithm achieves a dijet mass resolution of about 10\%
for $\mH = 125\GeV$.
The performance of the regression algorithm is checked with data,
where it is observed to improve the top quark mass scale and resolution in top
quark pair events and
to improve the \pT balance between
a $\cPZ$ boson and b jets in $\cPZ(\to\ell\ell)+\bb$ events.
Events with higher \pTV have smaller backgrounds and better dijet mass
resolution.
A cascade of MVA classifiers, trained to distinguish the signal from top quark
pairs, $\text{V}+\text{jets}$, and diboson events, is used to improve the
sensitivity in the \wlnbb, \wtnbb, and \znnbb channels.
The rates of the main backgrounds, consisting of $\text{V}+\text{jets}$ and
top quark pair events, are derived from signal-depleted data control samples.
The \wz and \zz backgrounds where $\cPZ\to\bb$,
as well as the single top quark background, are estimated from
simulated samples.
The MVA classifier output distribution is used as the final
discriminant in performing measurements.

At the time of publication of Ref.~\cite{CMSHbbLegacyRun1}, the simulation of
the \zh signal process included only $\Pq\Paq$-initiated diagrams.
Since then, a more accurate prediction of the \pTZ distribution has become available, taking
into account the contribution of the gluon-gluon initiated associated
production process \ggzh, which is included in the results presented in this
paper.
The calculation of the \ggzh contribution includes next-to-leading order (NLO)
effects~\cite{Altenkamp:2012sx,ggZH1,ggZH2,ggZH3} and is particularly important
given that the \ggzh process contributes to the most
sensitive categories of the analysis.
This treatment represents a significant improvement with respect to
Ref.~\cite{CMSHbbLegacyRun1}, as discussed in Section~\ref{sec:differences}.

\subsection{\texorpdfstring{\tth production}{Tags for Higgs boson
production in association with top quarks} }
\label{sec:tth}

Given its distinctive signature, the \tth production process can be tagged
using the decay products of the top quark pair.
The search for \tth production is performed in four main channels:
\hgg, \hbb, \hthth, and \hlep~\cite{CMSttHbb7TeV,CMSttHLegacyRun1}.
The \tth search in \hgg events is described in Section~\ref{sec:hgg};
the following focuses on the other three topologies.

In the analysis of \tth production with \hbb, two signatures for the top quark
pair decay are considered:
lepton+jets ($\ttbar\to \ell\cPgn \text{jj}\text{bb}$) and
dilepton ($\ttbar\to \ell\cPgn \ell\cPgn \text{bb}$).
In the analysis of \tth production with \hthth, the $\ttbar$ lepton+jets decay
signature is required.
In both channels, the events are further classified according to the numbers
of identified jets and b-tagged jets.
The major background is from top-quark pair production accompanied by extra
jets.
An MVA is trained to discriminate between background and signal
events using information related to reconstructed object kinematic properties, event shape, and the
discriminant output from the b-tagging algorithm.
The rates of background processes are estimated from simulated samples
 and are constrained through a
simultaneous fit to background-enriched control samples.

The analysis of \tth production with \hlep is mainly sensitive to Higgs boson
decays to \ww, \tt, and \zz, with subsequent decay to electrons and/or muons.
The selection starts by requiring the presence of at least two central jets and
at least one b jet.
It then proceeds to categorize the events
according to the number, charge, and flavour of the reconstructed leptons:
$2\ell$ SS,
$3\ell$ with a total charge of $\pm1$,
and $4\ell$.
A dedicated MVA lepton selection is used to suppress the reducible background
from nonprompt leptons, usually from the decay of b hadrons.
After the final selection,
the two main sources of background are nonprompt leptons, which is
evaluated from data, and associated production of top quark pairs and vector
bosons, which is estimated from simulated samples.
Measurements in the $4\ell$ event category are performed using the number of
reconstructed jets, $N_\text{j}$.
In the $2\ell$ SS and $3\ell$ categories, an MVA classifier is employed, which
makes use of $N_\text{j}$ as well as other kinematic and event shape variables
to discriminate between signal and background.

\subsection{Searches for Higgs boson decays into invisible particles}
\label{sec:hinv}

{\tolerance=500
The search for a Higgs boson decaying into particles that escape direct
detection, denoted as \hinv in what follows, is performed using \vbf-tagged
events and \zh-tagged events \cite{CMSHinvLegacyRun1}.
The \zh production mode is tagged via the $\cPZ\to\ell\ell$ or $\cPZ\to\bb$
decays.
For this combined analysis, only the \vbf-tagged and $\cPZ\to\ell\ell$ channels
are used; the event sample of the less sensitive $\cPZ\to\bb$ analysis overlaps with that
used in the analysis of \vh with \hbb decay described in Section~\ref{sec:vhbb}
and is not used in this combined analysis.
\par}

The \vbf-tagged event selection is performed only on the 8\TeV data and
 requires a dijet mass above 1100\GeV as well as a large
separation of the jets in pseudorapidity, $\eta$.
The \MET is required to be above 130\GeV and events with additional jets
with $\pT>30\GeV$ and a value of $\eta$ between those of the tagging jets are
rejected.
The single largest background is due to the production of $\cPZ(\nu\nu)+\text{jets}$ and
is estimated from data using a sample of events with visible $\cPZ\to\Pgm\Pgm$
decays that also satisfy the dijet selection requirements above.
To extract the results, a one bin counting experiment is performed in a
region where the expected signal-to-background ratio is 0.7, 
calculated assuming the Higgs boson is produced with the SM cross section but
decays only into invisible particles.

The event selection for \zh with $\cPZ\to\ell\ell$
rejects events with two or more jets with $\pT>30\GeV$.
The remaining events are categorized according to the
$\cPZ$ boson decay into \ee or \mumu and
the number of identified jets, zero or one.
For the 8\TeV data, the results are extracted from a two-dimensional
fit to the azimuthal angular difference between the leptons and the transverse mass of the
system composed of the dilepton and the missing transverse energy in the event.
Because of the smaller amount of data in the control samples used for modelling
the backgrounds in the signal region, the results for the 7\TeV data set are
based on a fit to the aforementioned transverse mass variable only.
For the \njet[0] categories the signal-to-background ratio
varies between 0.24 and 0.28, while for the \njet[1] categories it varies
between 0.15 and 0.18, depending on the $\cPZ$ boson decay channel and the
data set (7 or 8\TeV).
The signal-to-background ratio increases as a function of the
transverse mass variable.

The data from these searches are used for results in
Sections~\ref{sec:C2BSM} and \ref{sec:C6BSM}, where the partial widths
for invisible and/or undetected decays of the Higgs boson are
probed.

\subsection{\texorpdfstring{\hmm}{Higgs boson decay to muons}}
\label{sec:hmm}

The \hmm analysis~\cite{CMSHmmLegacyRun1} is a search in the distribution
of the dimuon invariant mass, $m_{\mumu}$, for a narrow signal peak
over a smoothly falling background dominated by Drell--Yan and top quark pair
production.
A sample of events with a pair of OS muons is
split into mutually exclusive categories of differing expected signal-to-background
ratios, based on the event topology and kinematic properties.
Events with two or more jets are assigned to \njet[2] categories, while the remaining
events are assigned to untagged categories.
The \njet[2] events are divided into three categories using
selection criteria based on the properties of the dimuon and the dijet systems:
a VBF-tagged category,
a boosted dimuon category, and
a category with the remaining \njet[2] events.
The untagged events are distributed among twelve categories based on the dimuon
\pT and the pseudorapidity of the two muons, which are directly
related to the $m_{\mumu}$ experimental resolution.

The $m_{\mumu}$ spectrum in each event category is fitted with parameterized
signal and background shapes to estimate the number of signal events, in a procedure similar to that of the
\hgg analysis, described in Section~\ref{sec:hgg}.
The uncertainty due to the choice of the functional form used to model the
background is incorporated in a different manner than in the \hgg analysis,
namely by introducing an additive systematic uncertainty in the number of
expected signal events.
This uncertainty is estimated by evaluating the bias of the signal function plus
nominal background function when fitted to pseudo-data generated from
alternative background functions.
The largest absolute value of this difference for all the
alternative background functions considered and Higgs boson mass hypotheses
between 120 and 150\GeV is taken as the systematic uncertainty and applied uniformly for
all Higgs boson mass hypotheses.
The effect of these systematic uncertainties on the final result is sizeable,
about 75\% of the overall statistical uncertainty.

The data from this analysis are used for the results in
Section~\ref{sec:mepsc5}, where the scaling of the couplings with the mass of
the involved particles is explored.

\section{Combination methodology}
\label{sec:method}

The combination of Higgs boson measurements
requires the simultaneous analysis of the data selected by all individual analyses,
accounting for all statistical uncertainties, systematic uncertainties, and
their correlations.

The overall statistical methodology used in this combination was developed
by the ATLAS and CMS Collaborations in the context of the LHC Higgs Combination Group
and is described in Refs.~\cite{LHC-HCG-Report, Chatrchyan:2012tx, CMSLong2013}.
The chosen test statistic, $q$, is based on the profile likelihood ratio
and is used to determine how signal-like or background-like the data are.
Systematic uncertainties are incorporated in the analysis
via nuisance parameters that are treated according to the frequentist paradigm.
Below we give concise definitions of statistical quantities
that we use for characterizing the outcome of the measurements.
Results presented herein are obtained using asymptotic
formulae~\cite{Cowan:2010st}, including routines available in
the \textsc{RooStats} package~\cite{RooStats}.

\subsection{Characterizing an excess of events: \texorpdfstring{\pval}{p-values} and significance}

To quantify the presence of an excess of events over the expected
background we use the test statistic where the likelihood appearing in the numerator
corresponds to the background-only hypothesis:
\begin{equation}
\label{eq:method_q0}
 q_{0}  =  - 2 \ln \frac {\mathcal{L}(\text{data} \, | \, b, \, \hat \theta_{0} ) }
                       {\mathcal{L}(\text{data} \, | \, \hat \mu \, s + b, \, \hat \theta ) }\text{, with }\hat{\mu}>0,
\end{equation}
where $s$ stands for the signal expected for the SM Higgs boson,
$\mu$ is a signal strength modifier
introduced to accommodate deviations from the SM Higgs boson predictions,
$b$ stands for backgrounds, and $\theta$ represents nuisance parameters
describing systematic uncertainties.
The value $\hat \theta_{0}$ maximizes the likelihood in
the numerator under the background-only hypothesis, $\mu=0$, while
$\hat \mu$ and $\hat \theta$ define the point at which the likelihood reaches its global maximum.

The quantity $p_0$, henceforth referred to as the local \pval, is defined as the probability,
under the background-only hypothesis,
to obtain a value of $q_0$ at least as large as that observed in data,
$q_0^\text{data}$:
\begin{equation}
p_0 = \mathrm{P}\left(q_0 \geq q_0^\text{data} \, \middle| \, b\right).
\end{equation}

The local significance $z$ of a signal-like excess is then computed
according to the one-sided Gaussian tail convention:
\begin{equation}
\label{eq:Z}
p_0  =  \int_{z}^{+\infty} \frac{1}{\sqrt{2\pi}} \exp(-x^2/2) \, \rd{}x.
\end{equation}
It is important to note that very small \pvals should be interpreted with
caution, since systematic biases and uncertainties in the underlying model are
only known to a given precision.

\subsection{Extracting signal model parameters}

Signal model parameters $a$, such as the signal strength modifier $\mu$,
are evaluated from scans of the profile likelihood ratio $q(a)$:
\begin{equation}
q(a) =  -2  \Delta \ln \mathcal{L}  = - 2  \ln \frac {\mathcal{L}(\text{data} \, | \, s(a) + b,      \, \hat \theta_{a} ) }
                      {\mathcal{L}(\text{data} \, | \, s(\hat a) + b, \, \hat \theta ) } .
\end{equation}
The parameter values $\hat a$ and $\hat \theta$ correspond to the global maximum
likelihood
and are called the best-fit set.
The post-fit model, obtained using the best-fit set, is used when deriving
expected quantities.
The post-fit model corresponds to the parametric bootstrap described in
the statistics literature and includes information gained in the fit regarding
the values of all parameters~\cite{Efron1979,Lee2005}.

The 68\% and 95\% confidence level (CL) confidence intervals for a given
parameter of interest, $a_i$, are evaluated from $q(a_i)=1.00$ and
$q(a_i)=3.84$, respectively, with all other unconstrained model parameters
treated in the same way as the nuisance parameters.
The two-dimensional (2D) 68\% and 95\%~CL confidence regions for pairs of
parameters are derived from $q(a_i, a_j) = 2.30$ and $q(a_i, a_j) = 5.99$,
respectively.
This implies that boundaries of 2D confidence regions projected on
either parameter axis are not identical to the one-dimensional (1D) confidence
interval for that parameter.
All results are given using the chosen test statistic, leading to approximate
CL confidence intervals when there are no large non-Gaussian
uncertainties~\cite{Wilks,Wald,Engle}, as is the case here.
If the best-fit value is on a physical boundary, the theoretical basis for
computing intervals in this manner is lacking.
However, we have found that for the results in this paper, the intervals in
those conditions are numerically similar to those obtained by the method of
Ref.~\cite{FeldmanCousins}.

\subsection{Grouping of channels by decay and production tags}
\label{sec:grouping}

The event samples selected by each of the different analyses are mutually
exclusive.
The selection criteria can, in many cases, define high-purity selections of
the targeted decay or production modes, as shown in Table~\ref{tab:channels}.
For example, the \tth-tagged event categories of the \hgg analysis are pure in
terms of \gg decays and are expected to contain less than 10\% of non-\tth events.
However, in some cases such purities cannot be achieved for both
production and decay modes.

Mixed production mode composition is common in \vbf-tagged event categories
where the \ggh contribution can be as high as 50\%, and in \vh tags where \wh
and \zh mixtures are common.

For decay modes, mixed composition is more marked for signatures involving
light leptons and \MET, where both the \hww and \htt decays may contribute.
This can be seen in Table~\ref{tab:channels}, where some \vh-tag analyses
targeting \hww decays have a significant contribution from \htt decays and
vice versa.
This is also the case in the \tetm channel in the \htt analysis, in particular
in the 2-jet \vbf tag categories, where the contribution from \hww decays is sizeable and concentrated at
low values of $m_{\tt}$, entailing a genuine sensitivity of these categories to
\hww decays.
On the other hand, in the \tete and \tmtm channels of the \htt analysis,
the contribution from \hww is large when integrated over the
full range of the MVA observable used, but
given that the analysis is optimized for \tt decays the
contribution from \hww is not concentrated in the regions with largest
signal-to-background ratio, and provides little added
sensitivity.

{\tolerance=500
Another case of mixed decay mode composition is present in the analyses
targeting \tth production, where the \hlep decay selection includes sizeable
contributions from \hww and \htt decays, and to a lesser extent also from \hzz decays.
The mixed composition is a consequence of designing the analysis
to have the highest possible sensitivity to the \tth production mode.
The analysis of \tth with \hthth decay has an expected signal composition that
is dominated by \htt decays, followed by \hww decays, and a smaller contribution of \hbb decays.
Finally, in the analysis of \tth with \hbb, there is an event category of the
$\text{lepton}+\text{jets}$ channel that requires six
or more jets and two b-tagged jets where the signal composition is expected to
be 58\% from \hbb decays, 24\% from \hww decays, and the remaining 18\% from
other SM decay modes; in the dilepton channel, the signal composition in the
event category requiring four or more jets and two b-tagged jets is expected to
be 45\% from \hbb decays, 35\% from \hww decays, and 14\% from \htt decays.
\par}

When results are grouped according to the decay tag, each individual category is
assigned to the decay mode group that, in the SM,
is expected to dominate the sensitivity in that channel.
In particular,
  \begin{description}
    \item [{\cmsBold$\hgg$} tagged] includes only categories from the \hgg
    analysis of Ref.~\cite{CMSHggLegacyRun1}.
    \item [{\cmsBold$\hzz$} tagged] includes only categories from the \hzz
    analysis of Ref.~\cite{CMSHzzLegacyRun1}.
    \item [{\cmsBold$\hww$} tagged] includes
    all the channels from the \hww analysis of Ref.~\cite{CMSHwwLegacyRun1} and
    the channels from the analysis of \tth with \hlep of Ref.~\cite{CMSttHLegacyRun1}.
    \item [{\cmsBold$\htt$} tagged] includes
    all the channels from the \htt analysis of Ref.~\cite{CMSHttLegacyRun1} and
    the channels from the analysis of \tth targeting $\PH\to\thth$
    of Ref.~\cite{CMSttHLegacyRun1}.
    \item [{\cmsBold$\hbb$} tagged] includes
    all the channels of the analysis of \vh with \hbb of Ref.~\cite{CMSHbbLegacyRun1} and
    the channels from the analysis of \tth targeting \hbb of Ref.~\cite{CMSttHLegacyRun1}.
    \item [{\cmsBold$\hmm$} tagged] includes only categories from the \hmm
    analysis of Ref.~\cite{CMSHmmLegacyRun1}.
  \end{description}

When results are grouped by the production tag, the same reasoning of assignment
by preponderance of composition is followed, using the information in
Table~\ref{tab:channels}.

In the combined analyses, all contributions in a given production tag or
decay mode group are considered as signal and scaled accordingly.

\subsection{Expected differences with respect to the results of input analyses}
\label{sec:differences}

The grouping of channels described in Section~\ref{sec:grouping} is among the
reasons why the results of the combination may seem to differ from those of the
individual published analyses.
In addition, the combined analysis takes into account
correlations among several sources of systematic uncertainty.
Care is taken to understand the post-fit behaviour of the parameters that are
correlated between analyses, both in terms of the post-fit parameter values and
uncertainties.
Finally, the combination is evaluated at a value of \mH
that is not the value that was used in some of the individual published
analyses, entailing changes to the expected production cross sections and
branching fractions of the SM Higgs boson.
Changes are sizeable in some cases:
\begin{itemize}

  \item In Refs.~\cite{CMSHzzLegacyRun1,CMSHwwLegacyRun1} the results for
  \hzzllll and \hwwlnln are evaluated for $\mH=125.6\GeV$,
  the mass measured in the \hzzllll analysis.
  In the present combination, the results are evaluated for $\mH=\mX$, the
  mass measured from the combined analysis of the \hgg and \hzzllll
  measurements, presented in Section~\ref{sec:mass}.
  For values of \mH in this region, the branching fractions for \hzz and \hww
  vary rapidly with \mH.
  For the change of \mH in question,
  $\mathcal{B}(\hzz,\mH=\mX)/\mathcal{B}(\hzz,\mH=125.6\GeV)=0.95$
  and
  $\mathcal{B}(\hww,\mH=\mX)/\mathcal{B}(\hww,\mH=125.6\GeV)=0.96$~\cite{LHCHXSWG3}.

\item
  The expected production cross sections for the SM Higgs boson depend on \mH.
  For the change in \mH discussed above, the total production cross sections for
  7 and 8\TeV collisions vary similarly:
  $\sigma_\text{tot}(\mH=\mX)/\sigma_\text{tot}(\mH=125.6\GeV)\sim1.01$.
  While the variation of the total production cross section is dominated by the
  \ggh production process, the variation is about
  1.005 for \vbf , around 1.016 for \vh, and around
  1.014 for \tth~\cite{LHCHXSWG3}.

  \item The \htt analysis of Ref.~\cite{CMSHttLegacyRun1} focused on
  exploring the coupling of the Higgs boson to the tau lepton.
  For this reason nearly all results in Ref.~\cite{CMSHttLegacyRun1}
  were obtained by treating the \hww contribution
  as a background, set to the SM expectation.
  In the present combined analysis, both the \htt and \hww contributions are
  considered as signal in the \tt decay tag analysis.
  This treatment leads to an increased sensitivity to the presence of a Higgs
  boson that decays into both \tt and \ww.

  \item The search for invisible Higgs decays of Ref.~\cite{CMSHinvLegacyRun1}
  includes a modest contribution to the sensitivity
  from the analysis targeting \zh production with $\cPZ\to\bb$
  decays.
  The events selected by that analysis overlap with those of the analysis of \vh
  production with \hbb decays, and are therefore not considered in this
  combination.
  Given the limited sensitivity of that search,
  the overall sensitivity to invisible decays
  is not significantly impacted.

  \item The contribution from the \ggzh process was not included in
  Ref.~\cite{CMSHbbLegacyRun1} as calculations for the cross section as a
  function of \pTZ were not available. Since then, the search for \vh
  production with \hbb has been augmented by the use of recent NLO calculations
  for the \ggzh contribution \cite{Altenkamp:2012sx,ggZH1,ggZH2,ggZH3}.
  In the \znnbb and \zllbb channels, the addition of this process leads to an
  increase of the expected signal yields by 10\% to 30\% for \pTZ around and
  above 150\GeV.
  When combined with the unchanged \wh channels, the overall
  expected sensitivity for \vh production with \hbb increases by about 10\%.

\end{itemize}

In all analyses used, the contribution from associated
production of a Higgs boson with a bottom quark pair, \bbh, is neglected; in
inclusive selections this contribution is much smaller than the
uncertainties in the gluon fusion production process, whereas in
exclusive categories it has been found that the jets associated with the bottom
quarks are so soft that the efficiency to select such events is low enough and
no sensitivity is lost.
In the future, with more data, it may be possible to devise experimental
selections that permit the study of the \bbh production mode as predicted by
the SM.

\section{Mass measurement and direct limits on the natural width}
\label{sec:masswidth}

In this section we first present a measurement of the mass of the new
boson from the combined analysis of the high-resolution \hgg and \hzzllll
channels.
We then proceed to set direct limits on its natural width.

\subsection{Mass of the observed state}
\label{sec:mass}

\begin{figure} [bht]
\centering
\includegraphics[width=0.49\textwidth]{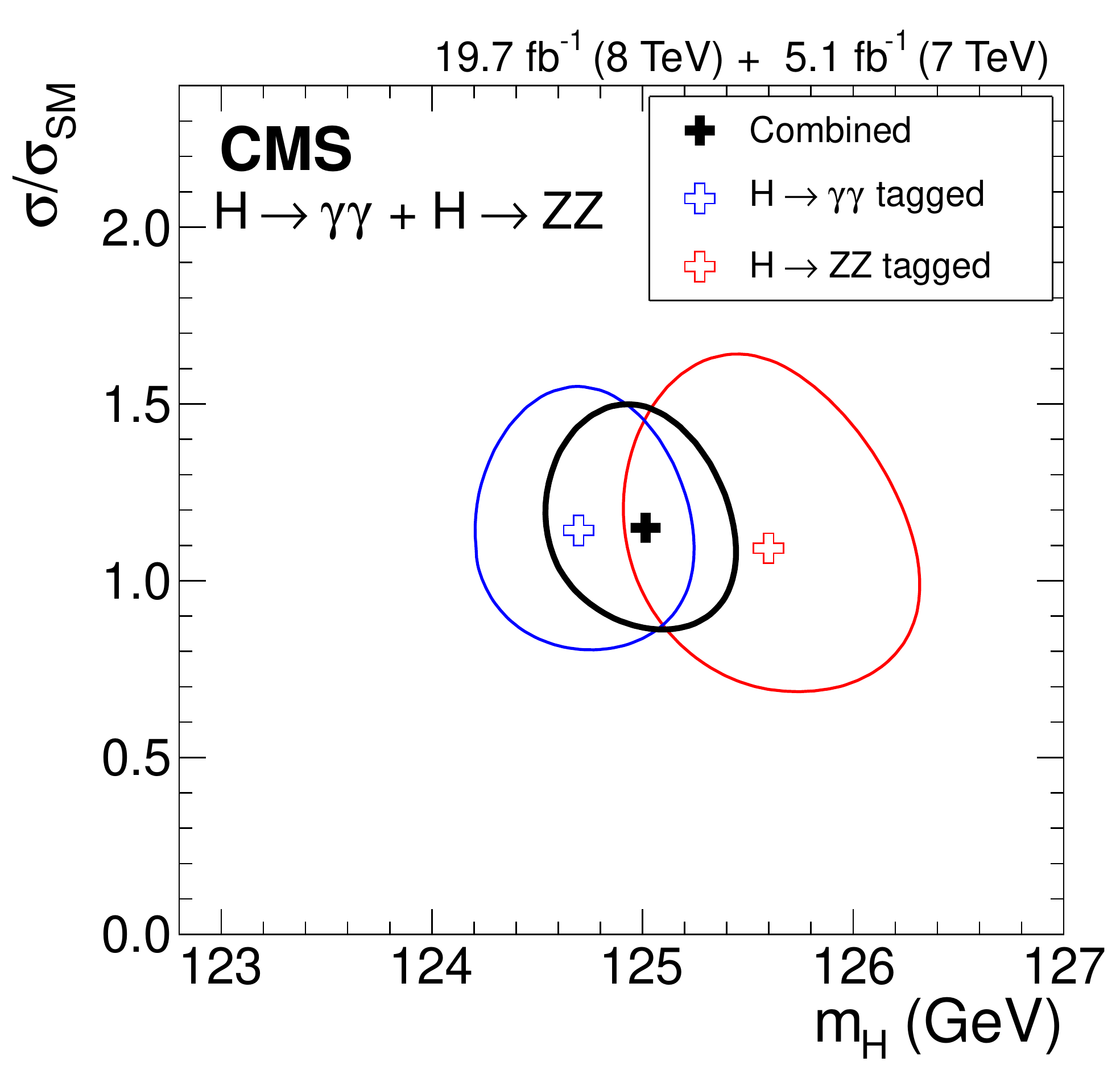}
\caption{
The 68\%~CL confidence regions for the signal strength $\sigma /
\sigma_{\text{SM}}$ versus the mass of the boson \mh for the \hgg and \hzzllll
   final states, and their combination.
The symbol $\sigma / \sigma_{\text{SM}}$ denotes the production cross section times the relevant
branching fractions, relative to the SM expectation. In this combination, the relative signal strength
for the two decay modes is set to the expectation for the SM Higgs boson.
}
\label{fig:fit_mass_2d}
\end{figure}

Figure~\ref{fig:fit_mass_2d} shows
the 68\%~CL confidence regions for two parameters of interest,
the signal strength relative to the SM expectation,
$\mu=\sigma/\sigma_\text{SM}$, and the mass, $\MX$, obtained from
the \hzzllll and \gg channels, which have excellent mass resolution.
The combined 68\%~CL confidence region, bounded by a black curve in
Fig.~\ref{fig:fit_mass_2d}, is calculated assuming the relative
event yield between the two channels as predicted by the SM, while the overall
signal strength is left as a free parameter.

\begin{figure*} [bht]
\centering
\includegraphics[width=0.49\textwidth]{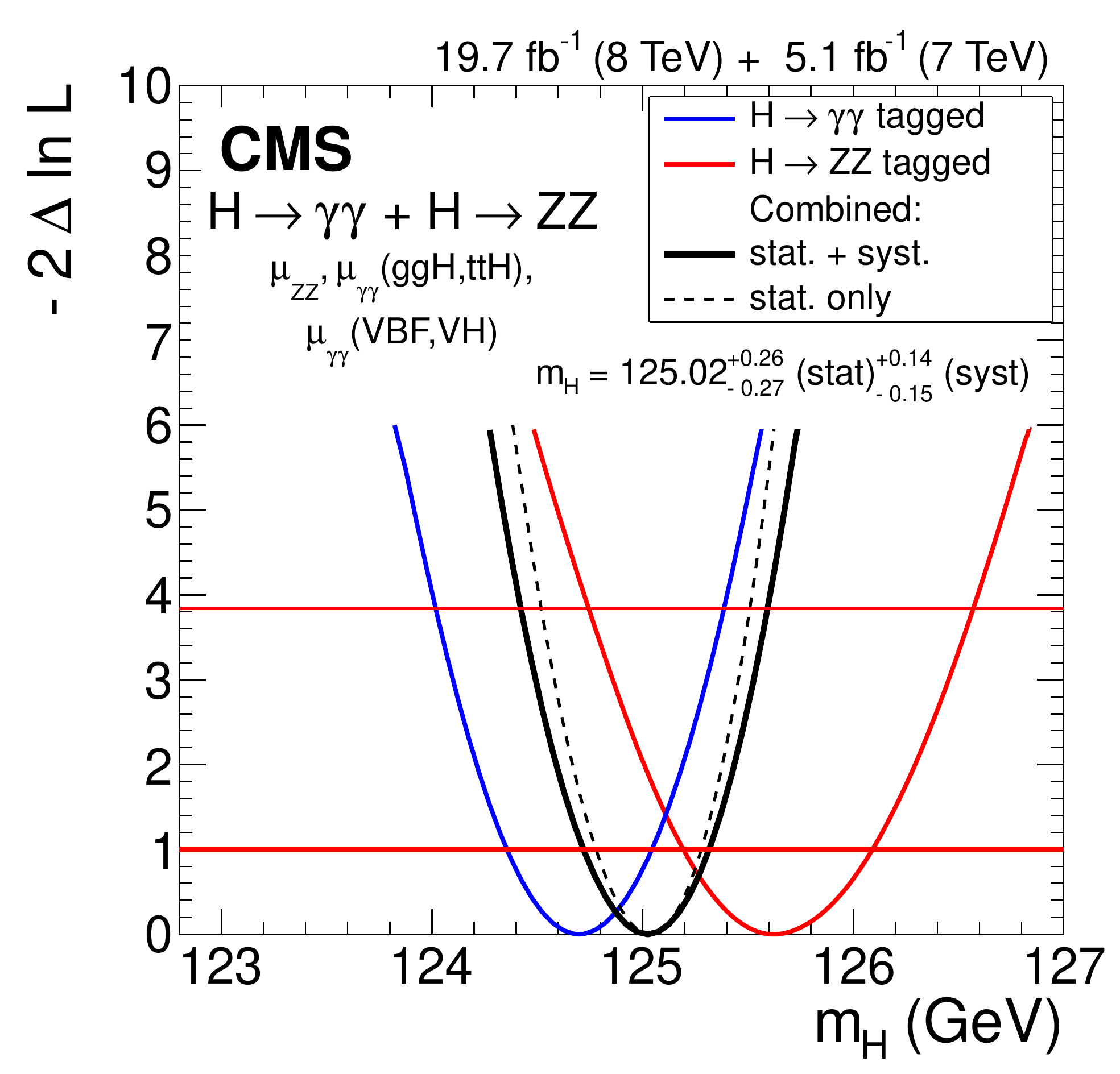}\hfill
\includegraphics[width=0.49\textwidth]{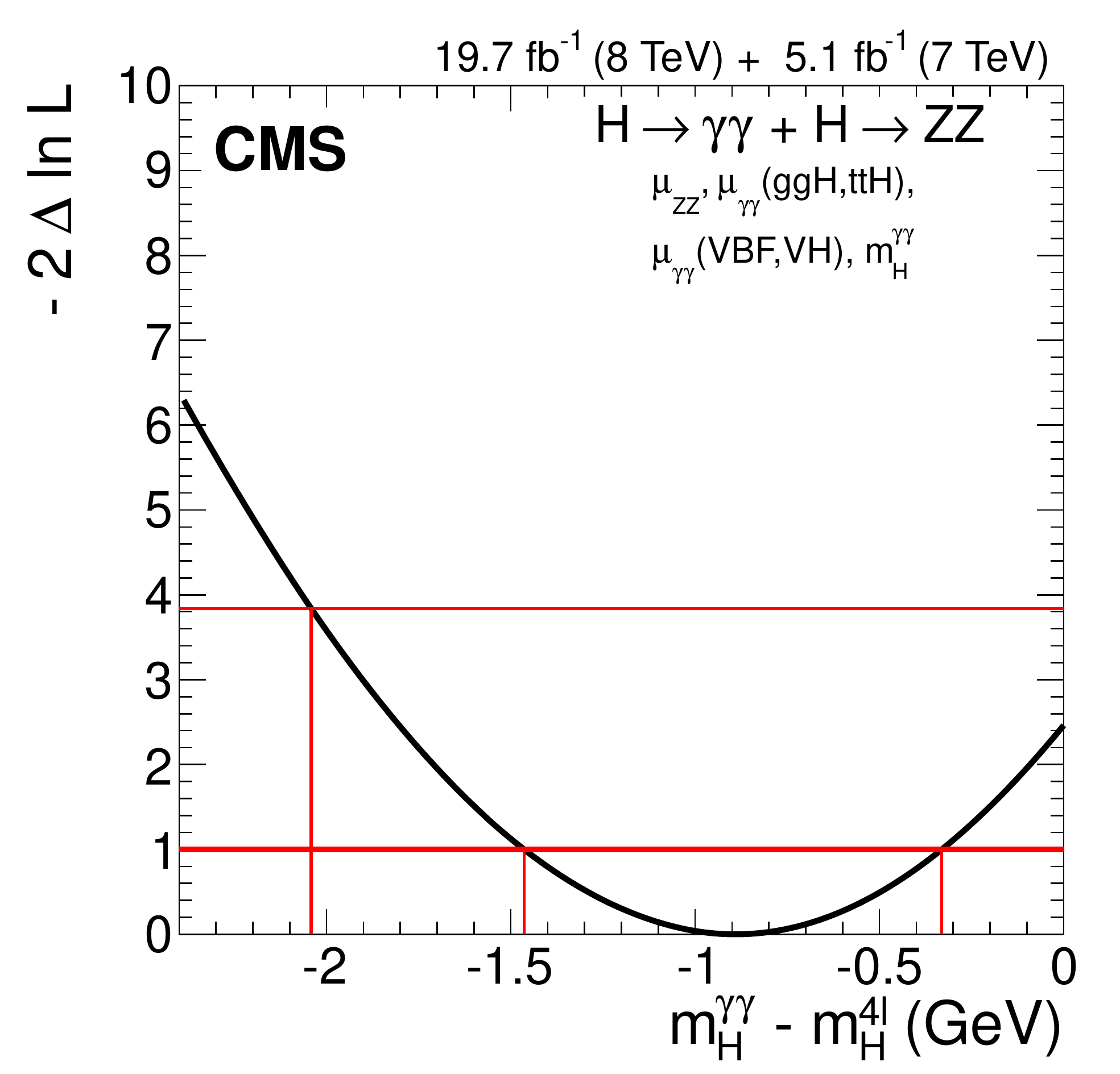}

\caption{
(Left)  Scan of the test statistic $q(\MX)=-2  \Delta \ln \mathcal{L} $
   versus the mass of the boson $\MX$
   for the \hgg and \hzzllll final states
   separately and for their combination.
   Three independent signal strengths, $(\ggh,\tth)\to
   \gg$, $(\vbf,\vh)\to \gg$, and $\text{pp}\to\hzzllll$, are profiled
   together with all other nuisance parameters.
(Right) Scan of the test statistic $q(m_{\PH}^{\gg} - m_{\PH}^{4\ell})$
versus the difference between two individual mass measurements for the same
model of signal strengths used in the left panel.
}
\label{fig:fit_mass}
\end{figure*}

To extract the value of \MX in a way that is not completely dependent on the
SM prediction for the production and decay ratios, the signal strength
modifiers for the $(\ggh,\tth)\to \gg$, $(\vbf,\vh)\to \gg$, and $\text{pp}\to\hzzllll$
processes are taken as independent, unconstrained, parameters.
The signal in all channels is assumed to be due to a single state with mass $\MX$.
The best-fit value of $\MX$ and its uncertainty are extracted from a scan of the combined
test statistic $q(\MX)$ with the three signal strength modifiers
profiled together with all other nuisance parameters; \ie the signal strength
modifiers float freely in the fits performed to scan $q(\MX)$.
Figure~\ref{fig:fit_mass}~(left) shows the scan of the test statistic
as a function of the mass $\MX$ separately for the \hgg and \hzzllll channels,
and for their combination.
The intersections of the $q(\MX)$ curves with the thick horizontal line at 1.00
and thin line at 3.84 define the 68\% and 95\%~CL confidence intervals for the
mass of the observed particle, respectively.
These intervals include both the statistical and systematic uncertainties.
The mass is measured to be \mbox{$\MX = \MASSH$}.
The less precise evaluations from
the \hww analysis~\cite{CMSHwwLegacyRun1}, $\MX=128^{+7}_{-5}\GeV$,
and from
the \htt analysis~\cite{CMSHttLegacyRun1}, $\MX=122\pm7\GeV$,
are compatible with this result.

To evaluate the statistical component of the overall uncertainty, we also perform a scan
of $q(\MX)$ fixing all nuisance parameters to their
best-fit values, except those related to the \hgg background models;
given that the \hgg background distributions are modelled from fits to data,
their degrees of freedom encode
fluctuations which are statistical in nature.
The result is shown by the dashed curve in Fig.~\ref{fig:fit_mass}~(left).
The crossings of the
dashed curve with the thick horizontal line define the
68\%~CL confidence interval for the statistical uncertainty
in the mass measurement: \MASSHstat\GeV.
We derive the systematic uncertainty assuming that the total uncertainty
is the sum in quadrature of the statistical and systematic components;
the full result is $\MX = \MASSHdetail$.
The median expected uncertainty is evaluated using an Asimov pseudo-data sample
\cite{Cowan:2010st} constructed from the best-fit values obtained when testing
for the compatibility of the mass measurement in the \hgg and \hzzllll channels.
The expected uncertainty thus derived is
\MASSHExpectedFullPostFit, in good agreement with the observation in data.
As a comparison, the median expected uncertainty is also derived by constructing
an Asimov pseudo-data sample as above except that the signal strength modifiers are
set to unity (as expected in the SM) and
$m_{\PH}^{\gg}=m_{\PH}^{4\ell}=125\GeV$, leading to an expected uncertainty of
\MASSHExpectedHybridSM.
As could be anticipated, the statistical uncertainty is slightly
larger given that the observed signal strength in the \hgg
channel is larger than unity, and the systematic uncertainty is slightly
smaller given the small mass difference between the two channels that is
observed in data.

To quantify the compatibility of
the \hgg and \hzz mass measurements with each other, we perform a scan of the
test statistic $q(m_{\PH}^{\gg} - m_{\PH}^{4\ell})$,
as a function of the difference between the two mass measurements.
Besides the three signal strength modifiers, there are two additional parameters
in this test: the mass difference and $m_{\PH}^{\gg}$.
In the scan, the three signal strengths and $m_{\PH}^{\gg}$ are profiled
together with all nuisance parameters.
The result from the scan shown in Fig.~\ref{fig:fit_mass}~(right) is
$m_{\PH}^{\gg} - m_{\PH}^{4\ell}=\DeltaMMeas$.
From evaluating $q(m_{\PH}^{\gg} - m_{\PH}^{4\ell}=0)$ it can be concluded
that the mass measurements in \hgg and \hzzllll agree at the $1.6\sigma$ level.

To assess the dependency of the result on the SM Higgs boson
hypothesis, the measurement of the mass is repeated using the same channels, but
with the following two sets of assumptions:
i) allowing a common signal strength modifier to float, which corresponds to
the result in Fig.~\ref{fig:fit_mass_2d}, and
ii) constraining the relative production cross sections and branching fractions
to the SM predictions, \ie $\mu=1$.
The results from these two alternative measurements differ by less than 0.1\GeV
from the main result,
both in terms of the best-fit value and the uncertainties.

\subsection{Direct limits on the width of the observed state}
\label{sec:total_width}

For $\mH \sim 125\GeV$ the SM Higgs boson is predicted to be narrow,
with a total width $\GSM \sim 4\MeV$.
From the study of off-shell Higgs boson production, CMS has previously set an
indirect limit on the total width, $\Gamma_{\text{tot}}/\GSM < 5.4$ $(8.0)$
observed (expected) at the 95\%~CL \cite{CMSOffShellWidth2014}.
While that result is about two orders of magnitude better than the experimental
mass resolution, it relies on assumptions on the underlying theory, such as the
absence of contributions to Higgs boson off-shell production from particles beyond the
standard model.
In contrast, a direct limit does not rely on such assumptions and is only
limited by the experimental resolution.

The best experimental mass resolution, achieved in the
\hgg and \hzzllll analyses, is typically
between 1\GeV and 3\GeV, as shown in Table~\ref{tab:channels}.
The resolution depends on the energy, rapidity, and azimuthal angle of the decay
products, and on the flavour of the leptons in the case of the \hzzllll decay.
If found inconsistent with the expected detector resolution, the total width
measured in data could suggest the production of a resonance with a greater
intrinsic width or the production of two quasi-degenerate states.

\begin{figure}[bpht]
\centering
\includegraphics[width=0.49\textwidth]{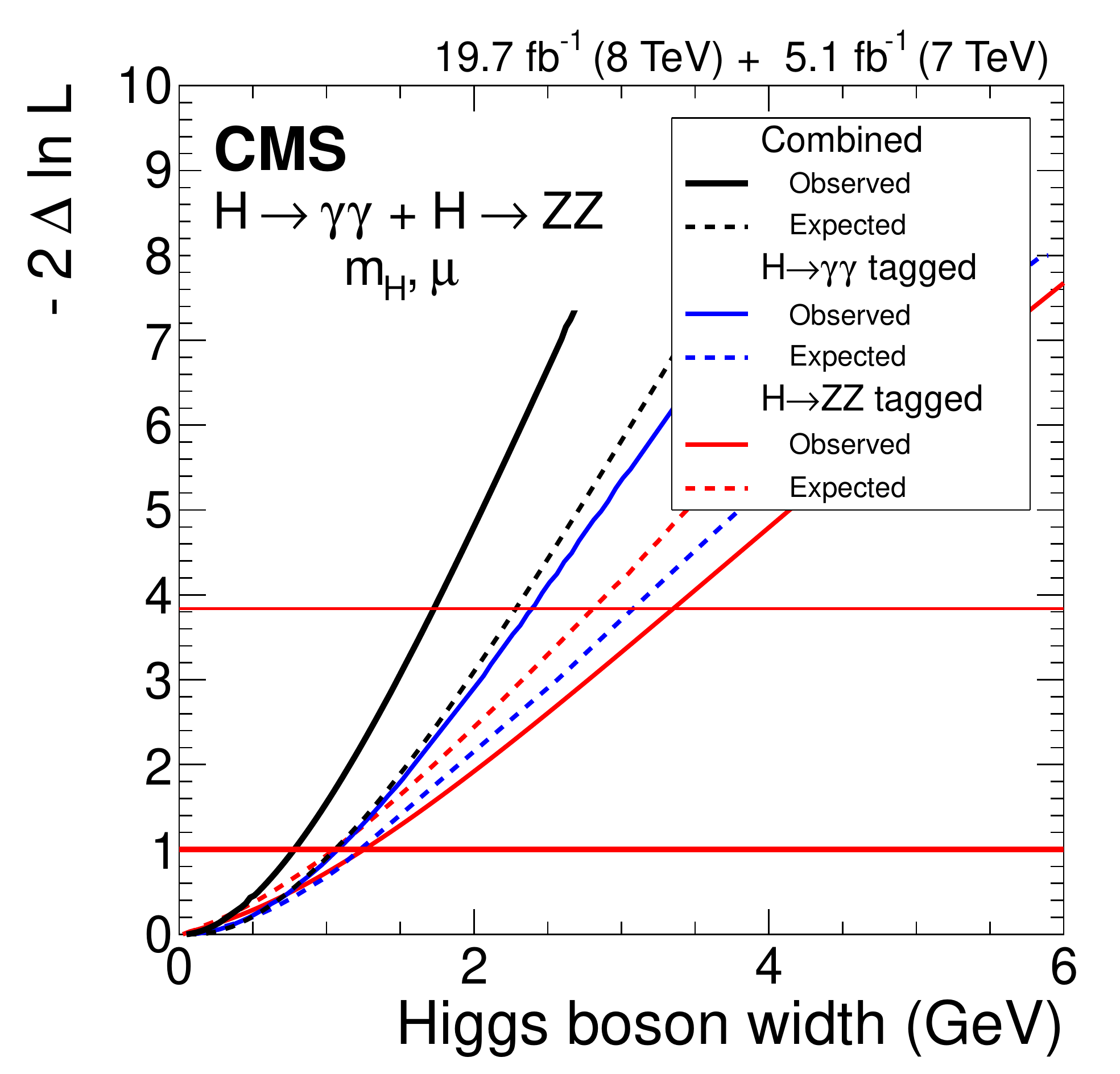}
\caption{
Likelihood scan as a function of the width of the boson.
The continuous (dashed) lines show the observed (expected) results for the \hgg
analysis, the \hzzllll analysis, and their combination.
The data are consistent with $\GSM \sim 4\MeV$ and
for the combination of the two channels the observed (expected) upper limit on
the width at the 95\%~CL is \GammaObsComb (\GammaExpComb)\GeV.}
\label{fig:width}

\end{figure}

To perform this measurement the signal models in the \hgg and \hzzllll analyses
allow for a natural width using the relativistic Breit--Wigner distribution, as
described in Refs.~\cite{CMSHzzLegacyRun1,CMSHggLegacyRun1}.
Figure~\ref{fig:width} shows the likelihood scan as a function of the
assumed natural width.
The mass of the boson and a common signal strength are
profiled along with all other nuisance parameters.
The dashed lines show the expected results for the SM Higgs boson.
For the \hgg channel
the observed (expected) upper limit at the 95\%~CL is \GammaObsHgg
(\GammaExpHgg)\GeV.
For the \hzzllll channel
the observed (expected) upper limit at the 95\%~CL is \GammaObsHzz
(\GammaExpHzz)\GeV.
For the combination of the two analyses,
the observed (expected) upper limit at the 95\%~CL is \GammaObsComb
(\GammaExpComb)\GeV.

\section{Significance of the observations in data}

\label{sec:significance}

This section provides an assessment of the significance of the observed excesses
at the best-fit mass value, $\mH=\mX$.

Table~\ref{tab:Signif} summarizes the median expected and observed local significance
for a SM Higgs boson mass of \mX
from the different decay mode tags, grouped as described in
Section~\ref{sec:grouping}.
The value of \MX is fixed to the best-fit combined measurement presented in
Section~\ref{sec:mass}.
The values of the expected significance are evaluated using the post-fit
expected background rates and the signal rates expected from the SM.
In the three diboson decay mode tags, the significance is close to, or above,
$5\sigma$.
In the \tt decay mode tag the significance is above $3\sigma$.

\begin{table}[tp]
\centering
\topcaption{
The observed and median expected significances of the excesses for each decay
mode group, assuming $\MX=\mX$.
The channels are grouped by decay mode tag as described in
Section~\ref{sec:grouping};
when there is a difference in the channels included with respect to
the published results for the individual channels, the result for the grouping used
in those publications is also given.
}
\label{tab:Signif}
\begin{tabular}{lcc}
\hline
\multirow{2}{*}{Channel grouping} & \multicolumn{2}{c}{Significance ($\sigma$)}  \\ 
 & Observed & Expected \\
\hline
\hzz tagged &  6.5  & 6.3 \\ %
\hgg tagged &  5.6  & 5.3  \\ %

\hww tagged &  4.7 &  5.4  \\ %
\quad\it Grouped as in Ref.~\cite{CMSHwwLegacyRun1}           & \it 4.3 & \it 5.4  \\ 

\htt tagged &  3.8 &  3.9  \\ %
\quad\it Grouped as in Ref.~\cite{CMSHttLegacyRun1}           & \it 3.9 & \it 3.9  \\ 

\hbb tagged &  2.0 &  2.6  \\ %
\quad\it Grouped as in Ref.~\cite{CMSHbbLegacyRun1}  & \it 2.1 & \it 2.5  \\ 
\hmm tagged &  $<0.1$ &  0.4  \\ %
\hline

\end{tabular}

\end{table}

Differences between the results in Table~\ref{tab:Signif} and the individual
publications are understood in terms of the discussion in
Sections~\ref{sec:grouping} and \ref{sec:differences}, namely the grouping of
channels by decay mode tag, the change of the \mH value at which the
significance of the \hzzllll and \hww analyses is evaluated, and the
treatment of \hww as part of the signal, instead of background, in the \htt analysis.

Finally, the observation of the \hgg and \hzzllll decay modes indicates that the
new particle is a boson, and the diphoton decay implies that its spin is
different from unity~\cite{Landau,Yang}.
Other observations, beyond the scope of this paper, disfavour spin-1 and
spin-2 hypotheses and, assuming that the boson has zero spin, are consistent
with the pure scalar hypothesis, while disfavouring the pure pseudoscalar
hypothesis~\cite{CMSHzzLegacyRun1,CMSHwwLegacyRun1,CMSAnomalousHVV2014}.

\section{Compatibility of the observed yields with the SM Higgs boson
hypothesis}
\label{sec:deviations}

The results presented in this section focus on the Higgs boson production and
decay modes, which can be factorized under the narrow-width approximation,
leading to $N_{ij}\sim\sigma_{i}\;\mathcal{B}_{j}$, where $N_{ij}$
represents the event yield for the combination of production mode $i$ and decay
mode $j$, $\sigma_{i}$ is the production cross section for production process
$i$, and $\mathcal{B}_{j}$ is the branching fraction into decay mode $j$.
Studies where the production and decay modes are interpreted in terms of
underlying couplings of particles to the Higgs boson are presented in
Section~\ref{sec:kappas}.

The size of the current data set permits many
compatibility tests between the observed excesses and the expected SM
Higgs boson signal.
These compatibility tests do not constitute measurements of any physics
parameters per se, but rather allow one to probe for deviations of the various
observations from the SM expectations.
The tests evaluate the compatibility of the data observed in the different
channels with the expectations for the SM Higgs boson with a mass equal to the
best-fit value found in Section~\ref{sec:mass}, $\MX=\mX$.

{\tolerance=500
This section is organized by increasing degree of complexity of the deviations
being probed.
In Section~\ref{sec:onemu} we assess the compatibility of the overall
signal strength for all channels combined with the SM Higgs hypothesis.
In Section~\ref{sec:mupergroups} the compatibility is assessed by
production tag group, decay tag group, and production and decay tag group.
We then turn to the study of production modes.
Using the detailed information on the expected SM Higgs production
contributions, Section~\ref{sec:rvrf} discusses, for each decay tag group, the
results of considering two signal strengths, one scaling the \ggh and \tth
contributions, and the other scaling the \vbf and \vh contributions.
Then, assuming the expected relative SM Higgs branching fractions,
Section~\ref{sec:muprod} provides a combined analysis for signal strengths
scaling the \ggh, \vbf, \vh, and \tth contributions individually.
Turning to the decay modes, Section~\ref{sec:decayratios} performs combined
analyses of signal strength ratios between different decay modes, where some
uncertainties from theory and some experimental uncertainties cancel out.
Finally, using the structure of the matrix of production and decay mode signal
strengths, Section~\ref{sec:matrixrank} tests for the possibility that the
observations are due to the presence of more than one state degenerate in mass.
\par}

\subsection{Overall signal strength}
\label{sec:onemu}

The best-fit value for the common signal strength modifier $\hat \mu = \hat
\sigma / \sigma_{\text{SM}}$, obtained from the combined analysis of all
channels, provides the simplest compatibility test.
In the formal fit, $\hat \mu$ is allowed to become negative if the observed number of events is
smaller than the expected yield for the background-only hypothesis.
The observed $\hat \mu$, assuming $\mH=\mX$, is \MUHAT,
consistent with unity, the expectation for the SM Higgs boson.
This value is shown as the vertical bands in the
three panels of Fig.~\ref{fig:muhat_compatibility}.

The total uncertainty can be broken down into
a statistical component (stat);
a component associated with the uncertainties related to renormalization and
factorization scale variations, parton distribution functions, branching
fractions, and underlying event description (theo);
and any other systematic uncertainties (syst).
The result is \MUHATdetail.
Evolution of the SM predictions may not only reduce the associated
uncertainties from theory, but also change the central value given above.

\subsection{Grouping by predominant decay mode and/or production tag}
\label{sec:mupergroups}

One step in going beyond a single signal strength modifier is to evaluate the
signal strength in groups of channels from different analyses.
The groups chosen reflect the different production tags, predominant decay
modes, or both.
Once the fits for each group are performed, a simultaneous fit to all groups is
also performed to assess the compatibility of the results with the SM Higgs boson
hypothesis.

\begin{figure*}[phtb]
\centering
\includegraphics[width=0.49\textwidth]{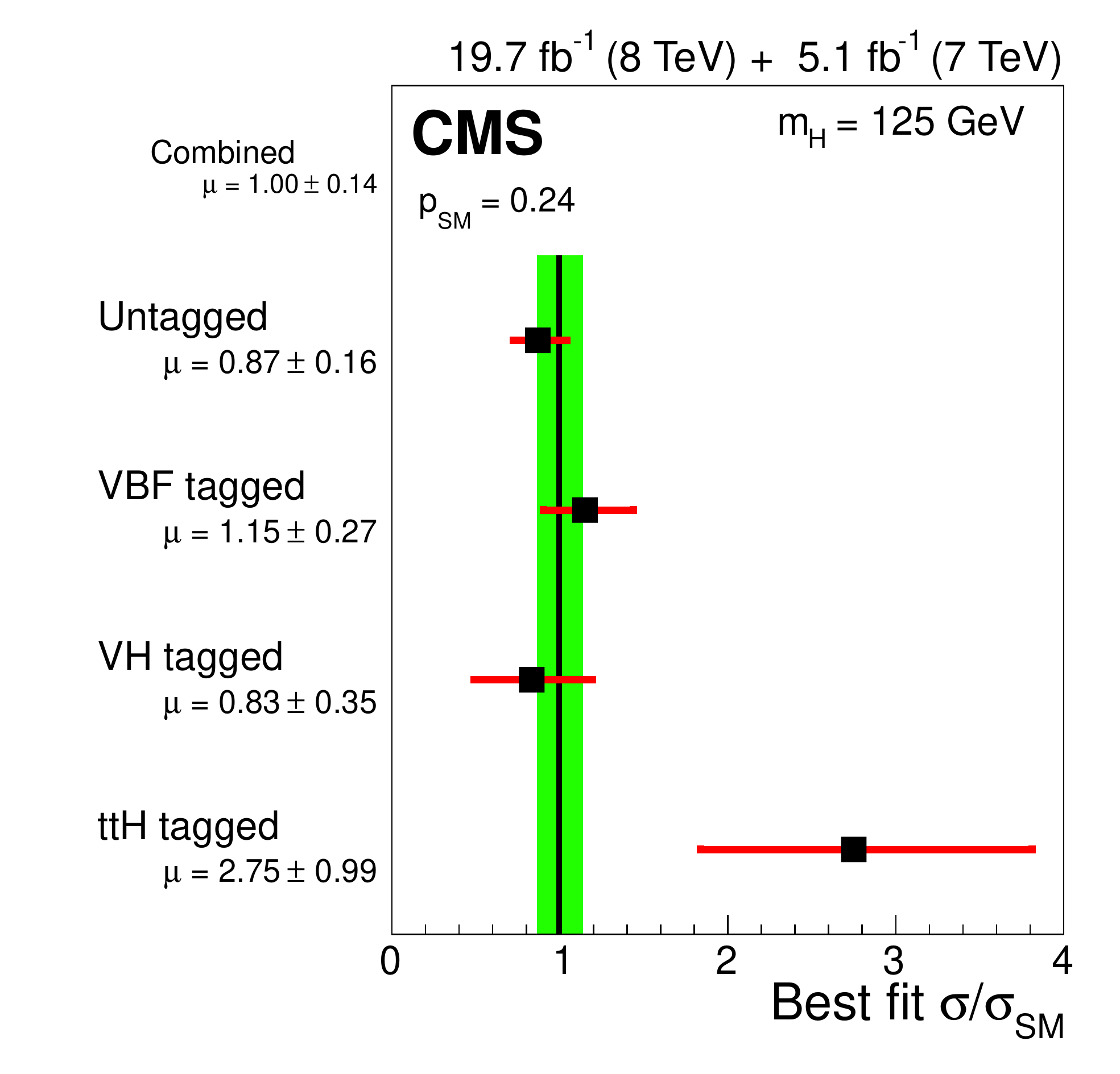}\hfill
\includegraphics[width=0.49\textwidth]{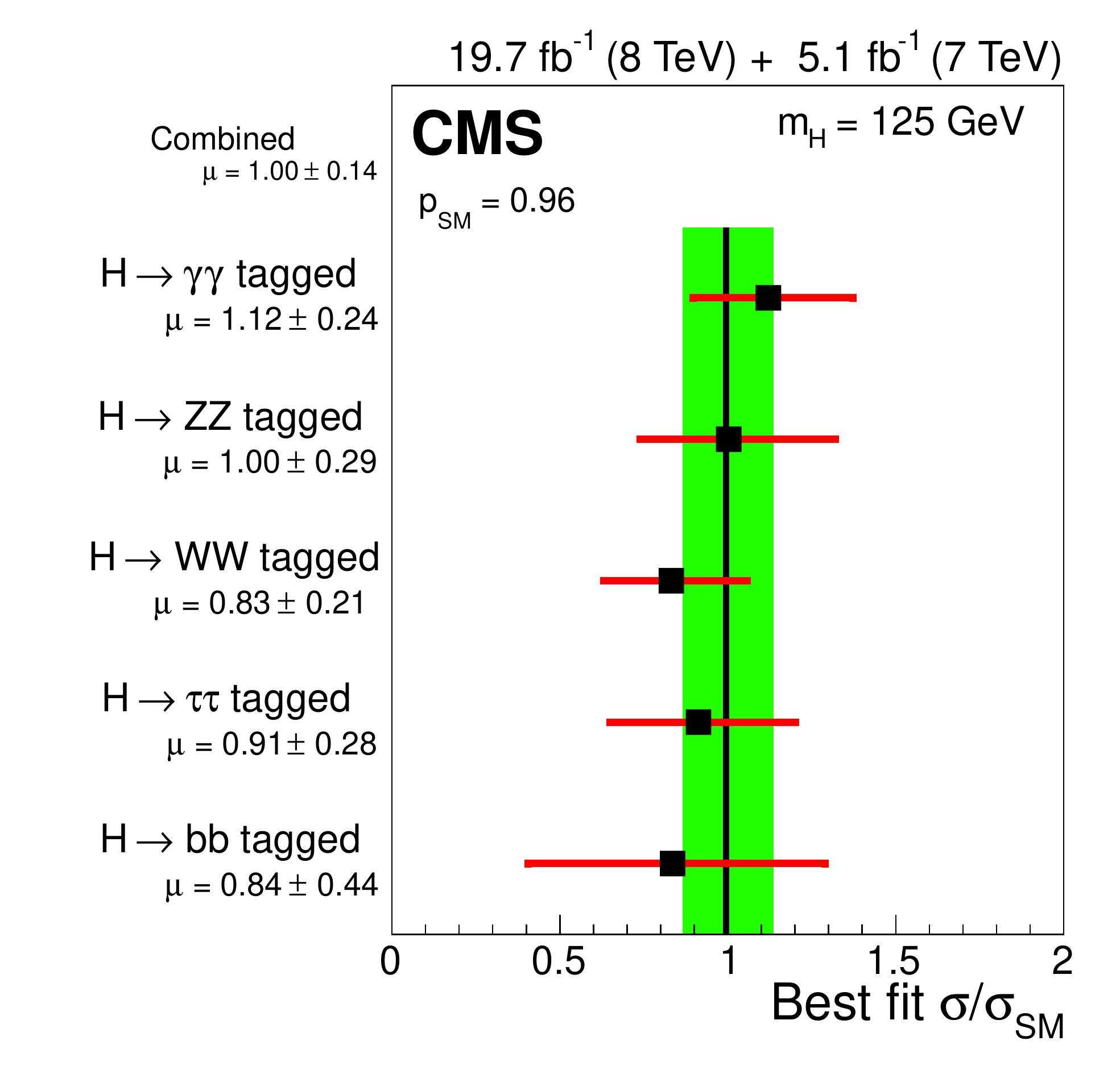}\\
\includegraphics[width=0.49\textwidth]{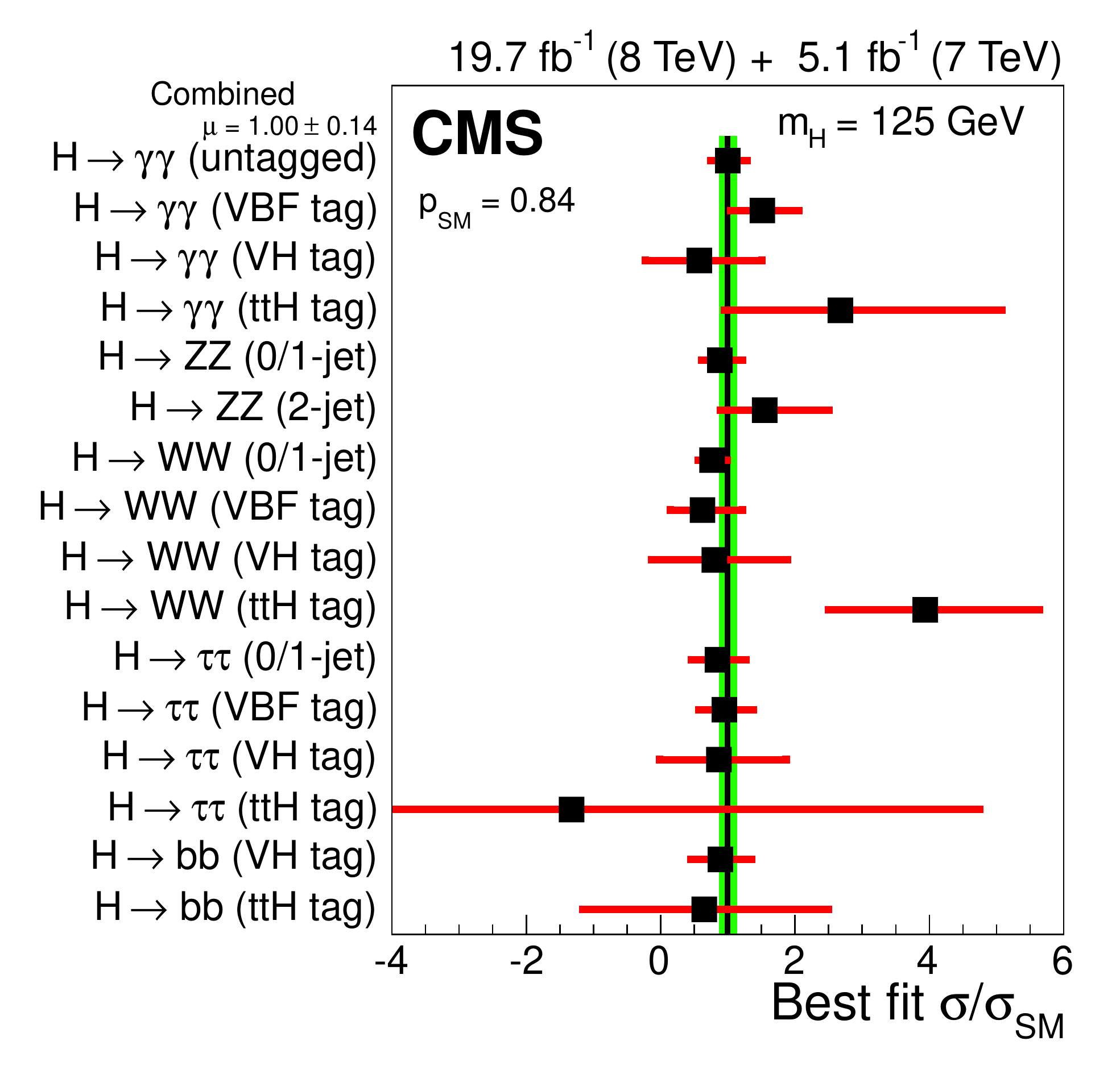}
\caption{
Values of the best-fit $\sigma / \sigma_\text{SM}$ for the overall combined
analysis (solid vertical line) and separate combinations grouped by production
mode tag, predominant decay mode, or both.
The $\sigma / \sigma_{\text{SM}}$ ratio denotes the production cross section
times the relevant branching fractions, relative to the SM expectation.
The vertical band shows the overall $\sigma / \sigma_\text{SM}$  uncertainty.
The horizontal bars indicate the $\pm 1 $ standard deviation uncertainties in
the best-fit $\sigma / \sigma_\text{SM}$ values for the individual combinations;
these bars include both statistical and systematic uncertainties.
(Top left) Combinations grouped by analysis tags targeting individual production
mechanisms; the excess in the \tth-tagged combination is largely driven by the
\tth-tagged \hgg and \hww channels as can be seen in the bottom panel.
(Top right) Combinations grouped by predominant decay mode.
(Bottom) Combinations grouped by predominant decay mode and additional tags
targeting a particular production mechanism.}
\label{fig:muhat_compatibility}
\end{figure*}

Figure~\ref{fig:muhat_compatibility} shows the $\hat \mu$ values obtained in
different independent combinations of channels for $\mH = \mX$, grouped by
additional tags targeting events from particular production mechanisms, by
predominant decay mode, or both.
As discussed in Section~\ref{sec:grouping}, the expected purities of the
different tagged samples vary substantially.
Therefore, these plots cannot be interpreted as compatibility tests for pure
production mechanisms or decay modes, which are studied in
Section~\ref{sec:muprod}.

For each type of grouping, the level of compatibility
with the SM Higgs boson cross section can be quantified by
the value of the test
statistic function of the signal strength parameters simultaneously
fitted for the $N$ channels considered in the group, $\mu_1,
\mu_2,\ldots,\mu_N$,
\begin{equation}
q_{\mu} =
-2  \Delta \ln \mathcal{L} =
-2  \ln  \frac
{ \mathcal{L}(\text{data} \, | \, \mu_{i}, \hat \theta_{\mu_{i}}) }
{\mathcal{L}(\text{data} \, | \, \hat \mu_{i}, \hat \theta)}
\end{equation}
evaluated for $\mu_1 = \mu_2 =\cdots=\mu_N = 1$.
For each type of grouping, the corresponding
$q_{\mu}(\mu_1=\mu_2=\cdots=\mu_N=1)$ from the simultaneous fit of $N$
signal strength parameters is expected to behave asymptotically as a \chisq
distribution with $N$ degrees of freedom (\dof).

{\tolerance=500
The results for the four independent combinations grouped by production mode tag
are depicted in Fig.~\ref{fig:muhat_compatibility}~(top left).
An excess can be seen for the \tth-tagged combination, due to the observations
in the \tth-tagged \hgg and \hlep analyses that can be appreciated from the
bottom panel.
The simultaneous fit of the signal strengths for each group of production
process tags results in $\chisqdof = \ChiFourProd$ and an asymptotic \pval of
\pvalFourProd, driven by the excess observed in the group of analyses
tagging the \tth production process.
\par}

{\tolerance=800
The results for the five independent combinations grouped by predominant decay
mode are shown in Fig.~\ref{fig:muhat_compatibility} (top right).
The simultaneous fit of the corresponding five signal strengths yields
$\chisqdof = \ChiFiveDec$ and an asymptotic \pval of
\pvalFiveDec.
\par}

The results for sixteen individual combinations grouped by production tag
and predominant decay mode are shown in
Fig.~\ref{fig:muhat_compatibility}~(bottom).
The simultaneous fit of the corresponding signal strengths gives a
$\chisqdof = \ChiSixteen$,
which corresponds to an asymptotic \pval of \pvalSixteen.

The \pvals above indicate that these different ways of splitting the overall
signal strength into groups related to the production mode tag, decay mode
tag, or both, all yield results compatible with the SM prediction for the Higgs
boson, $\mu=\mu_i=1$.
The result of the \tth-tagged combination is compatible with the
SM hypothesis at the $\ttHtagNonOneSignVal\sigma$ level.

\subsection{Fermion- and boson-mediated production processes and their ratio}
\label{sec:rvrf}

The four main Higgs boson production mechanisms can be associated with either
couplings of the Higgs boson to fermions (\ggh and \tth) or
vector bosons (\vbf and \vh).
Therefore, a combination of channels associated with a particular decay mode
tag, but explicitly targeting different production mechanisms,
can be used to test the relative strengths of the couplings to the vector bosons
and fermions, mainly the top quark, given its importance in \ggh
production.
The categorization of the different channels into production mode tags is
not pure.
Contributions from the different signal processes, evaluated from
Monte Carlo simulation and shown in Table~\ref{tab:channels}, are taken into
account in the fits, including theory and experimental uncertainties; the
factors used to scale the expected contributions from
the different production modes are shown in Table~\ref{tab:rvrfModel} and do
not depend on the decay mode.
For a given decay mode, identical deviations of \muv and \muf from unity may
also be due to a departure of the decay partial width from the SM expectation.

\begin{table*}[p]
\centering
\topcaption{
Parameterization used to scale the expected SM Higgs boson yields from the
different production modes when obtaining the results presented in
Table~\ref{tab:rvrf} and Fig.~\ref{fig:rvrf}~(left).
The signal strength modifiers \muf and \muv, common to all decay modes, are
associated with the \ggh and \tth and with the \vbf and \vh production
mechanisms, respectively.
}
\label{tab:rvrfModel}

\begin{tabular}{c|ccccc}
\hline
\multicolumn{6}{l}{Parameters of interest: \muf and \muv.}\\
\hline
\multicolumn{1}{c|}{\begin{tabular}[c]{@{}c@{}}Signal\\model\end{tabular}}

  &  \hgg  &  \hzz  &  \hww  &  \htt  &  \hbb \\
\hline
\ggh  & \muf & \muf & \muf & \muf & \muf \\
\vbf  & \muv & \muv & \muv & \muv & \muv \\
\vh   & \muv & \muv & \muv & \muv & \muv \\
\tth  & \muf & \muf & \muf & \muf & \muf \\

\hline
\end{tabular}

\end{table*}

\begin{figure*}[bpht]
\centering
\includegraphics[width=0.49\textwidth]{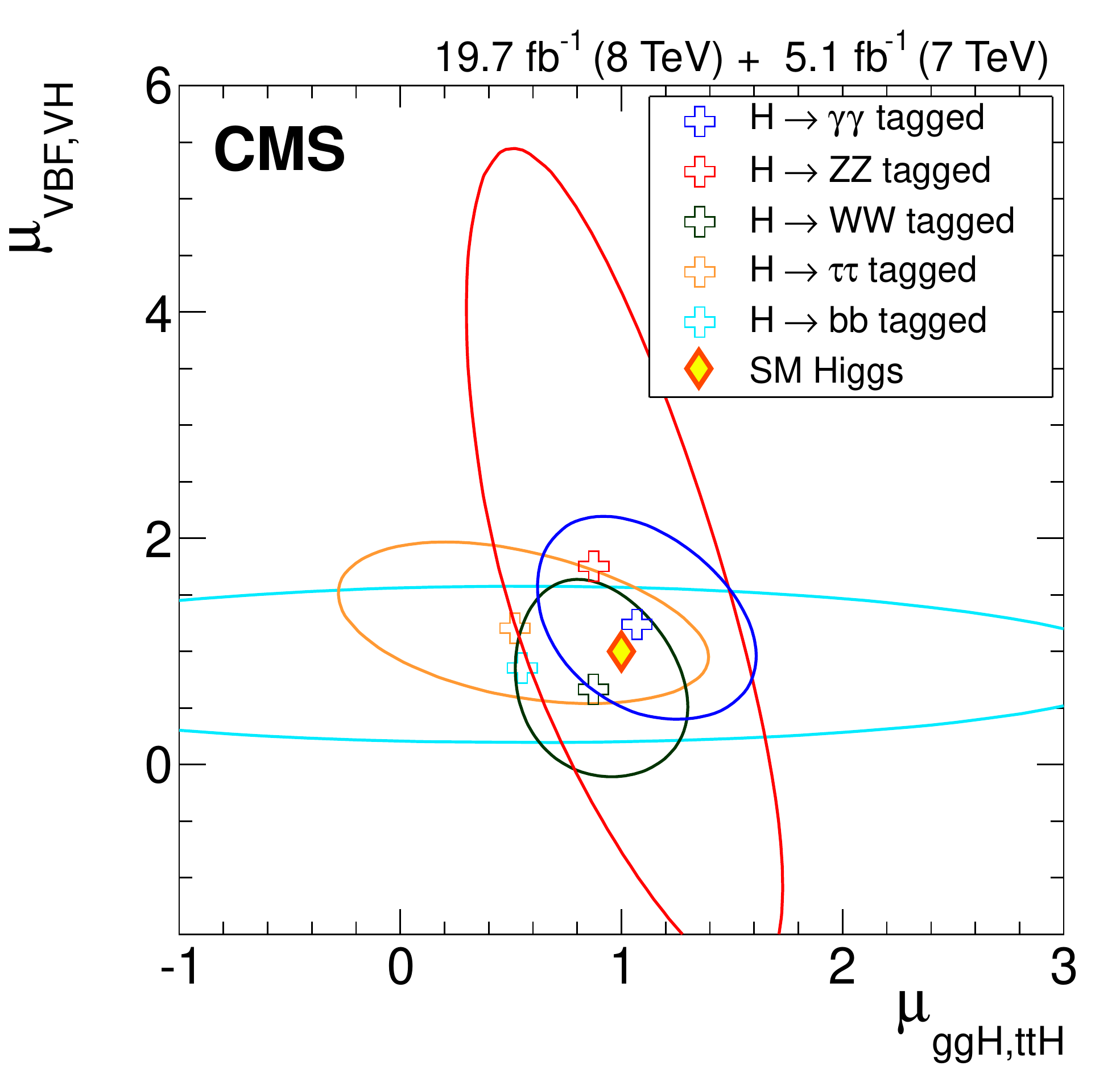} \hfill
\includegraphics[width=0.49\textwidth]{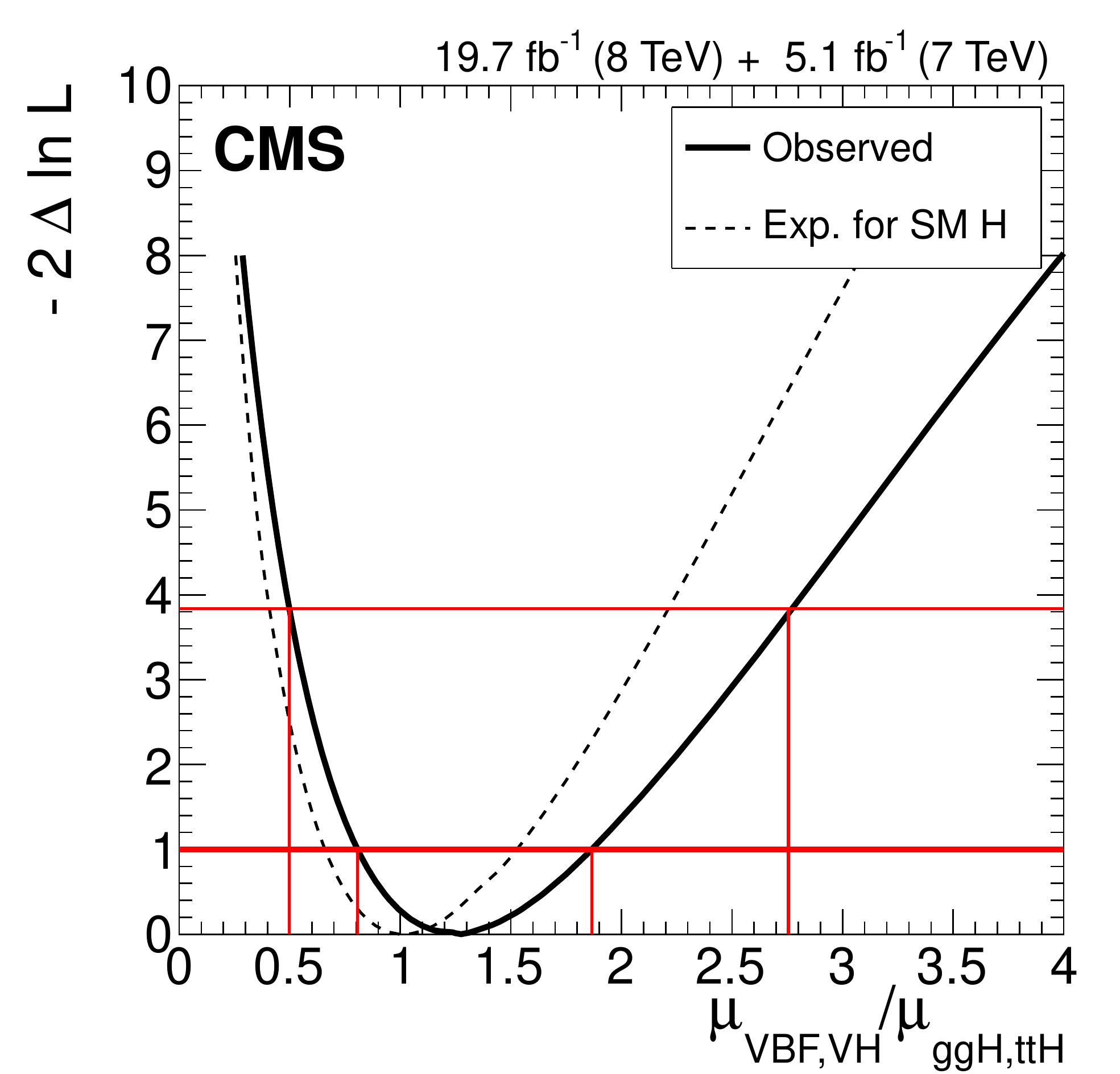}
\caption{
(Left) The 68\%~CL confidence regions (bounded by the solid curves) for the
signal strength of the \ggh and \tth and of the \vbf and \vh production mechanisms,
\muf and \muv, respectively.
The crosses indicate the best-fit values obtained in each group of predominant
decay modes: \gg, \zz, \ww, \tt, and \bb.
The diamond at $(1,1)$ indicates the expected values for the SM Higgs boson.
(Right) Likelihood scan versus the ratio \rvf, combined for all channels.
The fit for \rvf is performed while profiling the five \muf parameters, one per
visible decay mode, as shown in Table~\ref{tab:rvfModel}.
The solid curve represents the observed result in data while
the dashed curve indicates the expected median result in the presence of the SM Higgs boson.
Crossings with the
horizontal thick and thin lines denote the 68\%~CL and 95\%~CL confidence
intervals, respectively.
}
\label{fig:rvrf}
\end{figure*}

Figure~\ref{fig:rvrf}~(left) shows the 68\%~CL confidence regions for the signal
strength modifiers associated with the \ggh and \tth and with the \vbf and \vh
production mechanisms, \muf and \muv, respectively.
The five sets of contours correspond to the five
predominant decay mode groups, introduced in Section~\ref{sec:grouping}.
It can be seen in Figure~\ref{fig:rvrf}~(left) how the analyses in the \hbb
decay group constrain \muv more than \muf, reflecting the larger
sensitivity of the analysis of \vh production with \hbb with respect to
the analysis of \tth production with \hbb.
An almost complementary situation can be found for the \hzz analysis, where the
data constrain \muf better than \muv, reflecting the fact that
the analysis is more sensitive to \ggh, the most abundant production mode.
The SM Higgs boson expectation of $(1,1)$
is within the 68\%~CL confidence regions for all predominant decay groups.
The best-fit values for each decay tag group are given in Table~\ref{tab:rvrf}.

\label{sec:rvf}

The ratio of \muv and \muf provides a compatibility check with the
SM Higgs boson expectation that can be combined across all decay modes.
To perform the measurement of \rvf, the SM Higgs boson signal yields in the
different production processes and decay modes are parameterized according to
the scaling factors presented in Table~\ref{tab:rvfModel}.
The fit is performed simultaneously in all channels of all analyses and takes
into account, within each channel, the full detail of the expected
SM Higgs contributions from the different production processes and decay modes.

\begin{table*}[p]
\centering
\topcaption{
Parameterization used to scale the expected SM Higgs boson yields for the
different production processes and decay modes when obtaining the
\rvf results presented in Table~\ref{tab:rvf} and Fig.~\ref{fig:rvrf}~(right).
}
\label{tab:rvfModel}

\newcommand{\qstrut}{\rule[-7.5pt]{0pt}{18.2pt}}%
\begin{tabular}{c|ccccc}
\hline
\multicolumn{6}{l}{Parameter of interest: $R=\rvf$.}\\
\multicolumn{6}{l}{Other parameters: \mufgg, \qstrut\mufzz, \mufww, \muftt, and
\mufbb.\qstrut%
}\\

\hline
\multicolumn{1}{c|}{\begin{tabular}[c]{@{}c@{}}Signal\\model\end{tabular}}
     &  \hgg  &  \hzz  &  \hww  &  \htt  &  \hbb \\
\hline
\qstrut\ggh  & \mufgg & \mufzz & \mufww & \muftt & \mufbb \\
\qstrut\vbf  & $R\:\mufgg$ & $R\:\mufzz$ & $R\:\mufww$ & $R\:\muftt$ & $R\:\mufbb$ \\
\qstrut\vh   & $R\:\mufgg$ & $R\:\mufzz$ & $R\:\mufww$ & $R\:\muftt$ & $R\:\mufbb$ \\
\qstrut\tth  & \mufgg & \mufzz & \mufww & \muftt & \mufbb \\
\hline
\end{tabular}

\end{table*}

{\tolerance=500
Figure~\ref{fig:rvrf}~(right) shows the likelihood scan of the data for \rvf,
while the bottom part of Table~\ref{tab:rvrf} shows the corresponding
values; the best-fit \rvf is observed to be \rvfObs, compatible with the
expectation for the SM Higgs boson, $\rvf=1$.
\par}

\begin{table}[pt]
\centering
\topcaption{
The best-fit values for the signal strength of the \vbf and \vh and
of the \ggh and \tth production mechanisms, $\muv$ and
$\muf$, respectively, for $\MX=\mX$.
The channels are grouped by decay mode tag as described in
Section~\ref{sec:grouping}.
The observed and median expected results for the ratio of
\muv to \muf together with their uncertainties are also given
for the full combination.
In the full combination, \rvf is determined while profiling the five \muf
parameters, one per decay mode, as shown in Table~\ref{tab:rvfModel}.}
\label{tab:rvrf}
\label{tab:rvf}
\begin{tabular}{lcc}
\hline
Channel grouping  & Best fit $(\muf, \muv)$ \\
\hgg tagged &  $(1.07, 1.24)$  \\ %
\hzz tagged &  $(0.88, 1.75)$  \\ %
\hww tagged &  $(0.87, 0.66)$  \\ %
\htt tagged &  $(0.52, 1.21)$  \\ %
\hbb tagged &  $(0.55, 0.85)$  \\ %
\hline\hline
\multicolumn{2}{l}{ Combined best fit $\muv$/$\muf$ }\\
& {\small Observed (expected)} \\
 & \rvfObs $(\rvfExp)$ \\
\hline
\end{tabular}

\end{table}

\subsection{Individual production modes}
\label{sec:muprod}

While the production modes can be grouped by the type of interaction involved
in the production of the SM Higgs boson, as done in Section~\ref{sec:rvrf}, the
data set and analyses available allow us to explore signal strength
modifiers for different production modes, \muggh, \muvbf, \muvh, and \mutth.
These scaling factors are applied to the expected signal contributions from the
SM Higgs boson according to their production mode, as shown in
Table~\ref{tab:muprodModel}.
It is assumed that the relative values of the branching fractions are those
expected for the SM Higgs boson.
This assumption is relaxed, in different ways, in Sections~\ref{sec:decayratios}
and \ref{sec:matrixrank}.

\begin{table*}[tp]
\centering
\topcaption{
Parameterization used to scale the expected SM Higgs boson yields of the
different production and decay modes when obtaining the results presented in
Fig.~\ref{fig:muprod}.
}
\label{tab:muprodModel}

\begin{tabular}{c|ccccc}
\hline
\multicolumn{6}{l}{Parameters of interest: \muggh, \muvbf, \muvh, and \mutth.}\\
\hline
\multicolumn{1}{c|}{\begin{tabular}[c]{@{}c@{}}Signal\\model\end{tabular}}
      &  \hgg  &  \hzz  &  \hww  &  \htt  &  \hbb \\
\hline
\ggh  & \muggh & \muggh & \muggh & \muggh & \muggh \\
\vbf  & \muvbf & \muvbf & \muvbf & \muvbf & \muvbf \\
\vh   & \muvh  & \muvh  & \muvh  & \muvh  & \muvh \\
\tth  & \mutth & \mutth & \mutth & \mutth & \mutth \\
\hline
\end{tabular}

\end{table*}

\begin{figure}[bpht]
\centering
\includegraphics[width=0.49\textwidth]{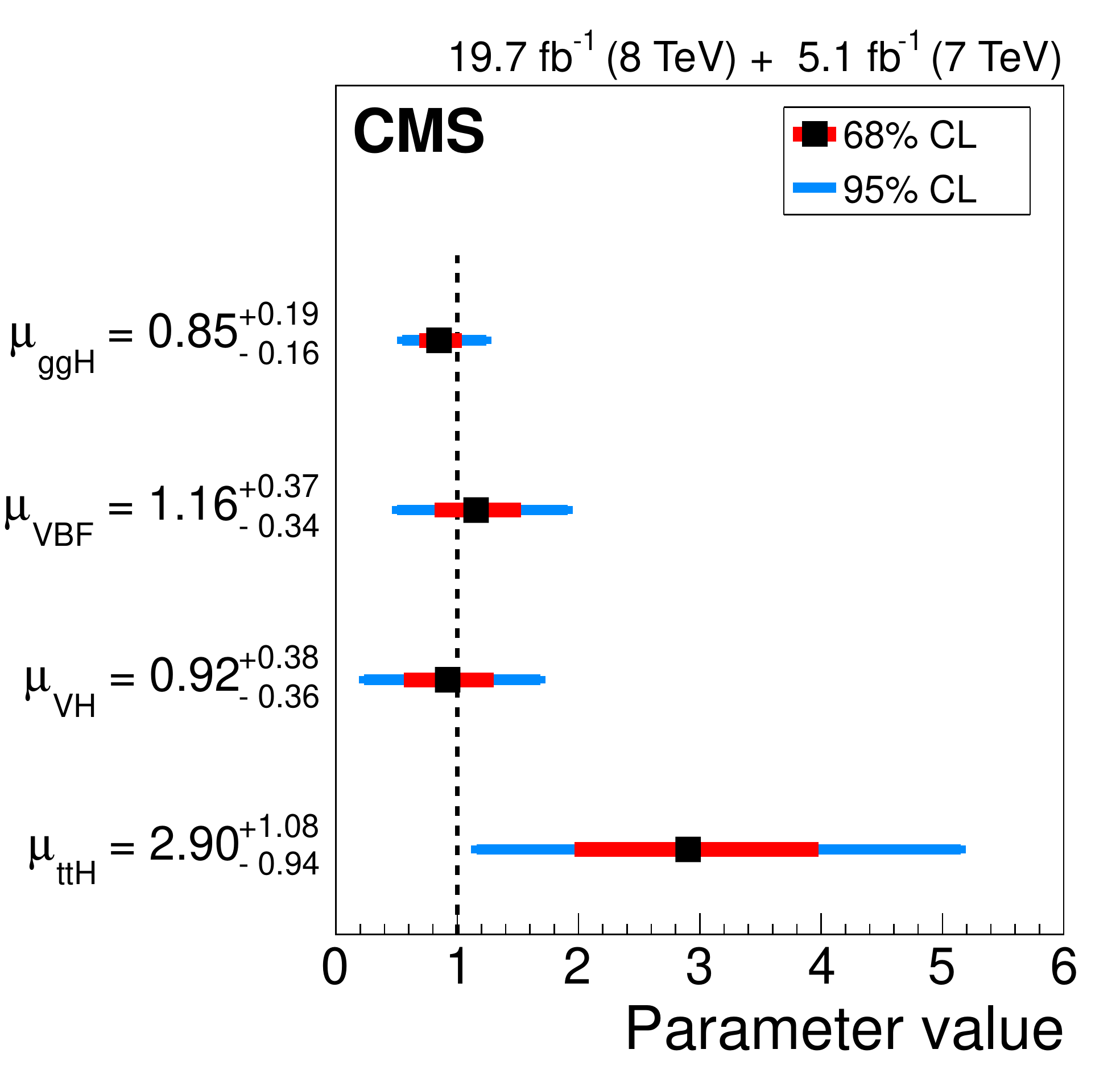}\hfill
\caption{
Likelihood scan results for \muggh, \muvbf, \muvh, and
\mutth.
The inner bars represent the 68\%~CL confidence intervals while the outer bars
represent the 95\%~CL confidence intervals.
When scanning each individual parameter, the three other parameters are
profiled.
The SM values of the relative branching fractions are assumed for the
different decay modes.
}
\label{fig:muprod}
\end{figure}

Figure~\ref{fig:muprod} summarizes the results of likelihood scans for
the four parameters of interest described in Table~\ref{tab:muprodModel} in terms
of the 68\%~CL (inner) and 95\%~CL (outer) confidence intervals.
When scanning the likelihood of the data as a function of one parameter, the
other parameters are profiled.

Table~\ref{tab:mu_prod} shows the best-fit results for the 7\TeV and 8\TeV
data sets separately, as well as for the full combined analysis.
Based on the combined likelihood ratio values for each parameter,
Table~\ref{tab:mu_prod} also shows the observed significance, the expected
significance, and the pull of the results with respect to the SM hypothesis.
The observed significance is derived from the observed likelihood ratio for the
background-only hypothesis, $\mu_{i}=0$, in data.
The expected significance is derived from the likelihood ratio for $\mu_{i}=0$
obtained using the median expected result for the SM Higgs boson.
The pull with respect to the SM hypothesis is derived from the observed
likelihood ratio for $\mu_{i}=1$; by definition, the expected pull with respect
to the SM hypothesis is zero.

{\tolerance=500
The \muggh best-fit value is found to be \MUggH.
After calculating the component of the uncertainty that is statistical in nature
(stat) and the component related to the theory inputs (theo), one can
subtract them in quadrature from the total uncertainty and assign the remainder
as the systematic uncertainty (syst), yielding \MUggHdetail.
Advances in the calculation of the \ggh cross section, \eg when considering
higher-order effects, may not only reduce the
uncertainty above, but also shift the central value.
The signal strengths for the \vbf and \vh production modes are assessed
independently.
Individual likelihood scans are performed as a function of \muvbf (or \muvh),
allowing the modifiers associated with the other production processes to float in
the fit together with the nuisance parameters.
In data, the best-fit result for \muvbf is \qqHobsOneSig, while for
\muvh it is \VHxobsOneSig.
For the \tth production mode, the best-fit value for \mutth is
found to be \ttHobsOneSig.
The results for \vbf, \vh, and \tth are driven by the corresponding
tagged categories, while the contribution from \ggh is constrained by the
\njet[0] and untagged categories.
\par}

The results in Table~\ref{tab:mu_prod} show a clear observation of
Higgs bosons produced through gluon fusion, and evidence for the
production of Higgs bosons through vector boson fusion, for which both the
expected and observed significances are above the $3\sigma$ level.
For \vh production, the expected significance is $\VHxNonZeroExpSign\sigma$ and
the observed significance is $\VHxNonZeroSign\sigma$.
The large best-fit value for \mutth is compatible with the
results presented and discussed in Section~\ref{sec:mupergroups}; the data are compatible with the
$\mutth=1$ hypothesis at the $\ttHNonOneSignVal\sigma$ level.
Because of the different parameterizations used, this significance is not
exactly the same as that found in Section~\ref{sec:mupergroups} when considering the
combination of \tth-tagged categories.

\begin{table*}[tp]
\centering
\topcaption{
The best-fit results for independent signal strengths scaling the
\ggh, \vbf, \vh, and \tth production processes; the expected and
observed significances with respect to the background-only hypothesis,
$\mu_{i}=0$; and the pull of the observation with respect to the SM
hypothesis, $\mu_{i}=1$.
The best-fit results are also provided separately for the 7\TeV and 8\TeV
data sets, for which the predicted cross sections differ.
These results assume that the relative values of the branching fractions are
those predicted for the SM Higgs boson.
}
\label{tab:mu_prod}
\renewcommand{\arraystretch}{1.1}
\begin{tabular}{ccc|cccc}
\hline
\multicolumn{1}{c|}{\multirow{2}{*}{Parameter}} & \multicolumn{3}{c|}{Best-fit result (68\%~CL)} & \multicolumn{2}{c}{Significance ($\sigma$)} & Pull to SM \\
\multicolumn{1}{c|}{}                           & 7\TeV & 8\TeV & \multicolumn{1}{c|}{Combined} & Observed & Expected &           ($\sigma$) \\
\hline
\muggh  & \ggHobsSevenOneSig & \ggHobsEightOneSig & \ggHobsOneSig & \ggHNonZeroSign & \ggHNonZeroExpSign & \ggHNonOneSign \\ %
\muvbf  & \qqHobsSevenOneSig & \qqHobsEightOneSig & \qqHobsOneSig & \qqHNonZeroSign & \qqHNonZeroExpSign & \qqHNonOneSign \\ %
\muvh   & \VHxobsSevenOneSig & \VHxobsEightOneSig & \VHxobsOneSig & \VHxNonZeroSign & \VHxNonZeroExpSign & \VHxNonOneSign \\ %
\mutth  & \ttHobsSevenOneSig & \ttHobsEightOneSig & \ttHobsOneSig & \ttHNonZeroSign & \ttHNonZeroExpSign & \ttHNonOneSign \\ %
\hline
\end{tabular}

\end{table*}

\subsection{Ratios between decay modes}
\label{sec:decayratios}

Some of the largest uncertainties in SM Higgs predictions are related to
the production cross sections.
In an attempt to evade those uncertainties, it has been
proposed~\cite{DjouadiRatios1,DjouadiRatios2} to perform measurements of ratios
of the signal strengths in different decay modes,
$\lambda_{yy,xx} = \beta_{yy}/\beta_{xx}$, where
$\beta_{xx}=\mathcal{B}(\PH\to xx)/\mathcal{B}(\PH\to
xx)_\text{SM}$ and $\mathcal{B}$ denotes a branching fraction.
In such $\beta_{xx}$ ratios, uncertainties related to the production and
decay predictions for the Higgs boson, as well as some experimental
uncertainties, may cancel out.
On the other hand, the uncertainty in a given ratio will reflect the combined
statistical uncertainties of both the $yy$ and $xx$ decay modes.

To probe the different $\lambda_{yy,xx}$, the expected signal yields
for the different production and decay modes are scaled by the factors shown in
Table~\ref{tab:doubleratioModel}.
To reduce the dependency of the results on the expected structure of the SM
Higgs production cross section, the \muf and \muv parameters are introduced and
allowed to float independently.
Therefore, these measurements only assume the SM ratio of \ggh and \tth
cross sections and the ratio of \vbf and \vh cross sections.

\begin{table*}[tp]
\centering
\topcaption{
Parameterization used to scale the expected SM Higgs boson yields of the
different production and decay modes when obtaining the results presented in
Table~\ref{tab:decayratio}.
The \muf and \muv parameters are introduced to reduce the
dependency of the results on the SM expectation.}
\label{tab:doubleratioModel}

\begin{tabular}{c|ccccc}
\hline
\multicolumn{6}{l}{Parameters of interest: $\lambda_{yy,xx}$, $\lambda_{ii,xx}$, $\lambda_{jj,xx}$, and $\lambda_{kk,xx}$.} \\
\multicolumn{6}{l}{Other parameters: \muf and \muv.}\\
\hline
\multicolumn{1}{c|}{\begin{tabular}[c]{@{}c@{}}Signal\\model\end{tabular}}
      &  $\PH\to xx$  & $\PH\to yy$ &  $\PH\to ii$  &  $\PH\to jj$  & $\PH\to kk$ \\
\hline
\ggh  & \muf & $\muf\:\lambda_{yy,xx}$ & $\muf\:\lambda_{ii,xx}$ & $\muf\:\lambda_{jj,xx}$ & $\muf\:\lambda_{kk,xx}$ \\
\vbf  & \muv & $\muv\:\lambda_{yy,xx}$ & $\muv\:\lambda_{ii,xx}$ & $\muv\:\lambda_{jj,xx}$ & $\muv\:\lambda_{kk,xx}$ \\
\vh   & \muv & $\muv\:\lambda_{yy,xx}$ & $\muv\:\lambda_{ii,xx}$ & $\muv\:\lambda_{jj,xx}$ & $\muv\:\lambda_{kk,xx}$ \\
\tth  & \muf & $\muf\:\lambda_{yy,xx}$ & $\muf\:\lambda_{ii,xx}$ & $\muf\:\lambda_{jj,xx}$ & $\muf\:\lambda_{kk,xx}$ \\
\hline
\end{tabular}

\end{table*}

Given the five decay modes that are currently accessible, four
ratios can be probed at a time.
For example, the choice of the \hgg decay as denominator, $xx=\gg$, fixes the
four ratio parameters to be \lzzgg, \lbbgg, \lwwgg, and \lttgg.
When scanning the likelihood for the data as a function of a given
$\lambda_{yy,xx}$ ratio, the production cross section modifiers \muf and
\muv, as well as the other three ratios, are profiled.
The best-fit results for each choice of denominator are presented as the
different rows in Table~\ref{tab:decayratio}.
While correlated uncertainties from theory and correlated experimental
uncertainties may cancel out to some extent in these ratios, each ratio includes the statistical uncertainties from
the two decay modes involved.
For the available data set and analyses, the resulting statistical uncertainty
dominates the total uncertainty.
It can be seen that the SM expectation, $\lambda_{yy,xx}=1$, is
inside the 68\%~CL confidence interval for all measurements.

\begin{table*}[tp]
\centering
\topcaption{
The best-fit results and 68\%~CL confidence intervals
for signal strength ratios of the decay mode in each column and the decay
mode in each row, as modelled by the parameterization in Table~\ref{tab:doubleratioModel}.
When the likelihood of the data is scanned as a function of each individual
parameter, the three other parameters in the same row, as well the production
cross sections modifiers \muf and \muv, are profiled.
Since each row corresponds to an independent fit to data, the relation
$\lambda_{yy,xx}=1/\lambda_{xx,yy}$ is only approximately satisfied.}
\label{tab:decayratio}
\renewcommand{\arraystretch}{1.1}
\begin{tabular}{c|ccccc}
\hline
Best-fit $\lambda_\text{col,row}$ &  \hgg  &  \hzz  &  \hww  &  \htt  &  \hbb  \\
\hline
\hgg  & 1       & \DRzzgg & \DRwwgg & \DRttgg & \DRbbgg \\
\hzz  & \DRggzz & 1       & \DRwwzz & \DRttzz & \DRbbzz \\
\hww  & \DRggww & \DRzzww & 1       & \DRttww & \DRbbww \\
\htt  & \DRggtt & \DRzztt & \DRwwtt & 1       & \DRbbtt \\
\hbb  & \DRggbb & \DRzzbb & \DRwwbb & \DRttbb & 1       \\
\hline
\end{tabular}

\end{table*}

\subsection{Search for mass-degenerate states with different coupling
structures}
\label{sec:matrixrank}

One assumption that is made in Section~\ref{sec:kappas} when studying
the couplings of the Higgs boson is that the observations are due to
the manifestation of a single particle.
Alternatively, a superposition of states with indistinguishable mass values is
expected in models or theories beyond the
SM~\cite{Gunion:2012gc,Grzadkowski:2012ng,Drozd:2012vf,Ferreira:2012nv}.
In this section we explore the validity of this assumption.

Taking advantage of the very good mass resolution in the \hgg analysis, the
presence of near mass-degenerate states has been previously probed down to mass
differences between 2.5\GeV and 4\GeV without evidence for the presence of a
second state~\cite{CMSHggLegacyRun1}.
Given the finite mass resolution, such searches are not sensitive
to a mixture of states with mass values closer than the resolution itself,
such that other reported measurements would integrate the
contributions from both states.

In the case of two or more states with masses closer to each
other than the experimental resolution, it becomes impossible to
discern them using the mass observables.
However, the distinction between states can
still be made, provided that the states have different coupling structures, \ie
different coupling strengths to the SM particles.
Using the measurements of the different production and decay tags, as well as
the detailed knowledge of their expected composition in terms of production
processes and decay modes, it is possible to test the compatibility of
the observations with the expectations from a single state.
Several authors discussed this possibility, proposing methods to look for
deviations assuming that, in the presence of more than one state, the individual
states would couple differently to the SM
particles~\cite{GunionDegenerate,GrossmanDegenerate}.

A general parameterization of the $5\times4$ matrix, \calM, of signal strengths
for the different production processes and decay modes is shown in
Table~\ref{tab:degenerateSaturatedModel}.
This parameterization has as many degrees of freedom as there are elements in
the matrix and is completely general.
Depending on whether there is one particle or more particles responsible for the
observations in data, the algebraic properties of \calM, namely its rank,
\rankM, will vary.

If there is only one state it follows that $\rankM = 1$, \ie
there should be one common multiplier per row and one common multiplier per
column.
A general matrix with $\rankM=1$ can be parameterized as shown in
Table~\ref{tab:degenerateRank1Model}.
This parameterization can also be obtained by taking the
most general $5\times4$ parameterization in Table~\ref{tab:degenerateSaturatedModel} and
assuming $\lambda_{i}^{j}=\lambda_{i}$, where $i$ runs through the production
processes except \ggh and $j$ runs through the decay modes.
Given this relationship, the model for a general matrix with $\rankM=1$
presented in Table~\ref{tab:degenerateRank1Model} is nested, in the statistics
sense, in the general parameterization of the $5\times4$ matrix presented in
Table~\ref{tab:degenerateSaturatedModel}.

The expectation for the SM Higgs boson is a particular case of a rank 1 matrix,
namely that for which $\lambda_{i}=\mu_{j}=1$, where $i$ runs
through the production processes except \ggh and $j$ runs through the decay modes.

{\tolerance=500
If there is more than one particle contributing to the
observations, the structure of \calM may be such that $\rankM > 1$ as a
consequence of the different interaction strengths of the individual, yet
mass-degenerate, states.
\par}

\begin{table*}[ptb]
\centering
\topcaption{
A completely general signal parameterization used to scale the
expected yields of the $5\times4$ different production and decay modes.
The particular choice of parameters is such that the single-particle
parameterization shown in Table~\ref{tab:degenerateRank1Model} is a nested model,
\ie it can be obtained by assuming $\lambda_{i}^{j}=\lambda_{i}$, where
$i$ runs through the production processes except \ggh and $j$ runs through the
decay modes.
The expectation for the SM Higgs boson is $\lambda_{i}^{j}=\mu_{j}=1$.
This parameterization is used in the denominator of the test statistic defined
in Eq.~(\ref{eq:qlambda}).
}
\label{tab:degenerateSaturatedModel}
\renewcommand{\arraystretch}{1.1}
\begin{tabular}{c|ccccc}
\hline
\multicolumn{6}{l}{All parameters constrained to be positive.}\\
\hline
\multicolumn{1}{c|}{\begin{tabular}[c]{@{}c@{}}Signal\\model\end{tabular}}
      &  \hgg  &  \hzz  &  \hww  &  \htt  &  \hbb \\
\hline
\ggh  & \mugg & \muzz & \muww & \mutt & \mubb \\
\vbf  & $\lvbfgg\:\mugg$ & $\lvbfzz\:\muzz$ & $\lvbfww\:\muww$ & $\lvbftt\:\mutt$ & $\lvbfbb\:\mubb$ \\
\vh   & $\lvhgg\:\mugg$ & $\lvhzz\:\muzz$ & $\lvhww\:\muww$ & $\lvhtt\:\mutt$ & $\lvhbb\:\mubb$ \\
\tth  & $\ltthgg\:\mugg$ & $\ltthzz\:\muzz$ & $\ltthww\:\muww$ & $\ltthtt\:\mutt$ & $\ltthbb\:\mubb$ \\
\hline
\end{tabular}

\end{table*}

\begin{table*}[ptb]
\centering
\topcaption{
A general single-state parameterization used to scale the
expected yields of the different production and decay modes.
For this parameterization the matrix has $\rankM=1$ by definition.
It can be seen that this parameterization is nested in the general one
presented in Table~\ref{tab:degenerateSaturatedModel}, and can be obtained by
setting $\lambda_{i}^{j}=\lambda_{i}$, where $i$ runs through the production
processes except \ggh and $j$ runs through the decay modes.
The expectation for the SM Higgs boson is $\lambda_{i}=\mu_{j}=1$.
This parameterization is used in the numerator of the test statistic defined
in Eq.~(\ref{eq:qlambda}).
}
\label{tab:degenerateRank1Model}

\begin{tabular}{c|ccccc}
\hline
\multicolumn{6}{l}{All parameters constrained to be positive.}\\
\hline
\multicolumn{1}{c|}{\begin{tabular}[c]{@{}c@{}}Signal\\model\end{tabular}}
      &  \hgg  &  \hzz  &  \hww  &  \htt  &  \hbb \\
\hline
\ggh  & \mugg & \muzz & \muww & \mutt & \mubb \\
\vbf  & $\lvbf\:\mugg$ & $\lvbf\:\muzz$ & $\lvbf\:\muww$ & $\lvbf\:\mutt$ & $\lvbf\:\mubb$ \\
\vh   & $\lvh \:\mugg$ & $\lvh \:\muzz$ & $\lvh \:\muww$ & $\lvh \:\mutt$ & $\lvh \:\mubb$ \\
\tth  & $\ltth\:\mugg$ & $\ltth\:\muzz$ & $\ltth\:\muww$ & $\ltth\:\mutt$ & $\ltth\:\mubb$ \\
\hline
\end{tabular}

\end{table*}

{\tolerance=1200
The procedure to test for the presence of mass-degenerate states proposed in
Ref.~\cite{MatrixRank} takes into account both the fact that there may be
missing matrix elements and the fact that there are uncertainties in the
measurements, including their correlations.
A profile likelihood ratio test statistic, $q_\lambda$, is built
using two different models for the structure of \calM, namely those presented in
Tables~\ref{tab:degenerateRank1Model} and~\ref{tab:degenerateSaturatedModel},
\begin{equation}
\label{eq:qlambda}
q_{\lambda}=-2 \ln\frac{
\mathcal{L}(\text{data}\,|\,\lambda^{j}_{i}=\hat{\lambda}_{i},\hat{\mu_{j}})
}{
\mathcal{L}(\text{data}\,|\,\hat{\lambda}^{j}_{i},\hat{\mu_{j}}^\prime)}.
\end{equation}
The test statistic $q_{\lambda}$ is a function of the 20 variables defined in
Table~\ref{tab:degenerateSaturatedModel}: $\lambda_i^j$ and $\mu_j$, where the
index $i$ runs through the \vbf, \vh, and \tth production processes and the
index $j$ runs through the decay modes.
In this likelihood ratio, the model in Table~\ref{tab:degenerateSaturatedModel}
is taken as the alternative hypothesis and corresponds to the so-called
``saturated model'' in statistics, as it contains as many degrees of freedom as
there are elements in \calM.
The null hypothesis model is the one presented in
Table~\ref{tab:degenerateRank1Model}, which parameterizes \calM as a general
rank 1 matrix, where all rows are multiples of each other, as expected for a single
particle.
If the observations are due to a single particle, the $\lambda_i$ do not depend
on the decay mode and the value of the $q_{\lambda}$ is not very large, since
both hypotheses fit the data equally well.
However, for a matrix with $\rankM \neq 1$, the most general $5\times4$
matrix model will fit the data better than the general
rank 1 matrix model and the value of $q_{\lambda}$ is expected to be large.
\par}

\begin{figure}[ptbh]
\centering
\includegraphics[width=0.49\textwidth]{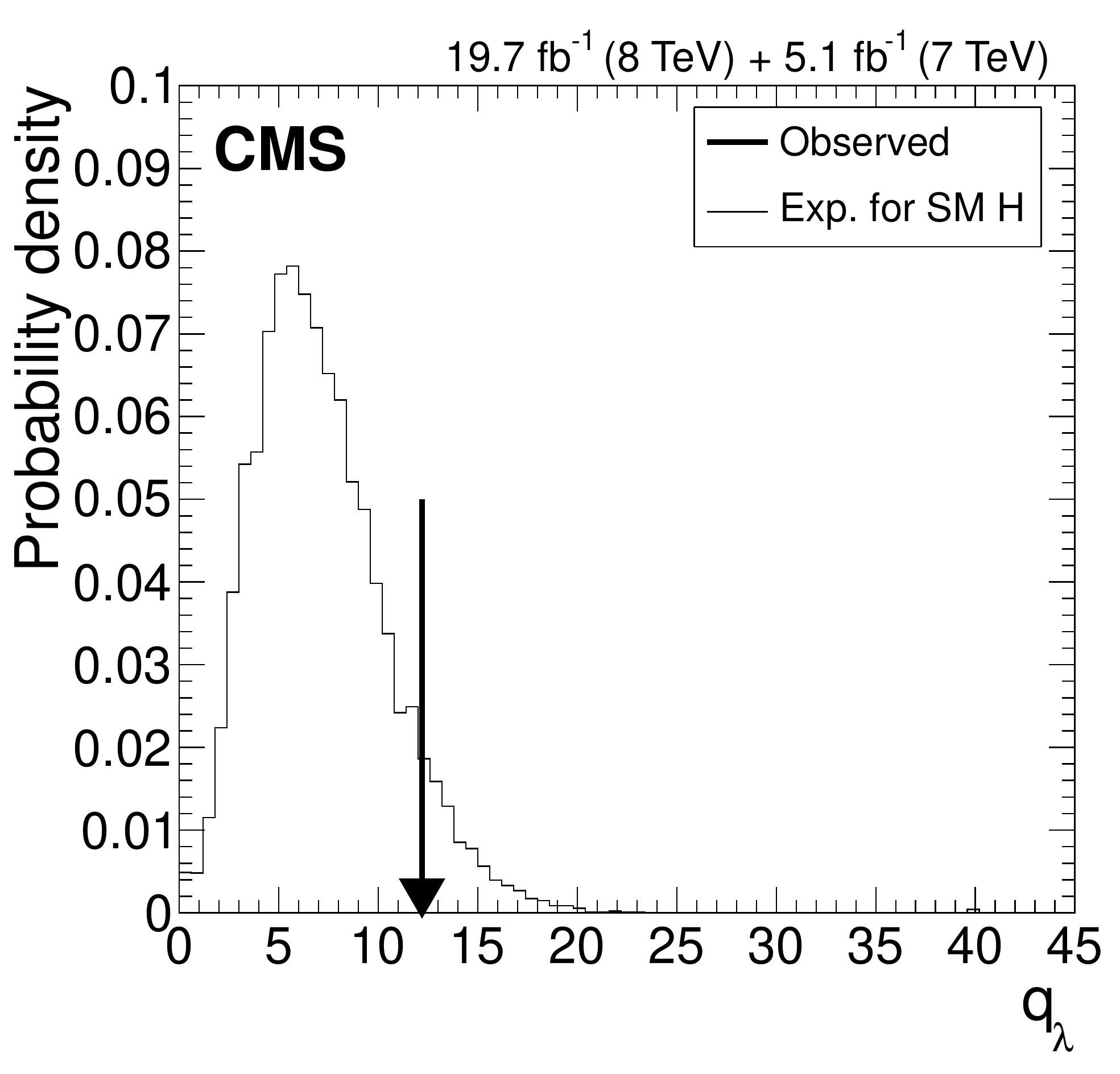}
\caption{Distribution of the profile likelihood ratio $q_\lambda$
between different assumptions for the structure of the matrix of signal
strengths for the production processes and decay modes
both for pseudo-data samples generated
under the SM hypothesis and the value observed in data.
The likelihood in the numerator is that for the data under a model of a general
rank 1 matrix, expected if the observations are due to a single particle
and of which the SM is a particular case.
The likelihood in the denominator is that for the data under a
``saturated model'' with as many parameters as there are matrix elements.
The arrow represents the observed value in data, $q_\lambda^\text{obs}$.
Under the SM hypothesis, the probability to find a value of
$q_\lambda \geq q_\lambda^\text{obs}$ is $(7.9\pm0.3)\%$, where the
uncertainty reflects only the finite number of pseudo-data samples generated.}
\label{fig:matrixrank}
\end{figure}

The compatibility of the value of the test statistic observed in data,
$q_\lambda^\text{obs}$, with the expectation from the SM is evaluated using
pseudo-data samples randomly generated under the SM hypothesis.
Figure~\ref{fig:matrixrank} shows the distribution of $q_\lambda$ for the
SM pseudo-data samples as well as the value observed in data,
$q_\lambda^\text{obs}=\rankQobs$.
Under the SM hypothesis, we find that the probability of observing a value of
$q_\lambda \geq q_\lambda^\text{obs}$ is $(7.9\pm0.3)\%$, where the uncertainty
reflects only the finite number of pseudo-data samples generated.
Such a \pval corresponds to a deviation from the SM expectation of about
$1.4\sigma$.
This small tension, not present in previous tests performed in this section, is
due to the observed data in the dijet-tagged channel of the \hzz analysis;
performing a fit to a model where the \vbf and \vh production modes are
floated separately shows that the data prefer a very large \vh contribution and
a very small \vbf contribution.
When \hzz analysis inputs are not considered, the \pval is found to be about
33\%.

\section{Compatibility of the observed data with the SM Higgs boson couplings}
\label{sec:kappas}

Whereas in Section~\ref{sec:deviations} the production and decay of the boson
were explored separately, the studies presented in this section simultaneously
investigate the couplings of the boson to SM particles in the production and
decay processes.
In this way, correlations are handled consistently between the production modes
and the decay modes.
For example, the coupling of the SM Higgs boson to the $\cPZ$ boson is involved
both in the \zh production mode and the \hzz decay mode, such that more
information can be extracted from a simultaneous modelling of the production
and decay modes in terms of the couplings involved.

Following the framework laid out in Ref.~\cite{LHCHXSWG3}, we assume the signal
arises from a single particle with $\JPC=0^{++}$ and a width such that the
narrow-width approximation holds, permitting its production and decay to be
considered independently.
These assumptions are supported by the results of Section~\ref{sec:matrixrank} on
the presence of more particles at the same mass, those of
Refs.~\cite{ATLASSpin2013,CMSAnomalousHVV2014} regarding alternative \JP
assignments and mixtures, and those of Ref.~\cite{CMSOffShellWidth2014}
concerning the width of the particle.

Under the assumptions above, the event yield in a given
(production)$\times$(decay) mode is related to the production cross section and
the partial and total Higgs boson decay widths via
\begin{equation}
\left(\sigma\;\mathcal{B}\right)(\mathit{x}\to\PH\to\mathit{yy}) =
\frac{\sigma_{\mathit{x}}\;\Gamma_{\mathit{yy}}}{\Gamma_{\text{tot}}},
\end{equation}
where $\sigma_{\mathit{x}}$ is the production cross section through
process $\mathit{x}$, which includes \ggh, \vbf, \wh, \zh,
and \tth;
$\Gamma_{\mathit{yy}}$ is the partial decay width into
the final state $\mathit{yy}$, such as \ww, \zz, \bb,
\tt, \gluglu, or \gg;
and $\Gamma_{\text{tot}}$ is the
total width of the boson.

Some quantities, such as $\sigma_{\ggh}$,
$\Gamma_{\gluglu}$, and $\Gamma_{\gg}$, are generated by loop diagrams and,
therefore, are sensitive to the presence of certain particles beyond the
standard model (BSM).
The possibility of Higgs boson decays to BSM particles,
with a partial width $\Gamma_{\mathrm{BSM}}$, can also
be accommodated by considering $\Gamma_{\text{tot}}$ as a dependent parameter so
that $\Gamma_{\text{tot}} = \sum \Gamma_{\mathit{yy}} +
\Gamma_{\mathrm{BSM}}$, where $\sum\Gamma_{\mathit{yy}}$ stands for
the sum over partial widths for all decays to SM particles.
With the data from the \hinv searches, $\Gamma_\mathrm{BSM}$ can be further
broken down as $\Gamma_\mathrm{BSM} = \Gamma_\mathrm{inv} +
\Gamma_\mathrm{undet}$, where $\Gamma_\mathrm{inv}$ can be constrained by
searches for invisible decays of the Higgs boson and $\Gamma_\mathrm{undet}$
corresponds to Higgs boson decays not fitting into the previous definitions.
The definition of $\Gamma_\mathrm{undet}$ is such that two classes of
decays can give rise to $\Gamma_\mathrm{undet}>0$:
i) BSM decays not studied in the analyses used in this paper, such as
hypothetical lepton flavour violating decays, \eg $\PH\to\Pgm\Pgt$, and
ii) decays that might not be detectable with the existing experimental
setup because of the trigger conditions of the experiment, such as hypothetical
decays resulting in a large multiplicity of low-\pT particles.

To test the observed data for possible deviations from the rates expected for
the SM Higgs boson in the different channels, we introduce coupling modifiers,
denoted by the scale factors $\kappa_{i}$ \cite{LHCHXSWG3}.
The scale factors are defined
for production processes by $\kappa_{i}^2=\sigma_{i}/\sigma_{i}^\text{SM}$,
for decay processes by $\kappa_{i}^2=\Gamma_{ii}/\Gamma_{ii}^\text{SM}$,
and for the total width by $\kappa_{\PH}^2=\Gamma_\text{tot}/\GSM$, where the SM
values are tabulated in Ref.~\cite{LHCHXSWG3}.
When considering the different $\kappa_{i}$, the index $i$ can represent many
ways to test for deviations:
\begin{itemize}
  \item For SM particles with tree-level couplings to the Higgs boson:
\kW ($\PW$ bosons),
\kZ ($\cPZ$ bosons),
\kb (bottom quarks),
\ktau (tau leptons),
\ktop (top quarks),
and \kmu (muons).
Unless otherwise noted, the scaling factors for other fermions are tied to those
that can be constrained by data.

  \item Particular symmetries of the SM make it interesting to test for
  deviations in whole classes of particles, leading to
  \kV (massive vector bosons),
  \kf (fermions),
  \kl (leptons),
\kq (quarks),
\ku (up-type fermions),
 and \kd (down-type fermions).

  \item For SM particles with loop-induced couplings, the scaling factors can
  be expressed in terms of the tree-level coupling modifiers, assuming the SM
  loop structure, but can also be taken as effective coupling modifiers:
\kglu (gluons)
and \kgam (photons).

\item The scaling factors for couplings to second generation
fermions are equal to those for the third generation: $\ks=\kb$, $\kmu=\ktau$,
and $\kc=\ktop$, except in Section~\ref{sec:mepsc5}, where \kmu is constrained
from the analysis of \hmm decays.

\end{itemize}
Given their small expected contributions, the couplings to electrons, up
quarks, and down quarks, are neglected.

In addition to the $\kappa_{i}$ parameters, the existence of BSM decays,
invisible decays, and undetectable decays of the Higgs boson is considered; the
corresponding branching fractions are denoted \BRBSM, \BRinv, and \BRundet, as
in Ref.~\cite{LHCHXSWG3}.

Significant deviations of any $\kappa$ parameter from unity or of any
$\mathrm{BR}$ parameter from zero would imply new physics beyond the SM Higgs
boson hypothesis.
The size of the current data set is insufficient to precisely quantify all
phenomenological parameters defining the Higgs boson production and decay rates.
Therefore, we present a set of combined analyses of different numbers of
parameters, where the remaining parameters are either set to the SM
expectations or profiled in the likelihood scans together with all other
nuisance parameters.
The value of \mH is fixed to the measured value of \mX, as determined in
Section~\ref{sec:mass}.
Since results for the individual channels are based on different assumed values
of the mass, differences should be expected when comparing the
previously published results from the individual channels with those in this
combined analysis.

This section is organized as follows.
In Section~\ref{sec:lwz} we explore whether \kW
and \kZ are compatible with each other and can be meaningfully used together as
\kV.
In Section~\ref{sec:kvkf} we test for deviations that would affect the
couplings of massive vector bosons and fermions differently.
The scaling factors among different types of fermions, leptons versus quarks and
up-type versus down-type, are investigated in Section~\ref{sec:ldullq}.
In Section~\ref{sec:mepsc5}, we consider the results of a fit for the
tree-level coupling scaling factors and the relation between the observations
and the corresponding particle masses.
We then turn to the study of models where BSM physics could manifest itself in
loops (\kglu, \kgam) or decays (\BRBSM, \BRinv, \BRundet).
In Section~\ref{sec:C2BSM} the tree-level couplings are constrained to those
expected in the SM, and the searches for \hinv are included.
This restriction is lifted in Section~\ref{sec:c6},
where a coupling scaling factor for the massive vector bosons and individual
fermion coupling scaling factors are allowed to float, while in
Section~\ref{sec:kgZ} the total width scaling factor is also left free to float.
In Section~\ref{sec:C6BSM}, the results from the searches for invisible decays
are included, and from the combination of the visible and invisible decays,
limits on \BRundet are set.
Closing this section, Table~\ref{tab:CouplingTests} summarizes the
results of the tests performed.

\subsection{Relation between the coupling to the \texorpdfstring{$\PW$ and
$\cPZ$}{W and Z} bosons}

\label{sec:lwz}

In the SM, the Higgs sector possesses an approximate $\mathrm{SU(2)_L
\times SU(2)_R}$ global symmetry, which is broken by the Higgs vacuum
expectation value to the diagonal subgroup $\mathrm{SU(2)_{L+R}}$.
As a result, the tree-level ratios of the $\PW$ and $\cPZ$ boson masses,
$m_{\PW} / m_{\cPZ}$, and the ratio of their couplings to the Higgs boson,
$g_{\PW} / g_{\cPZ}$, are protected against large radiative corrections,
a property known as ``custodial symmetry''~\cite{Veltman:1977kh,Sikivie:1980hm}.
However, large violations of custodial symmetry are possible in new physics models.
We focus on the two scaling factors
$\kW$ and $\kZ$ that modify
the couplings of the SM Higgs boson to the $\PW$ and $\cPZ$ bosons
and perform two different combined analyses to assess the consistency of
the ratio $\lWZ = \kW / \kZ$
with unity.

\begin{figure*}[bpht]
\centering
\includegraphics[width=0.49\textwidth]{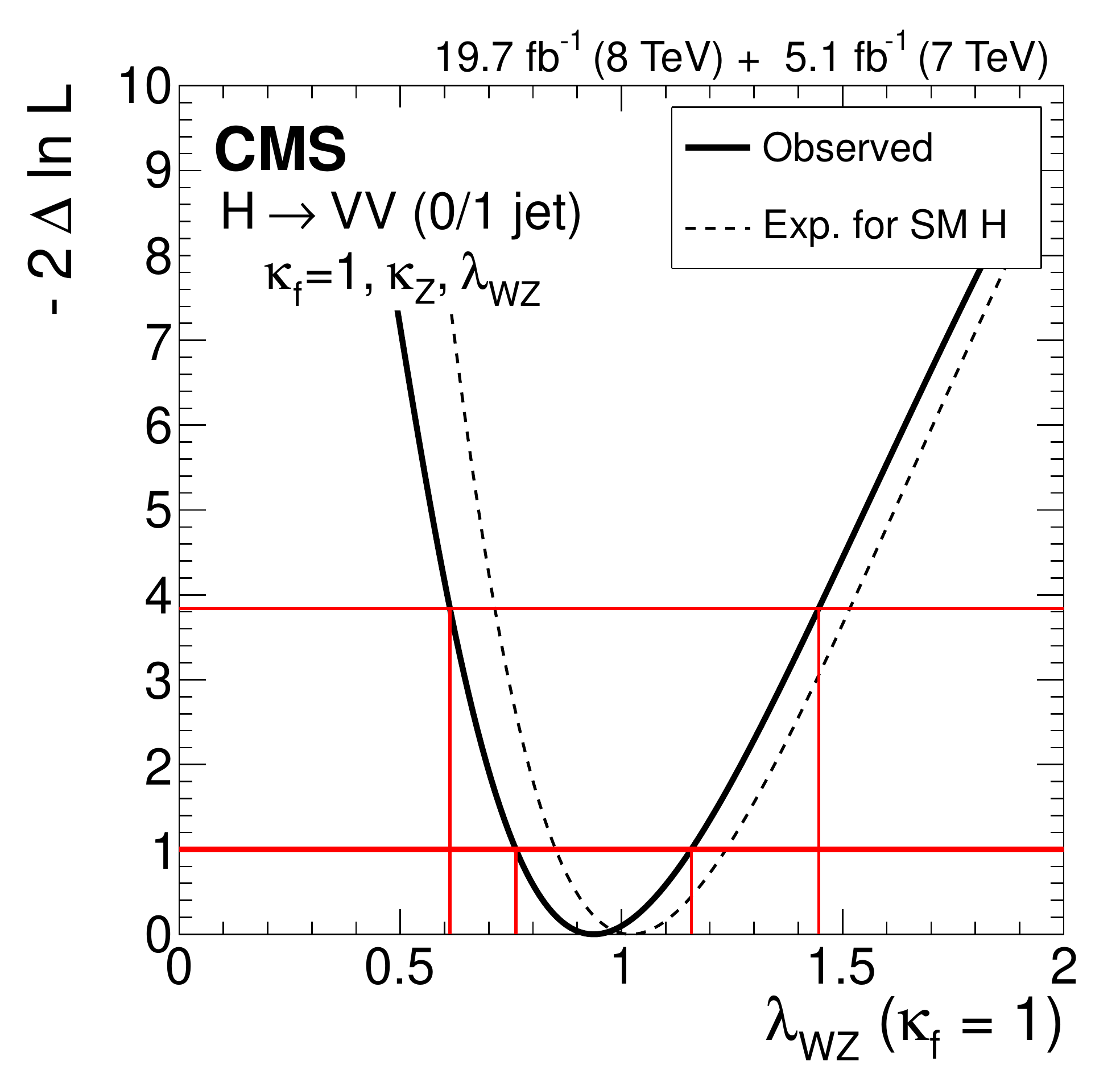} \hfill
\includegraphics[width=0.49\textwidth]{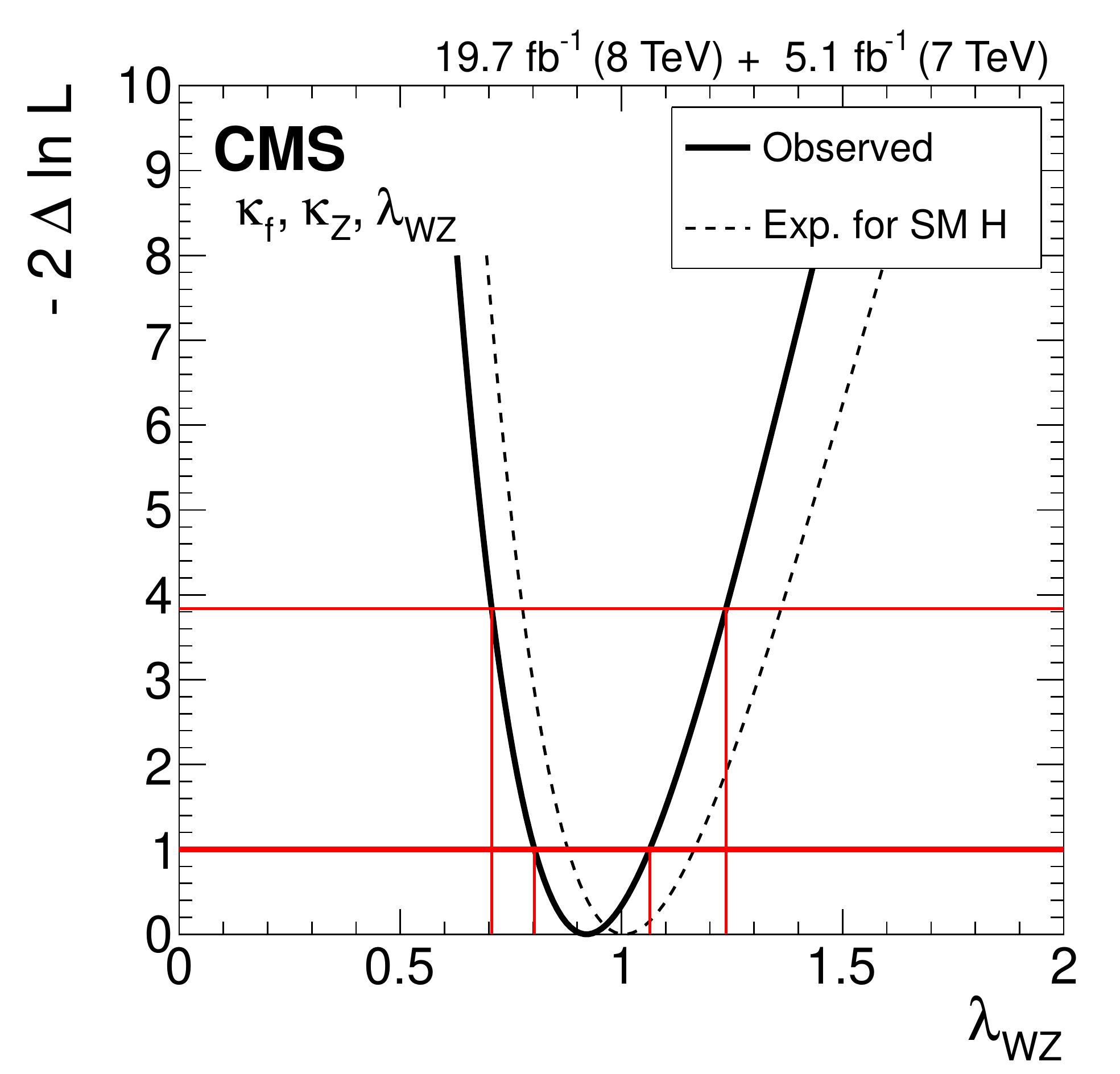}
\caption{
Likelihood scans versus $\lWZ$, the ratio of the coupling scaling factors to
$\PW$ and $\cPZ$ bosons:
(left) from untagged $\text{pp} \to \hww$ and $\text{pp} \to \hzz$ searches,
assuming the SM couplings to fermions, $\kf=1$;
(right) from the combination of all channels, profiling the coupling to fermions.
The solid curve represents the observation in data. The dashed curve
indicates the expected median result in the presence of the SM Higgs boson.
Crossings with the horizontal thick and thin lines denote the 68\%~CL and
95\%~CL confidence intervals, respectively.
}
\label{fig:fit_rwz_scan}
\end{figure*}

The dominant production mechanism populating the 0-jet and 1-jet channels of the
\hwwlnln analysis and the untagged channels of the \hzzllll analysis
is \ggh.
Therefore,
the ratio of event yields in these channels provides
a nearly model-independent measurement of $\lWZ$.
We perform a combined analysis of these two channels with two free parameters,
$\kZ$ and $\lWZ$.
The likelihood scan versus $\lWZ$ is shown in
Fig.~\ref{fig:fit_rwz_scan}~(left).
The scale factor $\kZ$ is treated as a nuisance parameter.
The result is $\lWZ=\lwzONEOneSig$, \ie
the data are consistent with the SM expectation ($\lWZ=1$).

We also extract $\lWZ$ from the combined analysis of all
channels.
In this approach, we introduce three parameters: $\lWZ$,
\kZ, and \kf.
The BSM Higgs boson width $\Gamma_{\mathrm{BSM}}$ is set to zero.
The partial width $\Gamma_\text{gg}$, induced by top and bottom quark loops,
scales as $\kf^2$.
The partial width $\Gamma_{\gg}$ is induced via
loop diagrams, with the $\PW$ boson and top quark being the dominant
contributors, and is scaled with $\kgam^2(\kb,\ktau,\ktop,\kW)$, a function
defined in Eq.~(113) of Ref.~\cite{LHCHXSWG3}.
In the likelihood scan as a function of \lWZ,
both $\kZ$ and $\kf$ are profiled together with all
other nuisance parameters.
The introduction of \kf carries with it the assumption that the coupling to all
fermions is common, but possibly different from the SM expectation.
The likelihood scan is shown in Fig.~\ref{fig:fit_rwz_scan}~(right)
with a solid curve.
The dashed curve indicates the median expected result for the SM Higgs boson,
given the current data set.
The measured value from the combined analysis of all channels is $\lWZ =
\lwzTWOOneSig$ and is consistent with the expectation from the SM.

Given these results, and unless otherwise noted, in all subsequent measurements
we assume $\lWZ=1$ and use a common factor $\kV$ to modify the couplings to
$\PW$ and $\cPZ$ bosons, while preserving their ratio.

\subsection{Test of the couplings to massive vector bosons and fermions}

\label{sec:kvkf}

In the SM, the nature of the coupling of the Higgs boson to fermions,
through a Yukawa interaction, is different from the nature of the
Higgs boson coupling to the massive vector bosons, a result of electroweak
symmetry breaking.
Some BSM models predict couplings to fermions and massive vector bosons
different from those in the SM.

We compare the observations in data with the expectation for the SM
Higgs boson by fitting two parameters, \kV and \kf,
where $\kV=\kW=\kZ$ is a common scaling factor for massive vector bosons, and
$\kf=\kb=\ktop=\ktau$ is a common scaling factor for fermions.
We assume that $\Gamma_{\mathrm{BSM}}=0$.
At leading order, all partial widths scale either as
$\kV^2$ or $\kf^2$,
except for $\Gamma_{\gg}$.
As discussed in Section~\ref{sec:lwz}, the partial width $\Gamma_{\gg}$ is
induced via loops with virtual $\PW$ bosons or top quarks and scales as a
function of both $\kV$ and $\kf$.
For that reason, the \hgg channel is the only channel being combined that is
sensitive to the relative sign of $\kV$ and $\kf$.

\begin{figure*}[bpht]
\centering
\includegraphics[width=0.49\textwidth]{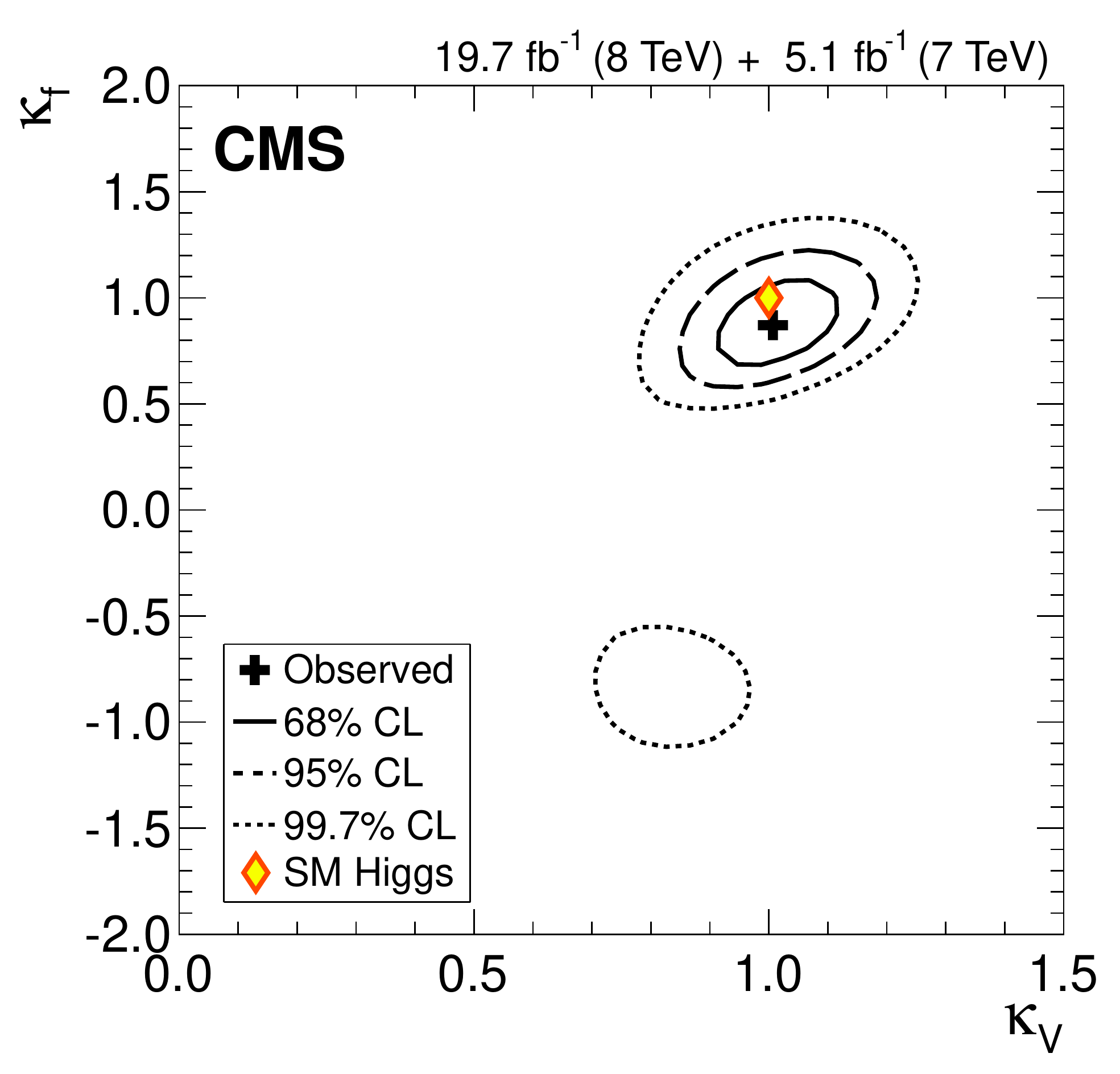} \hfill
\includegraphics[width=0.49\textwidth]{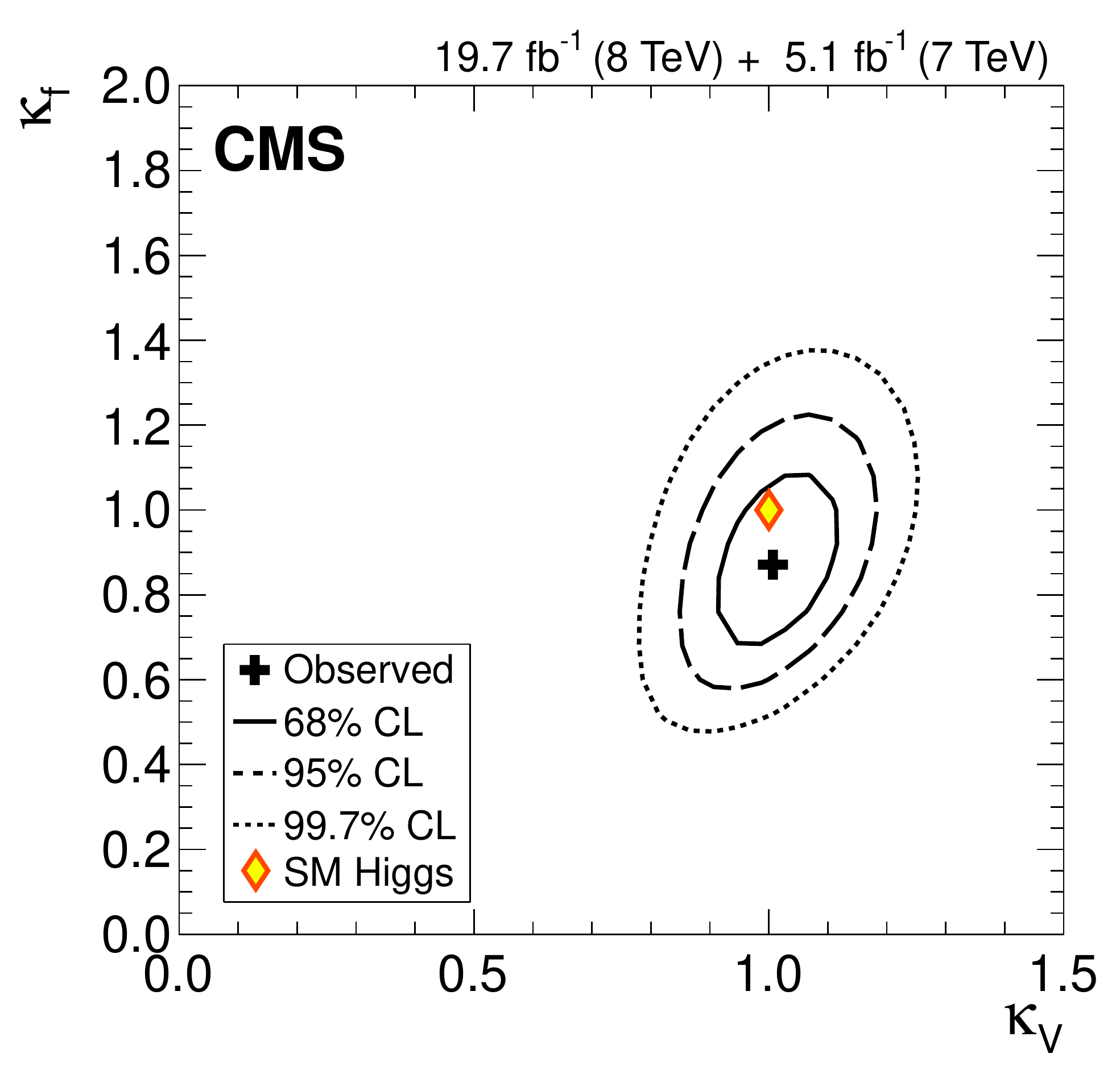}
\caption{
Results of 2D likelihood scans for the $\kV$ and $\kf$ parameters.
The cross indicates the best-fit values. The solid, dashed, and dotted contours show the
68\%, 95\%, and 99.7\%~CL confidence regions, respectively. The diamond shows
the SM point $(\kV, \kf)=(1,1)$. The left plot shows the likelihood scan in two quadrants, $(+,+)$ and $(+,-)$.
The right plot shows the likelihood scan constrained to the $(+,+)$ quadrant.
}
\label{fig:cVcF_2D}
\vspace{10 mm}
\centering
\includegraphics[width=0.49\textwidth]{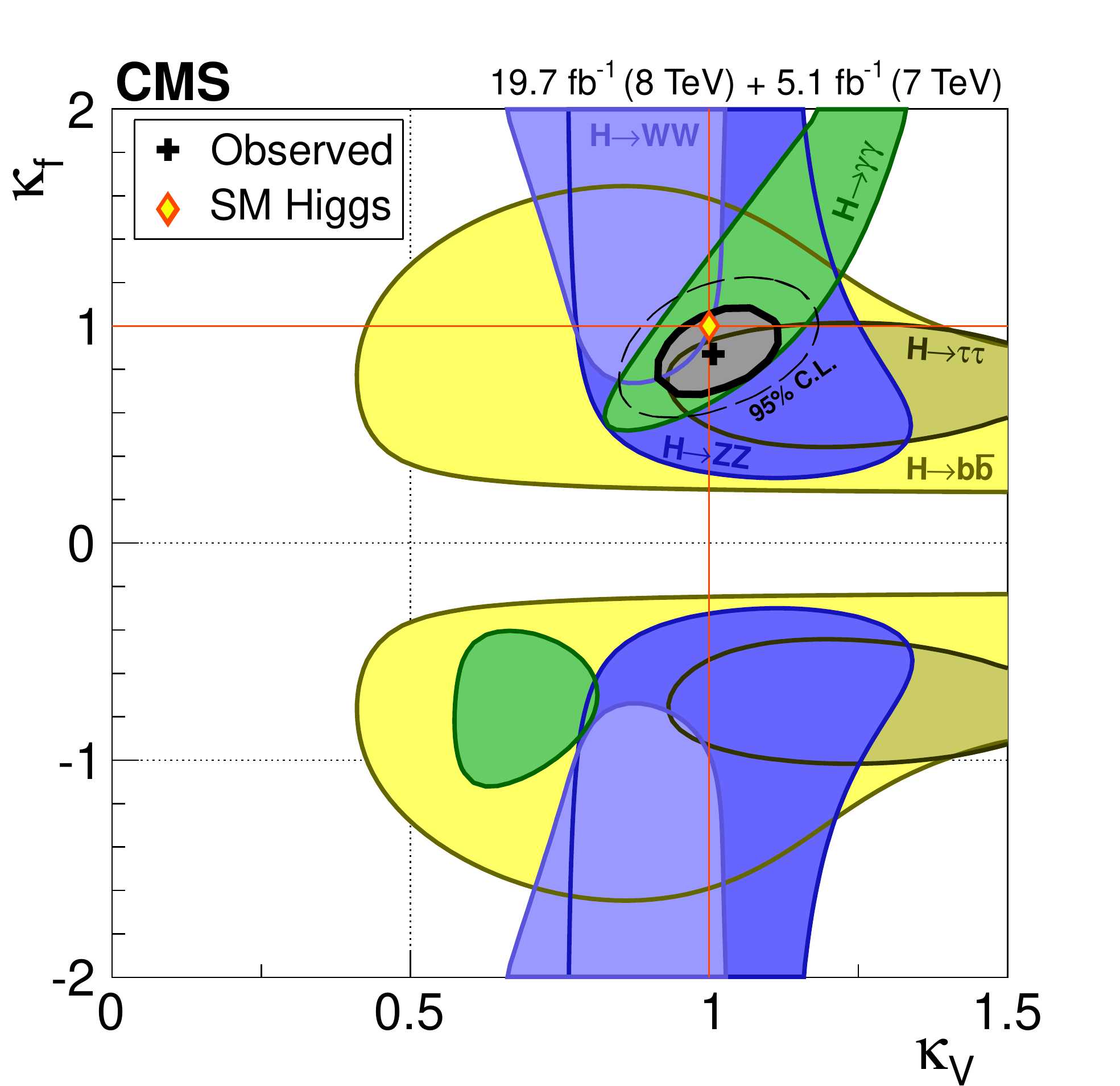}\hfill
\includegraphics[width=0.49\textwidth]{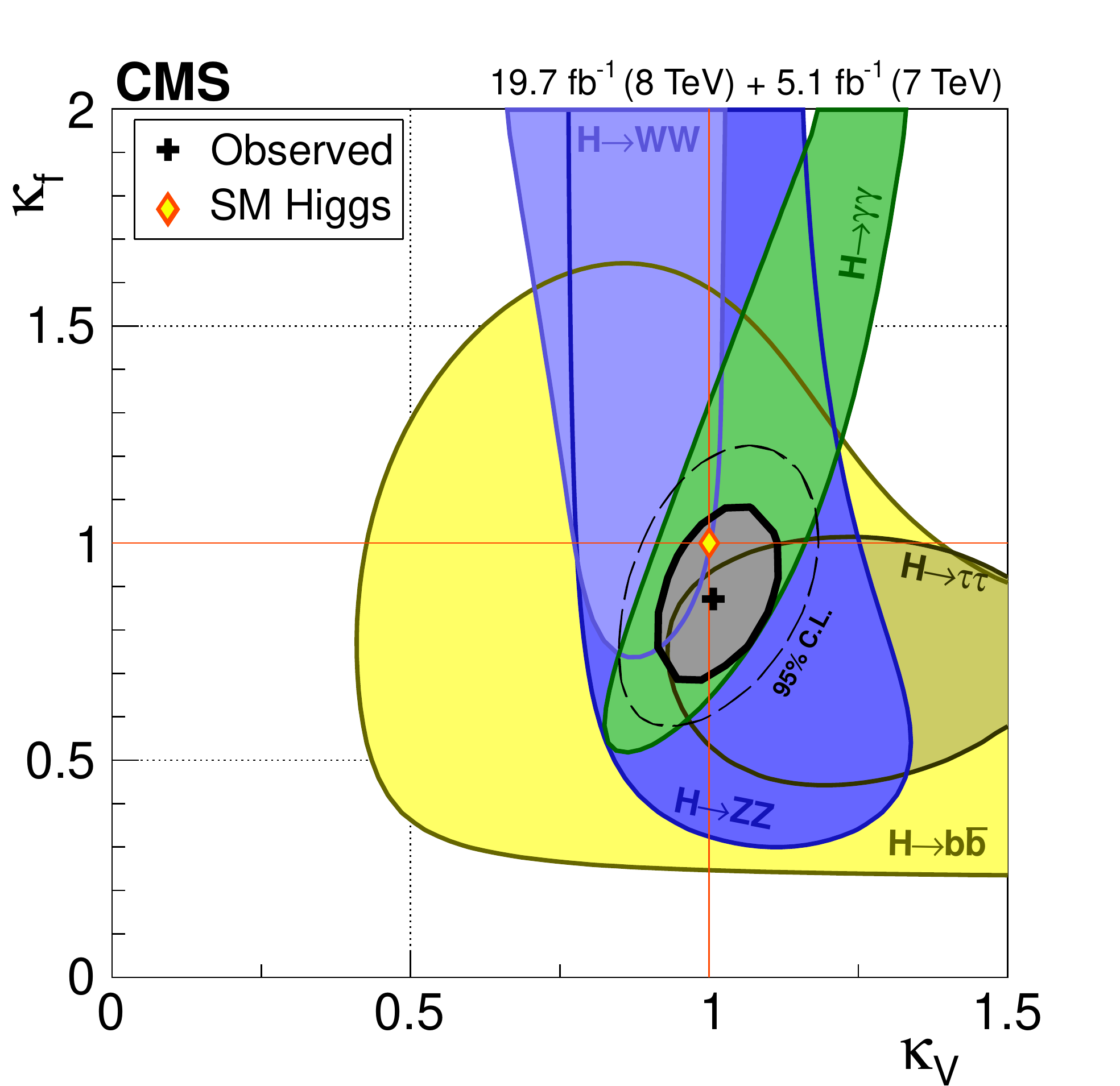}

\caption{
The 68\%~CL confidence regions for individual channels (coloured swaths) and for
the overall combination (thick curve) for the \kV and \kf parameters. The
cross indicates the global best-fit values. The dashed contour bounds the 95\%~CL
confidence region for the combination.
The diamond represents the SM expectation,
$(\kV, \kf)=(1,1)$.
The left plot shows the likelihood scan in two quadrants $(+,+)$ and $(+,-)$,
the right plot shows the positive quadrant only.
}
\label{fig:cVcF_subchannels}
\end{figure*}

Figure~\ref{fig:cVcF_2D} shows the 2D likelihood scan
over the $(\kV,\kf)$ parameter space.
While Fig.~\ref{fig:cVcF_2D}~(left) allows for different signs of $\kV$
and $\kf$, Fig.~\ref{fig:cVcF_2D}~(right) constrains the scan to
the $(+,+)$ quadrant that contains the SM expectation $(1,1)$.
The $(-,-)$ and $(-,+)$ quadrants are not shown since they are degenerate with
respect to the ones studied, with the implication that with the available
analyses we can only probe whether \kV and \kf have the same sign or different
signs.
Studies of the production of a Higgs boson associated with a single top quark
can, in principle, lift that degeneracy.

In Fig.~\ref{fig:cVcF_2D} the 68\%, 95\%, and 99.7\%~CL confidence regions
for $\kV$ and $\kf$ are shown with solid, dashed, and dotted curves, respectively.
The data are compatible with the expectation for the
standard model Higgs boson: the point $(\kV,\kf)=(1,1)$ is within the 68\%~CL
confidence region defined by the data.
Because of the way these compatibility tests are constructed, any
significant deviations from $(1,1)$ would not have a straightforward
interpretation within the SM and would imply BSM physics; the scale and sign of
the best-fit values in the case of significant deviations would guide us in
identifying the most plausible BSM scenarios.

Figure~\ref{fig:cVcF_subchannels} shows the results of this combined analysis in
the different decay mode groups.
The role and interplay of different channels is important. For example,
Fig.~\ref{fig:cVcF_2D}~(left) shows a region in the $(+,-)$ quadrant,
where \kV and \kf have opposite signs, which is excluded at the 95\%~CL but not
at the 99.7\%~CL; it can be seen in Fig.~\ref{fig:cVcF_subchannels}~(left) how
the combined exclusion in the $(+,-)$ quadrant is foremost due to the ability of
the \hgg decay to discern the relative sign between \kV and
\kf.
This is due to the destructive interference between the amplitudes of
the $\PW$ loops and top quark loops in the \hgg decay: $\kgam^2 \sim
1.59\;\kV^2 - 0.66\;\kV\kf+0.07\;\kf^2$; if \kV and \kf have
opposite signs, the interference becomes constructive, leading to a larger
\hgg branching fraction.
The shapes of the confidence regions for other decay channels are also
interesting: the analyses of decays to massive vector bosons
constrain \kV better than \kf, whereas the analyses of decays to fermions constrain \kf
better than \kV.
In the model used for this analysis, the total width scales as
$\kH^2 \sim 0.75\;\kf^2 + 0.25\;\kV^2$, reflecting the large expected
contributions from the bottom quark and $\PW$ boson.

The 95\%~CL confidence intervals for $\kV$ and $\kf$, obtained from a scan
where the other parameter is floated, are \kVBEST and \kFBEST,
respectively.

\subsection{Test for asymmetries in the couplings to fermions}

\label{sec:ldullq}

In models with two Higgs doublets (2HDM) \cite{2HDM:Lee:1973}, the couplings of
the neutral Higgs bosons to fermions can be substantially modified with respect to the couplings
predicted for the SM Higgs boson.
For example, in the minimal supersymmetric standard model
\cite{MSSM:Dimopoulos:1981}, the couplings of neutral Higgs bosons to up-type
and down-type fermions are modified, with the modification being the same for
all three generations and for quarks and leptons.
In more general 2HDMs, leptons can be made to virtually decouple from one Higgs
boson that otherwise behaves in a SM-like way with respect to
the $\PW$ bosons, $\cPZ$ bosons, and quarks.
Inspired by the possibility of such modifications to the fermion couplings,
we perform two combinations in which we allow for
different ratios of the couplings to down-type fermions and up-type fermions
($\ldu = \kd / \ku$)
or different ratios of the couplings to leptons and quarks
($\llq = \kl / \kq $).

\begin{figure*}[bpht]
\centering
\includegraphics[width=0.49\textwidth]{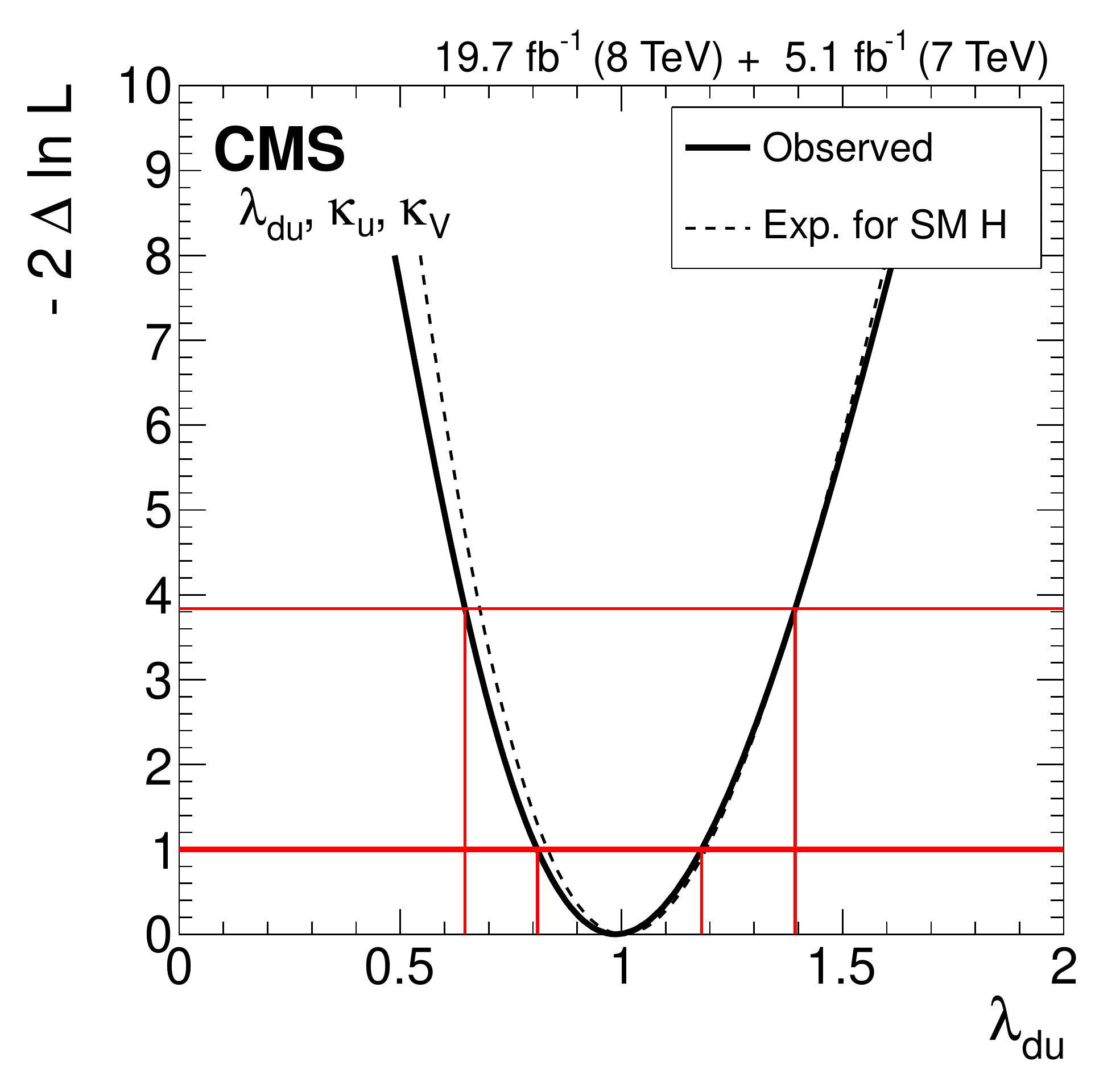} \hfill
\includegraphics[width=0.49\textwidth]{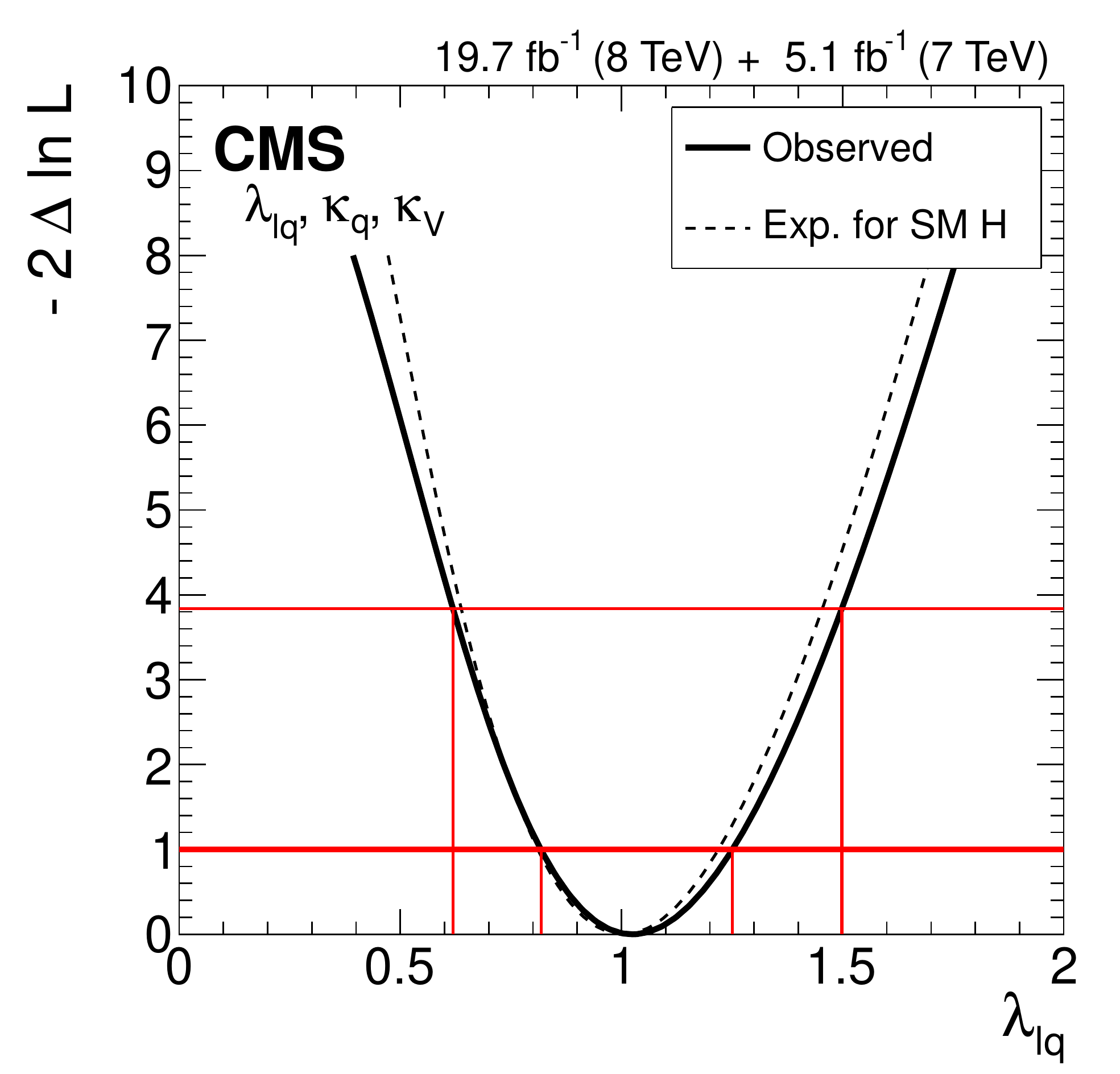}
\caption{
(Left)  Likelihood scan versus ratio of couplings to down/up fermions, \ldu,
with the two other free coupling modifiers, \kV and \ku,
profiled together with all other nuisance parameters.
(Right) Likelihood scan versus ratio of couplings to leptons and quarks, \llq,
with the two other free coupling modifiers, \kV and \kq,
profiled together with all other nuisance parameters.
}
\label{fig:fit_ldu_llq_scan}
\end{figure*}

Figure~\ref{fig:fit_ldu_llq_scan}~(left) shows the likelihood scan versus
$\ldu$, with $\kV$ and $\ku$ profiled together with all other
nuisance parameters.
Figure~\ref{fig:fit_ldu_llq_scan}~(right) shows the likelihood scan
versus $\llq$, with $\kV$ and $\kq$
profiled.
Assuming that both \ldu and \llq are positive, the 95\%~CL confidence
intervals are found to be \lduTwoSig and \llqTwoSig, respectively.
There is no evidence that different classes of fermions have different scaling
factors.

\subsection{Test of the scaling of couplings with the masses of SM
particles}
\label{sec:mepsc5}

\begin{figure*} [bpht]

\centering
\includegraphics[width=0.49\textwidth]{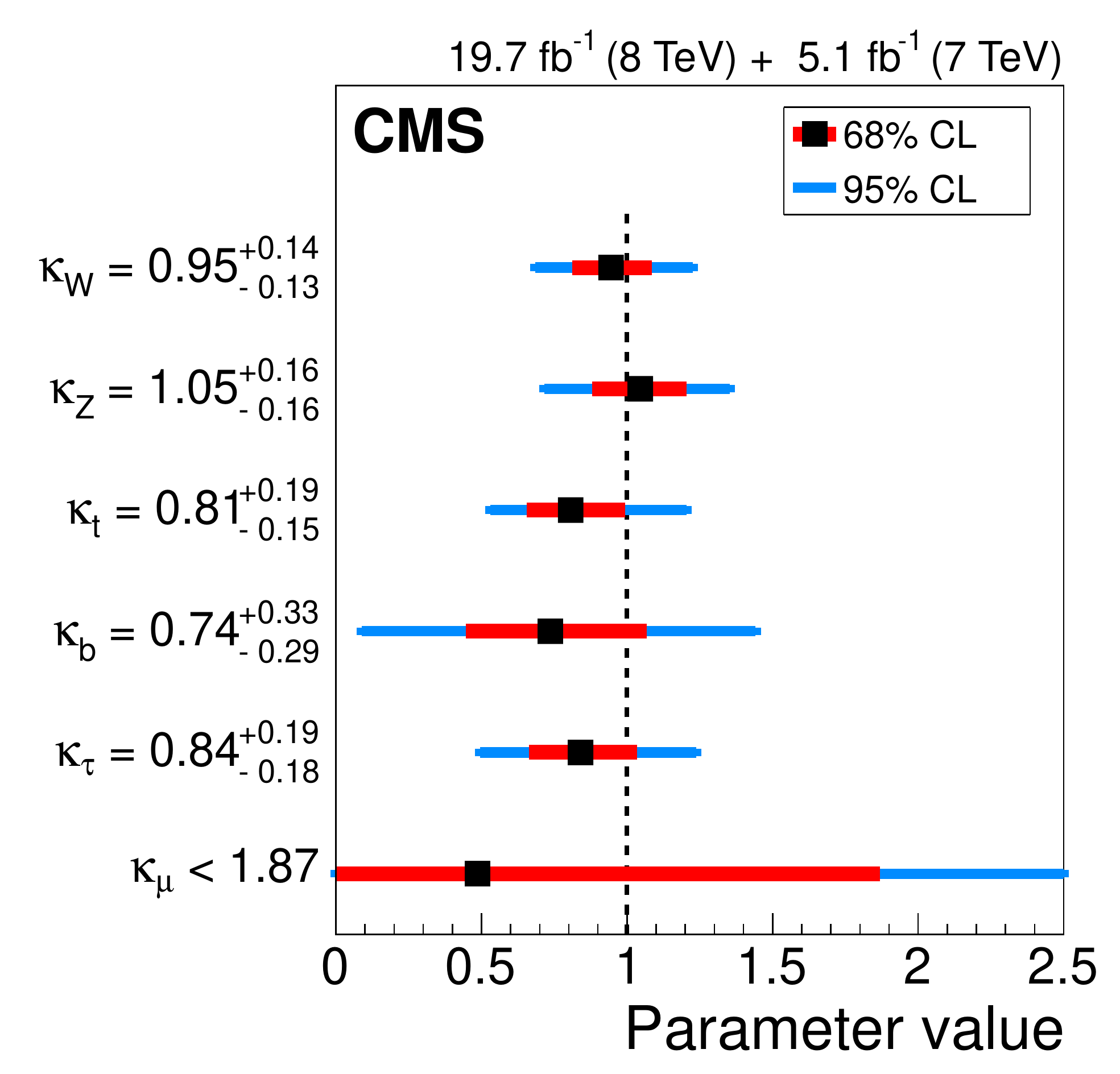}\hfill
\includegraphics[width=0.49\textwidth]{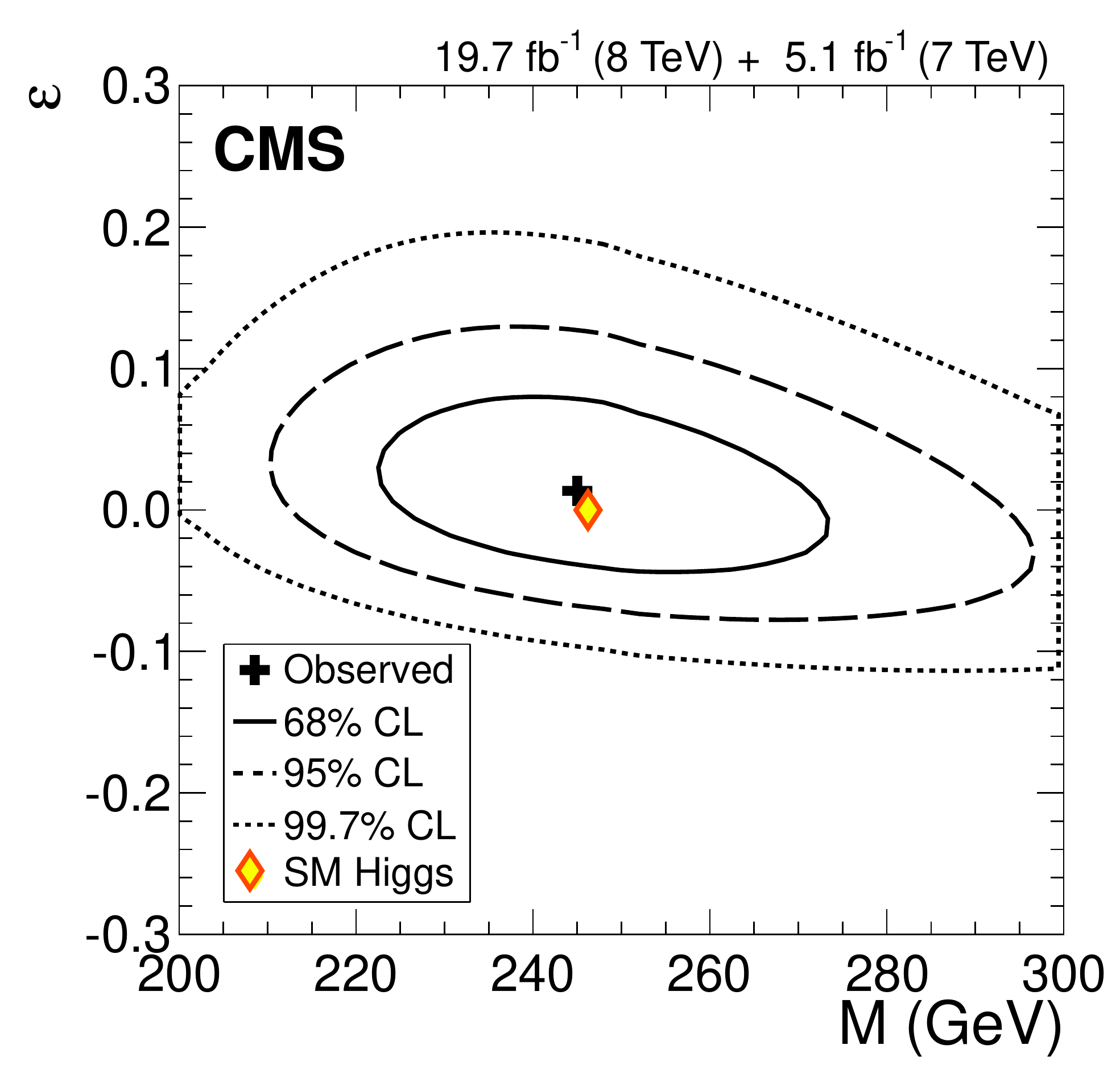}\\
\caption{
(Left) Results of likelihood scans for a model where the gluon and photon
loop-induced interactions with the Higgs boson are resolved in terms of the
couplings of other SM particles.
The inner bars represent the 68\%~CL confidence intervals while the
outer bars represent the 95\%~CL confidence intervals.
When performing the scan for one parameter, the other parameters in the
model are profiled.
(Right) The 2D likelihood scan for the $M$ and $\epsilon$ parameters of the
model detailed in the text.
        The cross indicates the best-fit values. The solid, dashed, and
        dotted contours show the
        68\%, 95\%, and 99.7\%~CL confidence regions, respectively.
        The diamond represents the SM expectation,
        $(M, \epsilon)=(v,0)$, where $v$ is the SM Higgs vacuum expectation
        value, $v=\SMVeV$.
}
\label{fig:mepsc5}
\end{figure*}

Under the assumption that there are no interactions of the Higgs boson other
than to the massive SM particles, the data allow a fit for deviations in \kW,
\kZ, \kb, \ktau, \ktop, and \kmu.
In this fit, the loop-induced processes ($\sigma_{\ggh}$,
$\Gamma_{\gluglu}$, and $\Gamma_{\gg}$) are expressed in terms of the above
tree-level $\kappa$ parameters and are scaled according to their SM loop
structure.
The result for this fit is displayed in Fig.~\ref{fig:mepsc5}~(left) and shows
no significant deviations from the SM expectation.
The small uncertainty in the \ktop parameter directly reflects the fact that
in this model, the \ggh production mode is being described in terms of \ktop and
\kb, $\kglu^2 \sim
1.11\;\ktop^2 +0.01\;\ktop\kb -0.12\;\kb^2$, such that \kb has a
small contribution.

In the SM, the Yukawa coupling between the Higgs boson
and the fermions, $\lambda_\mathrm{f}$, is proportional to the mass of the
fermion, $m_\mathrm{f}$.
This is in contrast with the coupling to weak bosons, $g_\mathrm{V}$,
which involves the square of the mass of the weak boson, $m_\mathrm{V}$.
With these differences in mind, it is possible to motivate a phenomenological
parameterization relating the masses of the fermions and weak bosons to the
corresponding $\kappa$ modifiers using two
parameters, $M$ and $\epsilon$~\cite{EllisYou2012,EllisYou2013}.
In such a model one has for each fermion
$\kf = v \; m_\mathrm{f}^{\epsilon} / M^{1+\epsilon}$
and for each weak boson
$\kV = v \; m_\mathrm{V}^{2\epsilon} / M^{1+2\epsilon}$,
where $v$ is the SM Higgs boson vacuum expectation value,
$v=\SMVeV$~\cite{RPP2014}.
The SM expectation, $\kappa_{i}=1$, is recovered when
$(M,\epsilon)=(v,0)$.
The parameter $\epsilon$ changes the
power with which the coupling scales with the particle mass; if the couplings
were independent of the masses of the particles, one would expect to find
$\epsilon=-1$.
To perform a fit to data, the
particle mass values need to be specified.
For leptons and weak bosons we have taken the values from Ref.~\cite{RPP2014}.
For consistency with theoretical calculations used in setting the SM
expectations, the top quark mass is taken to be \mtopTheo.
The bottom quark is evaluated at the scale of the Higgs boson mass,
$m_\mathrm{b}(\mH=\mX)=\mbAtmH$.
In the fit, the mass parameters are treated as constants.
The likelihood scan for $(M,\epsilon)$ is shown in Fig.~\ref{fig:mepsc5}~(right).
It can be seen that the data do not significantly deviate from the SM
expectation.
The 95\%~CL confidence intervals for the $M$ and $\epsilon$ parameters are
\mepsMTwoSig\GeV and \mepsETwoSig, respectively.

\begin{figure} [bpht]

\centering
\includegraphics[width=0.49\textwidth]{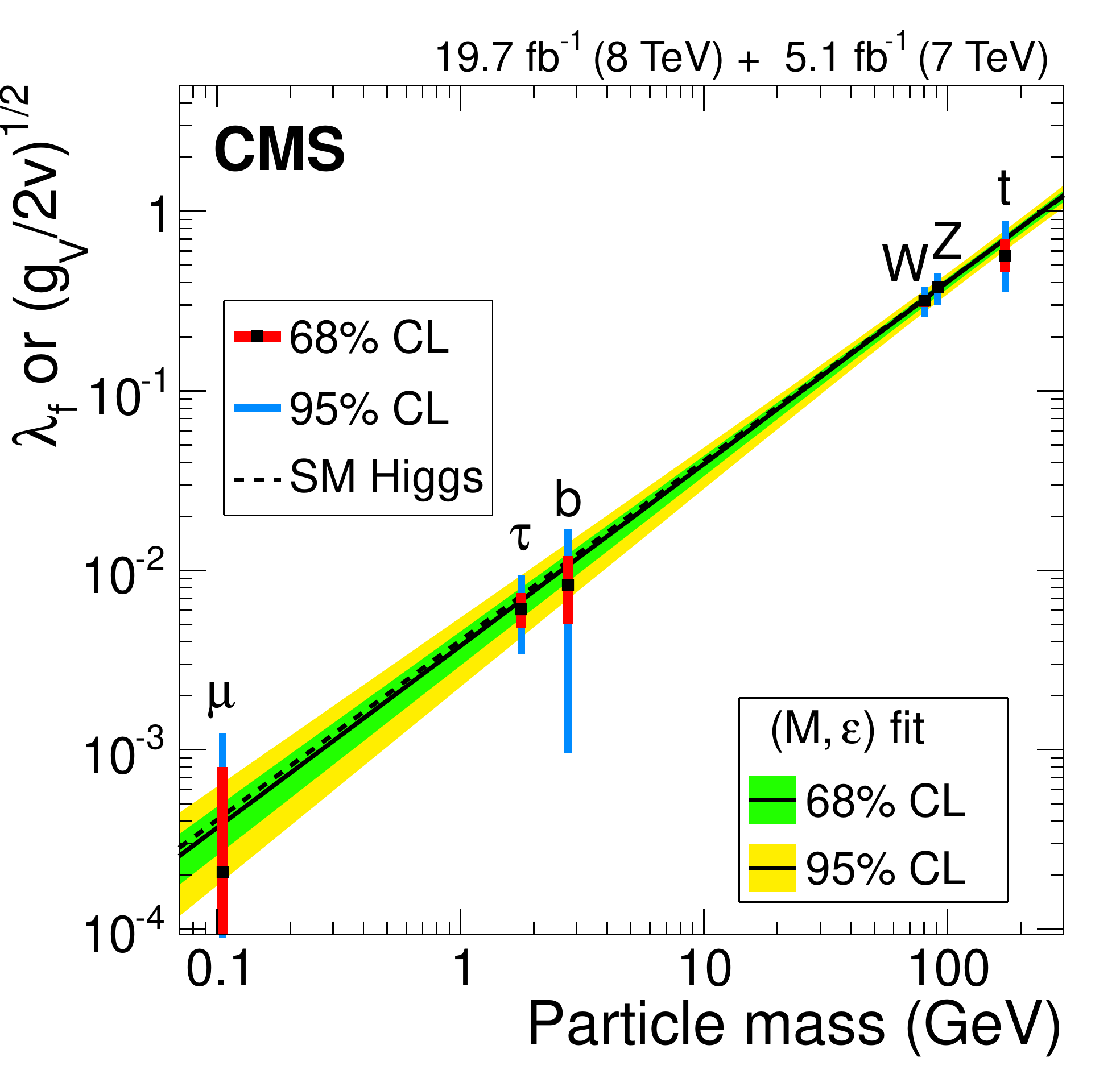}
\caption{
Graphical representation of the results obtained for the models
considered in Fig.~\ref{fig:mepsc5}.
The dashed line corresponds to the SM expectation.
The points from the fit in Fig.~\ref{fig:mepsc5}~(left) are placed at particle
mass values chosen as explained in the text.
The ordinates are different for fermions and massive vector bosons to take
into account the expected SM scaling of the coupling with mass, depending on the
type of particle.
The result of the $(M,\epsilon)$ fit from Fig.~\ref{fig:mepsc5}~(right) is shown
as the continuous line while the inner and outer bands represent the 68\% and
95\%~CL confidence regions.
}
\label{fig:mepsc5sum}
\end{figure}

The results of the two fits above are plotted versus the particle masses in
Fig.~\ref{fig:mepsc5sum}.
While the choice of the mass values for the abscissas is discussed above,
to be able to show both Yukawa and weak boson couplings in the same plot
requires a transformation of the results of the $\kappa$ fit.
Since $g_V \sim \kV 2m_{\mathrm{V}}^2/v$ and $\lambda_{\mathrm{f}} \sim
\kf m_\mathrm{f}/v$, we have chosen to plot a ``reduced'' weak boson coupling,
$\sqrt{g_\mathrm{V}/(2v)}=\kV^{1/2} m_\mathrm{V}/v$.
This choice allows fermion and weak boson results to be plotted
together, as shown in Fig.~\ref{fig:mepsc5sum}, but implies that
the uncertainties for \kW and \kZ will seem to be reduced.
This simply reflects the square root in the change of variables and not any gain
of information with respect to the $\kappa$ fit shown
Fig.~\ref{fig:mepsc5}~(left).
The result of the $(M,\epsilon)$ fit is shown in
Fig.~\ref{fig:mepsc5sum} as the band around the dashed line that
represents the SM expectation.
While the existing measurement of the scaling factor for the coupling of the
boson with muons is clearly imprecise, the picture that arises from covering
more than three orders of magnitude in particle mass is that the boson couples
differently to the different particles and that those couplings are
related to the mass of each particle.
This is further supported by upper limits set in searches for \hee decays: when
assuming the production cross sections predicted in the SM, the branching
fraction is limited to be $\mathcal{B}(\hee)<1.9\ten{-3}$ at the
95\%~CL~\cite{CMSHmmLegacyRun1}.

\subsection{Test for the presence of BSM particles in loops}
\label{sec:C2BSM}

The manifestation of BSM physics can considerably modify the Higgs boson
phenomenology even if the underlying Higgs boson sector in the model remains
unaltered.
Processes that are loop-induced at leading order, such as the \hgg decay and
\ggh production, can be particularly sensitive to the presence of new
particles.
Therefore, we combine and fit the data for the scale factors
for these two processes, $\kgam$ and $\kglu$.
The partial widths associated
with the tree-level production processes and decay modes are assumed to be
those expected in the SM, and the total width scales as $\kH^2\sim 0.0857
\;\kglu^2 +0.0023\;\kgam^2 + 0.912$.

\begin{figure}[bht]
\centering
\includegraphics[width=0.49\textwidth]{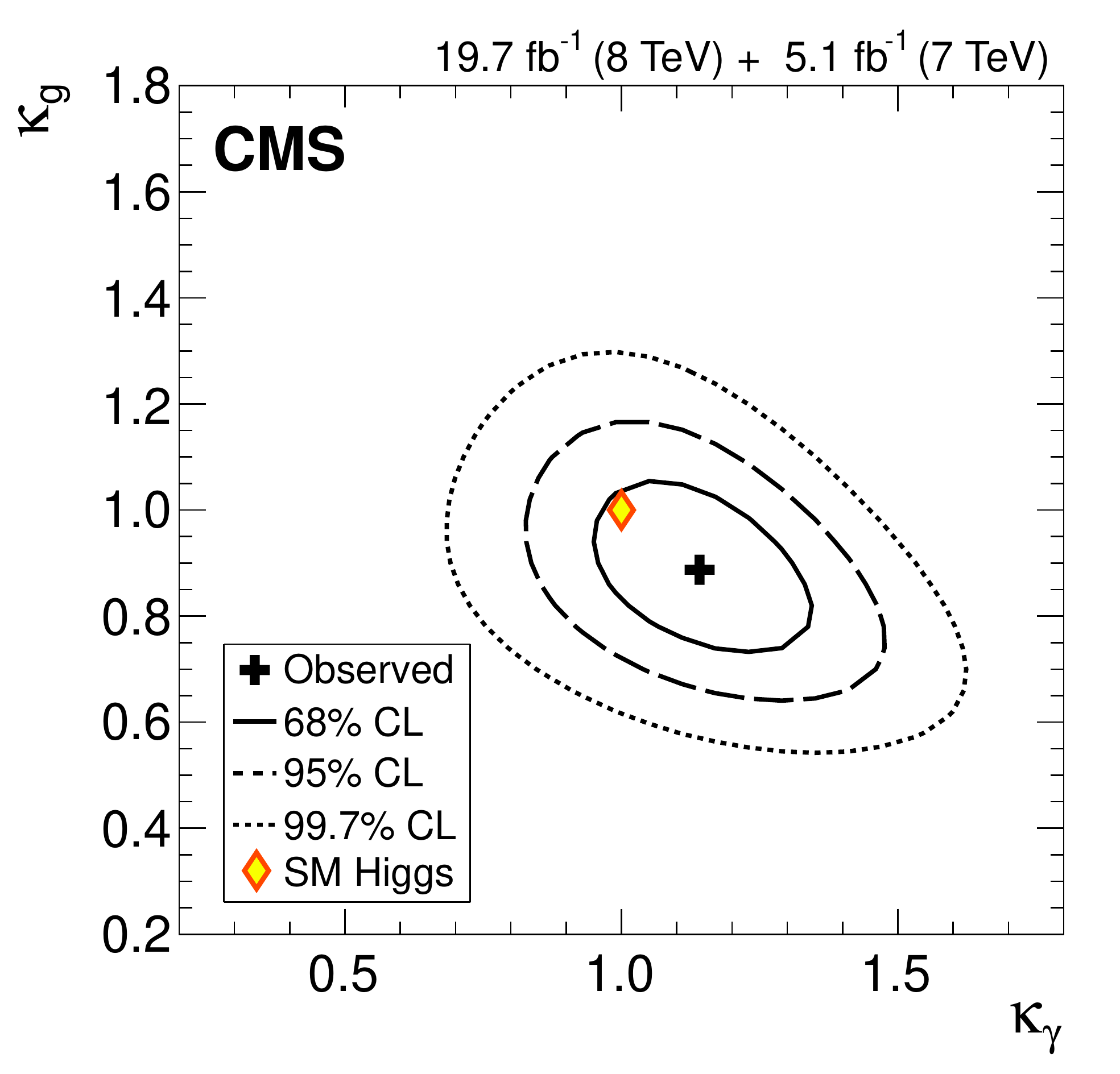}

\caption{
The 2D likelihood scan for the $\kglu$ and $\kgam$ parameters,
        assuming that $\Gamma_{\mathrm{BSM}}=0$.
        The cross indicates the best-fit values. The solid, dashed, and
        dotted contours show the
        68\%, 95\%, and 99.7\%~CL confidence regions, respectively.
        The diamond represents the SM expectation,
        $(\kgam, \kglu)=(1,1)$.
The partial widths associated with the tree-level production processes
and decay modes are assumed to be unaltered ($\kappa = 1$).
}
\label{fig:BSM1}
\end{figure}

Figure~\ref{fig:BSM1} shows
the 2D likelihood scan for the $\kglu$ and $\kgam$ parameters, assuming
that \mbox{$\Gamma_{\mathrm{BSM}}=0$}.
The results are compatible with the expectation for the SM Higgs boson, with the
point $(\kgam, \kglu)=(1,1)$ within the 68\%~CL confidence region defined by
the data.
The best-fit point is $(\kgam, \kglu)=(\CTWOkgamBEST, \CTWOkgluBEST)$.
The 95\%~CL confidence interval for $\kgam$, when profiling $\kglu$ and all
nuisance parameters, is \CTWOkgam.
For $\kglu$, the 95\%~CL confidence interval is \CTWOkglu, when profiling
$\kgam$ and all other nuisance parameters.

Another way in which BSM physics may manifest itself is through the
decay of the boson into BSM particles.
To explore this possibility, we consider a further parameter that allows for a
partial decay width into BSM particles,
$\BRBSM=\Gamma_{\mathrm{BSM}}/\Gamma_{\text{tot}}$.
In this case, the total width scales as $\kH^2\sim(0.0857\;\kglu^2 +
0.0023\;\kgam^2 + 0.912)/(1-\BRBSM)$.

\begin{figure*}[bht]

\includegraphics[width=0.329\textwidth]{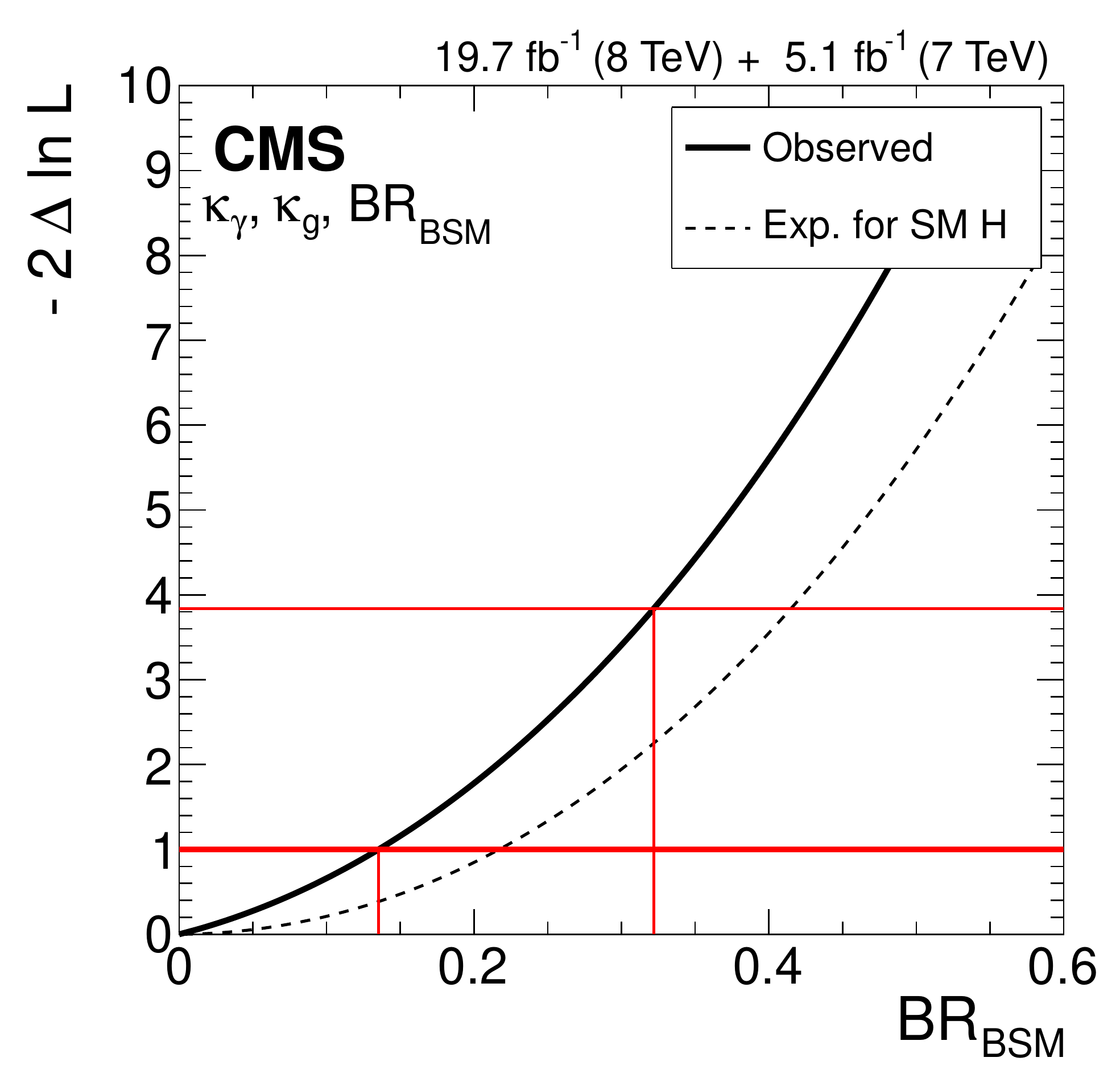}\hfill
\includegraphics[width=0.329\textwidth]{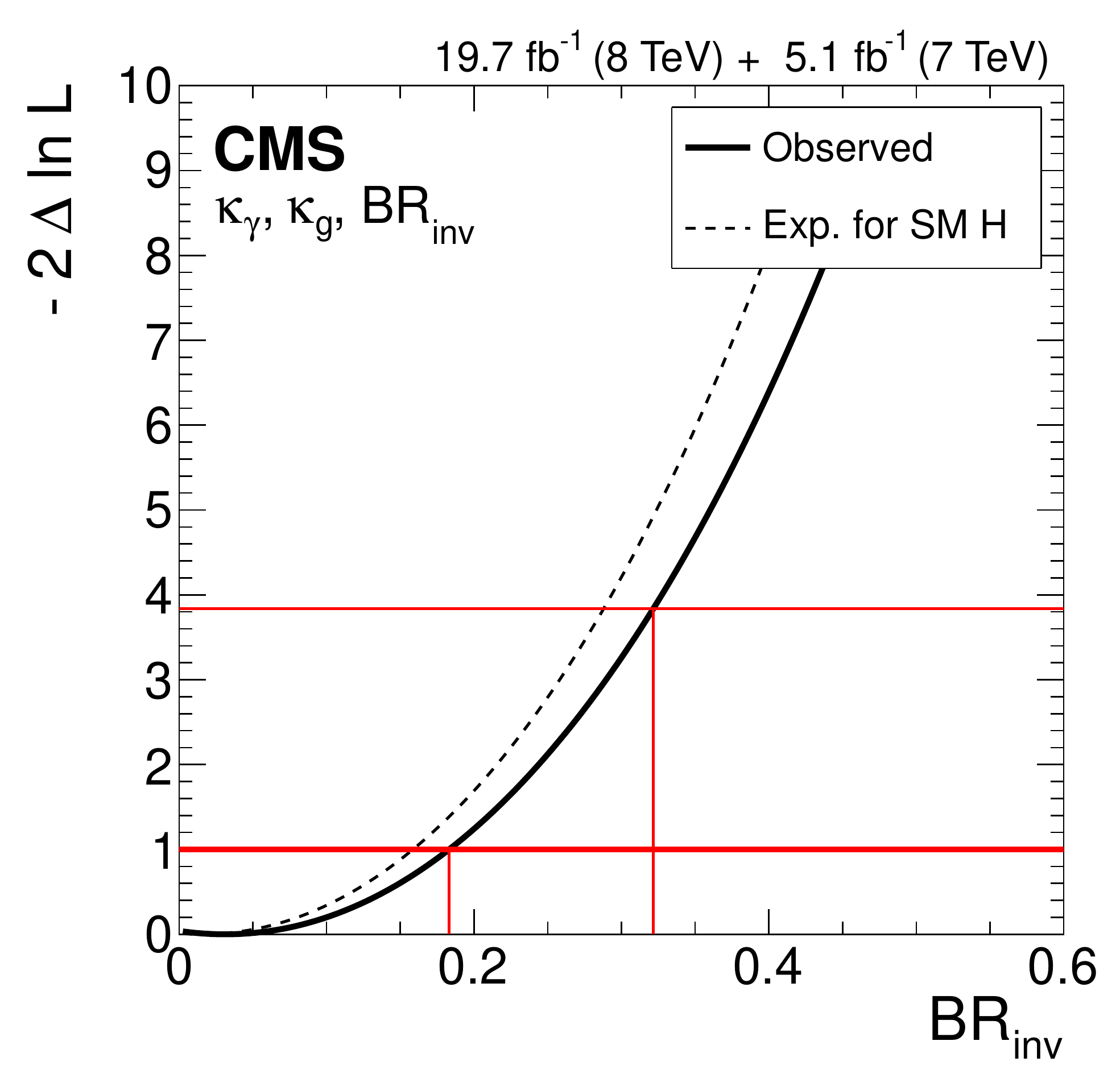}\hfill
\includegraphics[width=0.329\textwidth]{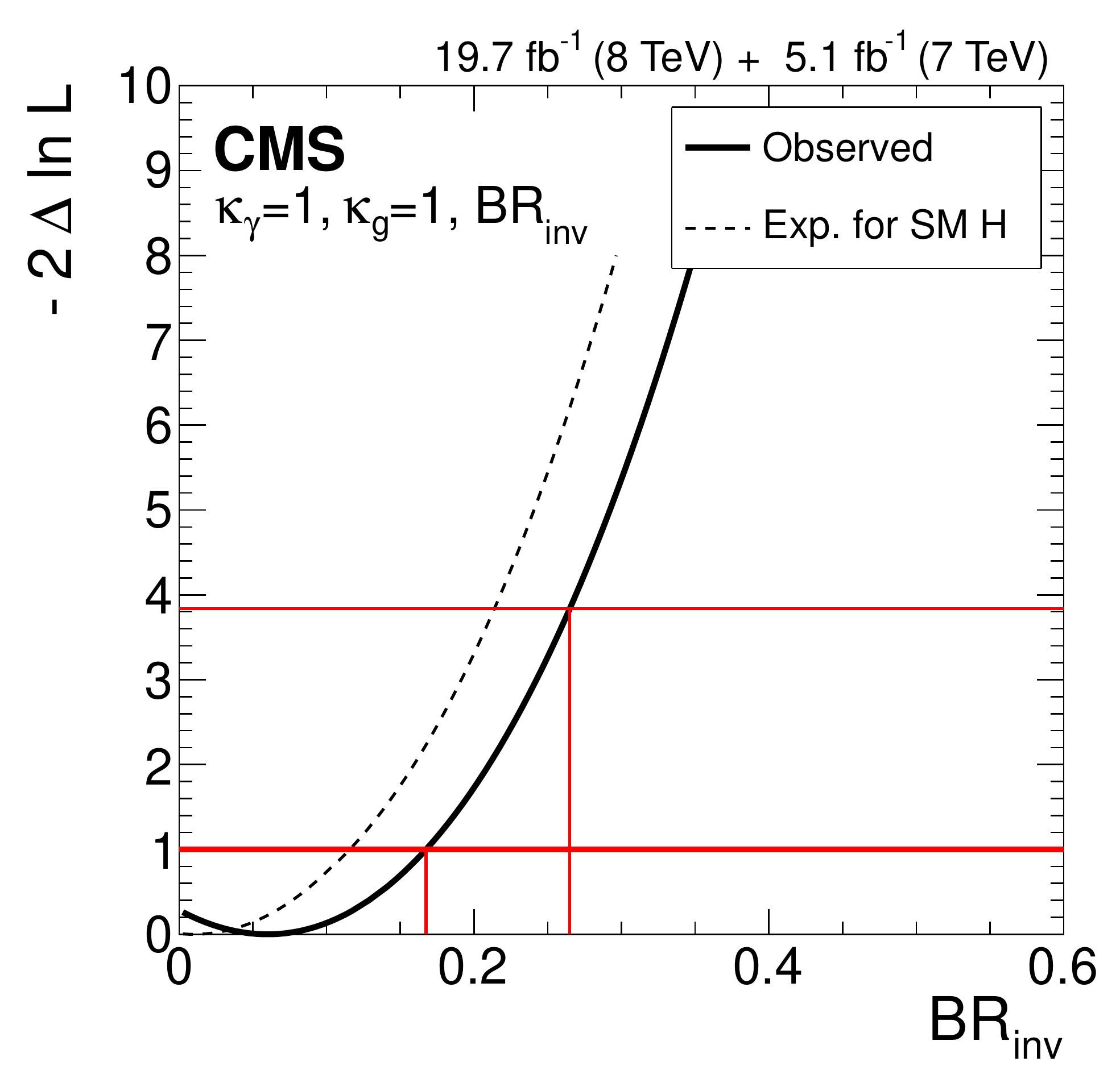}

\caption{
(Left) The likelihood scan versus $\BRBSM=\Gamma_{\mathrm{BSM}}/\Gamma_{\text{tot}}$.
The solid curve represents the observation and the dashed curve
indicates the expected median result in the presence of the SM Higgs boson.
The partial widths associated with the tree-level production processes
and decay modes are assumed to be as expected in the SM.
(Middle) Result when also combining with data from the \hinv searches,
thus assuming that $\BRBSM=\BRinv$, \ie that there are no undetected decays,
$\BRundet=0$.
(Right) Result when further assuming that $\kglu=\kgam=1$ and combining with the
data from the \hinv searches.}
\label{fig:BSM2}

\end{figure*}

Figure~\ref{fig:BSM2}~(left) shows the likelihood scan versus
\BRBSM, with $\kglu$ and $\kgam$ constrained to be positive and profiled
together with all other nuisance parameters.
While under the SM hypothesis the expected 95\%~CL confidence interval for
\BRBSM is \BRBSMTwoSigEXP, the data are such that the 95\%~CL confidence
interval for \BRBSM is \BRBSMTwoSig, narrower than the expectation.
The best fit in data also takes into account variations in \kglu and \kgam,
particularly the preference for \kglu smaller than unity in data, which
influences the observed limit on \BRBSM.

{\tolerance=500
A further step can be taken by also including the data from the
searches for \hinv.
The \hinv searches reported an observed (expected) upper limit on \BRinv
of 0.58 (0.44) at the 95\%~CL~\cite{CMSHinvLegacyRun1}.
When including the \hinv search results in the combined analysis, one can only
obtain bounds assuming that there are no undetected decay modes, $\BRundet=0$,
\ie that $\BRBSM=\BRinv$.
The results for the likelihood scan as a function of $\BRinv(\BRundet=0)$ when including the
data from the \hinv searches is shown in Fig.~\ref{fig:BSM2}~(middle).
The expected 95\%~CL confidence interval for
$\BRinv(\BRundet=0)$ under the SM hypothesis is \BRBSMHinvTwoSigEXP, 31\%
narrower than in the above case studied without the \hinv data, a
reflection of the added power of the \hinv analysis.
On the other hand, the 95\%~CL confidence interval for $\BRinv(\BRundet=0)$ in
data is \BRBSMHinvTwoSig, similar to the result obtained without including the
\hinv data, because the observed upper limit on $\BRinv$ was found to be larger
than expected in those searches.
It should be noted that the shape of the observed curve changes
substantially and the inclusion of the \hinv data leads to a very shallow
minimum of the likelihood when $\BRinv(\BRundet=0)=\BRBSMHinvBestFit$.
\par}

{\tolerance=500
Finally, one may further set $\kglu=\kgam=1$, which effectively
implies $\kappa_i=1$, \ie assumes that the couplings to all SM
particles with mass are as expected from the SM.
From the combined analysis including the data from the \hinv searches, we
can thus obtain bounds on $\BRinv(\BRundet=0,\kappa_i=1)$.
The likelihood scan results are shown in Fig.~\ref{fig:BSM2}~(right).
The expected 95\%~CL confidence interval for $\BRinv(\BRundet=0,\kappa_i=1)$
under the SM hypothesis is \BRBSMHinvKSMTwoSigEXP, which is 28\% narrower than
in the previous paragraph, a reflection of the total width now being fixed to the SM expectation.
The 95\%~CL confidence interval for $\BRinv(\BRundet=0,\kappa_i=1)$
in data is \BRBSMHinvKSMTwoSig, showing again a shallow minimum of the likelihood when
$\BRinv(\BRundet=0,\kappa_i=1)=\BRBSMHinvKSMBestFit$.
\par}

The results obtained from the different combined analyses presented in
Fig.~\ref{fig:BSM2} show the added value from combining the \hinv searches with
the visible decay measurements, with the expected 95\%~CL combined upper limit
on \BRinv being up to a factor of two smaller than either, depending on the
assumptions made.


\subsection{Test of a model with scaling factors for SM particles}

\label{sec:c6}

After having examined the possibility for BSM physics to manifest itself in
loop-induced couplings while fixing all the other scaling factors, we now
release the latter assumption.
For that, we explore a model with six independent coupling modifiers
and make the following assumptions:
\begin{itemize}
\item The couplings to $\PW$ and $\cPZ$ bosons scale with a common parameter
$\kV=\kW=\kZ$.
\item The couplings to third generation fermions, \ie the bottom quark,
tau lepton, and top quark, scale independently with \kb, \ktau, and
\ktop, respectively.
\item The effective couplings to gluons and photons, induced by loop diagrams,
scale with free parameters $\kglu$ and $\kgam$, respectively.
\item The partial width $\Gamma_{\mathrm{BSM}}$ is zero.
\end{itemize}

\begin{figure}[bpht]
\centering
\includegraphics[width=0.49\textwidth]{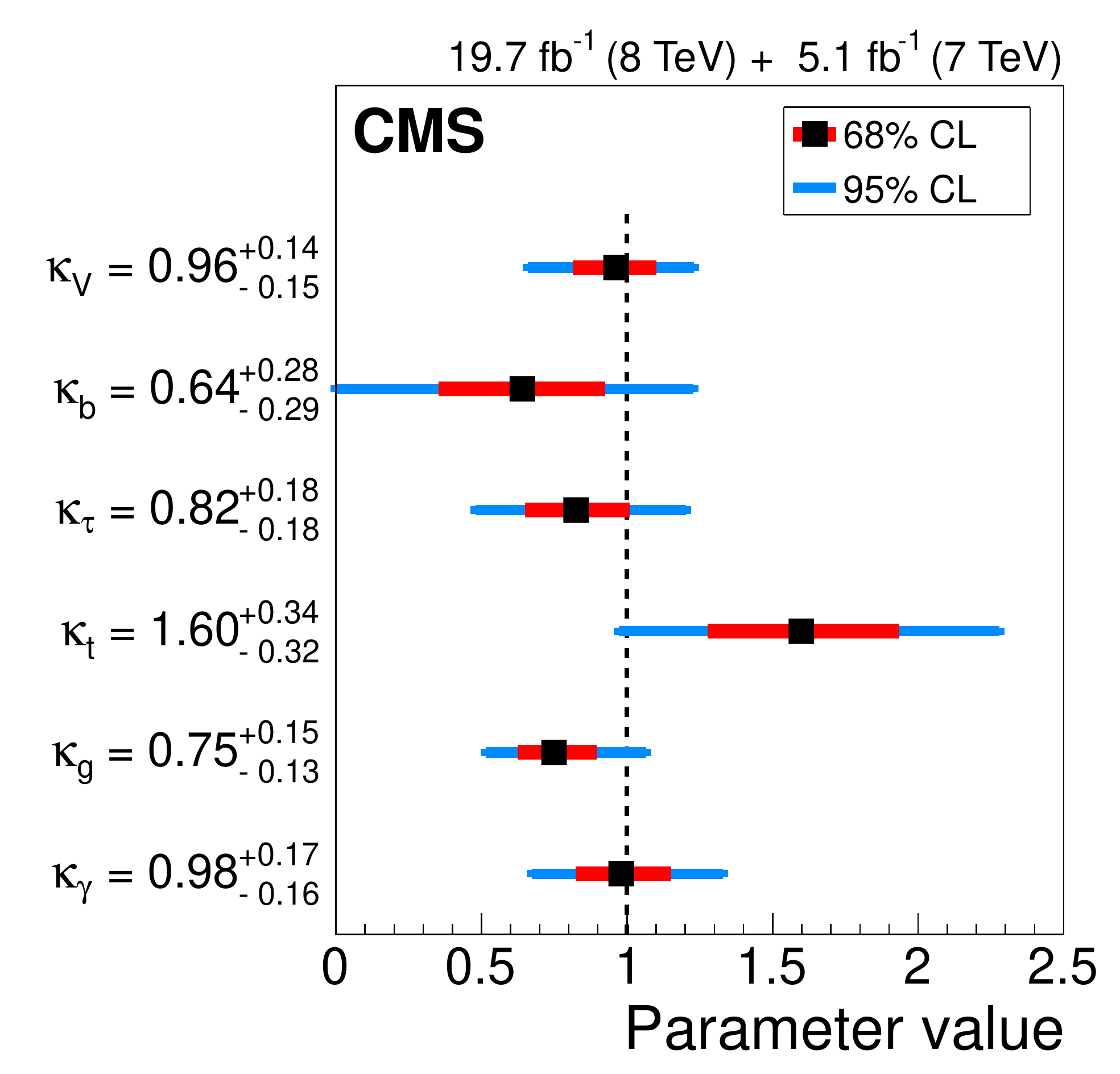}
\caption{
Likelihood scans for parameters in a model with coupling scaling
factors for the SM particles, one coupling at a time while profiling the
remaining five together with all other nuisance parameters; from top to bottom:
\kV ($\PW$ and $\cPZ$ bosons),
\kb (bottom quarks),
\ktau (tau leptons),
\ktop (top quarks),
\kglu (gluons; effective coupling), and
\kgam (photons; effective coupling).
The inner bars represent the 68\%~CL confidence intervals
while the outer bars represent the 95\%~CL confidence intervals.
}
\label{fig:C6}
\end{figure}

A likelihood scan for each of the six coupling modifiers is performed while
profiling the other five, together with all other nuisance parameters; the
results are shown in Fig.~\ref{fig:C6}.
With this set of parameters, the \ggh-production measurements will constrain
\kglu, leaving the measurements of \tth production to constrain \ktop, which
explains the best-fit value, $\ktop=\CSIXktopOneSig$.
The current data do not show any statistically significant deviation
with respect to the SM Higgs boson hypothesis.
For every $\kappa_i$ probed, the measured 95\%~CL confidence interval contains
the SM expectation, $\kappa_i=1$.
A goodness-of-fit test between the parameters measured in this model and the SM
prediction yields a {$\chisqdof = 7.5 / 6$},
which corresponds to an asymptotic \pval of {0.28}.  

\subsection{Test of a general model without assumptions on the total
width}

\label{sec:kgZ}

Given the comprehensiveness of the set of analyses being combined, we can
explore the most general model proposed in Ref.~\cite{LHCHXSWG3}, which makes
no assumptions on the scaling of the total width.
In this model, the total width is not rescaled according to the different
$\kappa_{i}$ values as a dependent parameter, but is rather left as a free
parameter, embedded in $\kgluZ=\kglu \kZ/\kH$.
All other parameters of interest are expressed as
ratios between coupling scaling factors, $\lambda_{ij}=\kappa_{i}/\kappa_{j}$.

\begin{figure}[bpht]
\centering
\includegraphics[width=0.49\textwidth]{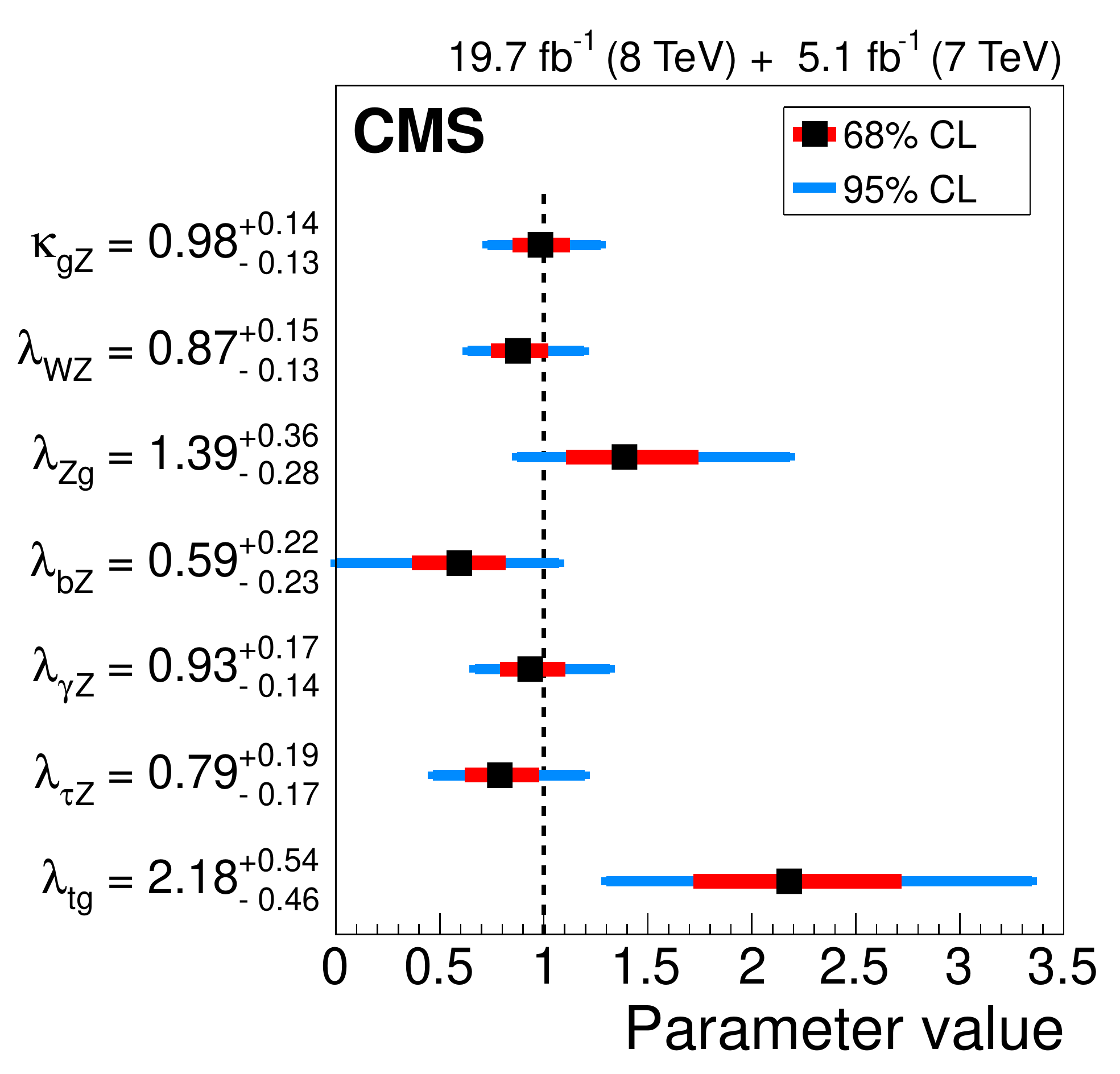}
\caption{
Likelihood scans for parameters in a model without assumptions on the total
width and with six coupling modifier ratios, one parameter at a time while
profiling the remaining six together with all other nuisance parameters; from top to bottom:
\kgluZ($=\kglu \kZ/\kH$),
\lWZ($=\kW/\kZ$),
\lZglu($=\kZ/\kglu$),
\lbZ($=\kb/\kZ$),
\lgamZ($=\kgam/\kZ$),
\ltauZ($=\ktau/\kZ$), and
\ltopglu($=\ktop/\kglu$).
The inner bars represent the 68\%~CL confidence intervals
while the outer bars represent the 95\%~CL confidence intervals.
}
\label{fig:ratios}
\end{figure}

A likelihood scan for each of the parameters \kgluZ, \lWZ, \lZglu, \lbZ, \lgamZ,
\ltauZ, and \ltopglu is performed while profiling the other six, together with
all other nuisance parameters.
The results are shown in Fig.~\ref{fig:ratios} and are in line with those
found in Section~\ref{sec:c6}.

\subsection{Constraints on \texorpdfstring{\BRBSM}{BR(BSM)} in a scenario with free couplings}
\label{sec:C6BSM}

An alternative and similarly general scenario can be built by allowing for
$\Gamma_{\mathrm{BSM}}>0$.
In order to avoid the degeneracy through which the total width and the
coupling scaling factors can compensate each other, we constrain $\kV \le 1$, a
requirement that holds in a wide class of models, namely in any model with an
arbitrary number of Higgs doublets, with and without additional Higgs singlets
\cite{LHCHXSWG3}.
The model has the following parameters: \kV, \kb, \ktau, \ktop, \kglu, \kgam,
and \BRBSM.
This is a much more general treatment than that performed in
Section~\ref{sec:C2BSM}, where only the loop-induced couplings to photons and
gluons were allowed to deviate from the SM expectation.
As in Section~\ref{sec:C2BSM}, this model also allows for
a combined analysis with the data from the \hinv searches.

\begin{figure*}[bpht]
\centering
\includegraphics[width=0.49\textwidth]{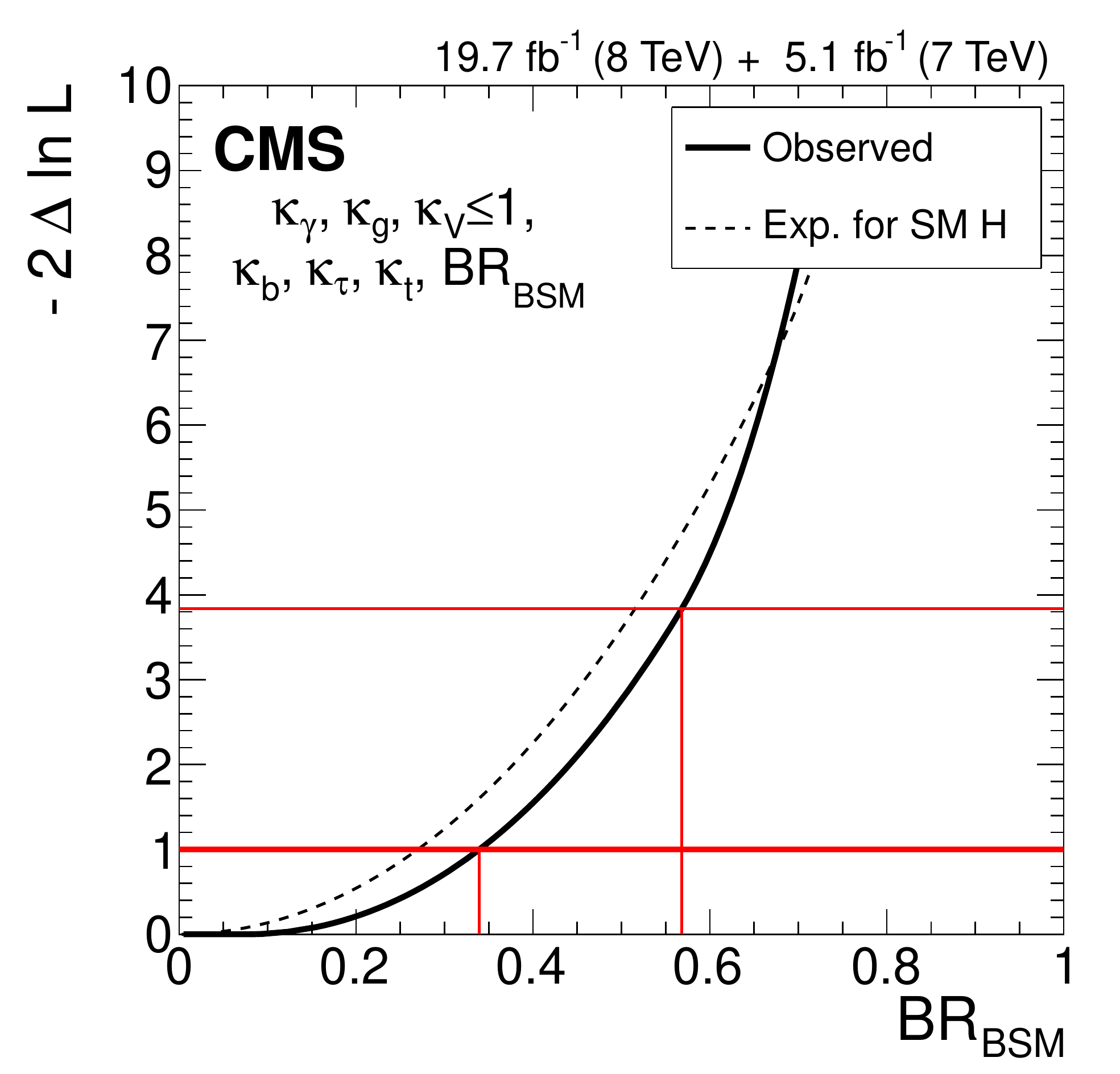}\hfill
\includegraphics[width=0.49\textwidth]{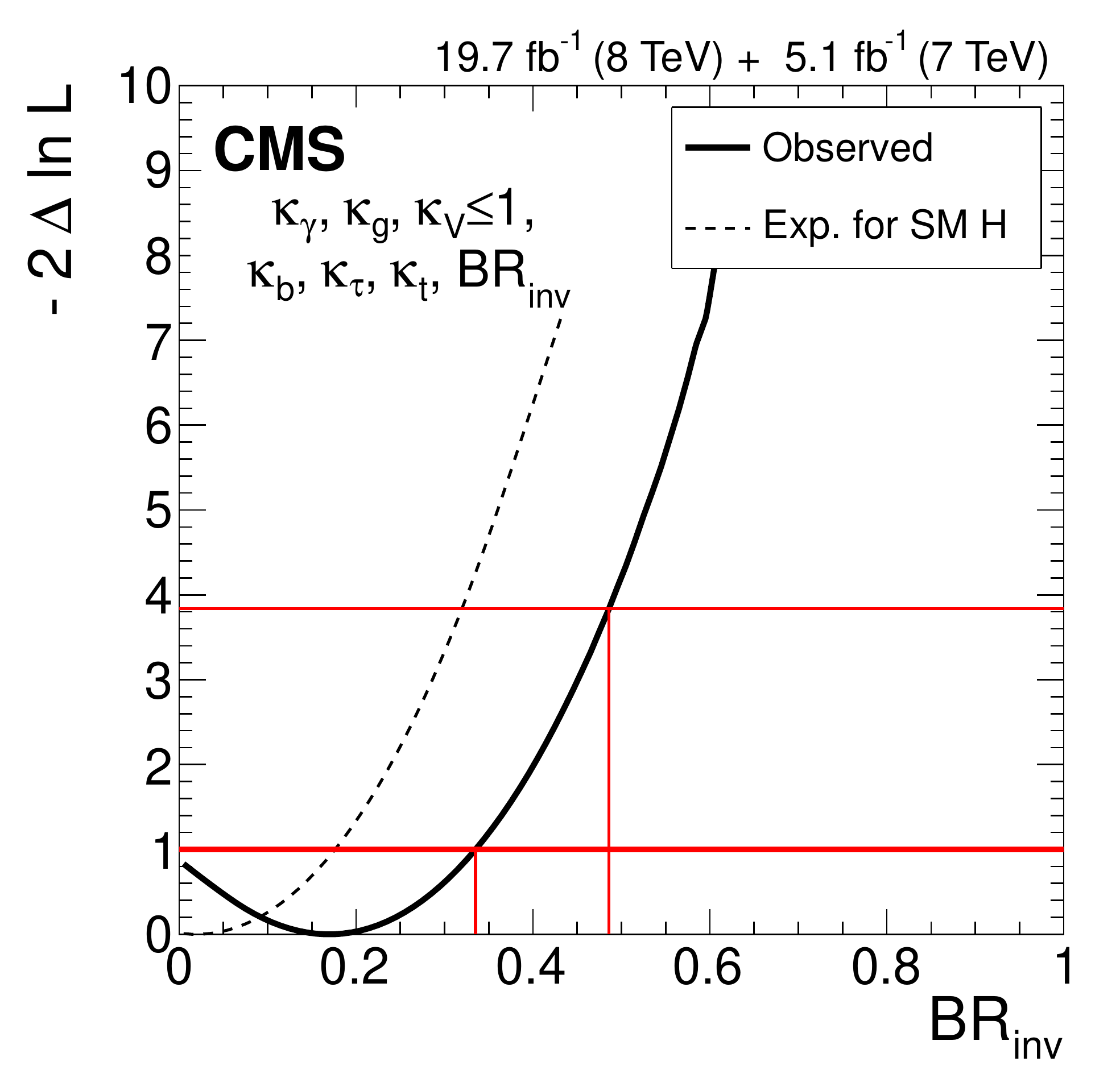}
\caption{
(Left)
Likelihood scan versus
$\BRBSM=\Gamma_{\mathrm{BSM}}/\Gamma_{\text{tot}}$.
The solid curve represents the observation in data and the dashed curve
indicates the expected median result in the presence of the SM Higgs boson.
The modifiers for both the tree-level and loop-induced couplings are profiled,
but the couplings to the electroweak bosons are assumed to be bounded by the SM
expectation ($\kV \le 1$).
(Right)
Result when also combining with data from the \hinv searches, thus
assuming that $\BRBSM=\BRinv$, \ie $\BRundet=0$.}
\label{fig:BSM3}
\end{figure*}

Figure~\ref{fig:BSM3}~(left) shows the likelihood scan versus \BRBSM derived
in this scenario, while profiling all the other coupling modifiers and nuisance
parameters.
Within these assumptions, the 95\%~CL confidence interval for \BRBSM in data
is \BRBSMCSeven, while the expected interval for the SM hypothesis is
\BRBSMCSevenEXPTwoSig.

Assuming that there are no undetected decay modes, $\BRundet=0$, it follows that
$\BRBSM=\BRinv$ and the data from the searches for \hinv can be combined with
the data from the other channels to set bounds on \BRinv.
The likelihood scan for such a model and combination is shown in
Fig.~\ref{fig:BSM3}~(right).
The 95\%~CL confidence interval for \BRinv in data
is \BRinvHinvCSeven, while the expected interval for the SM hypothesis is
\BRinvHinvCSevenEXPTwoSig.
The difference between the expected and observed confidence intervals reflects
the results of the \hinv analysis that reported an observed (expected) upper
limit on \BRinv of 0.58 (0.44) at the 95\%~CL~\cite{CMSHinvLegacyRun1}.

\begin{figure*}[bpht]
\centering
\includegraphics[width=0.49\textwidth]{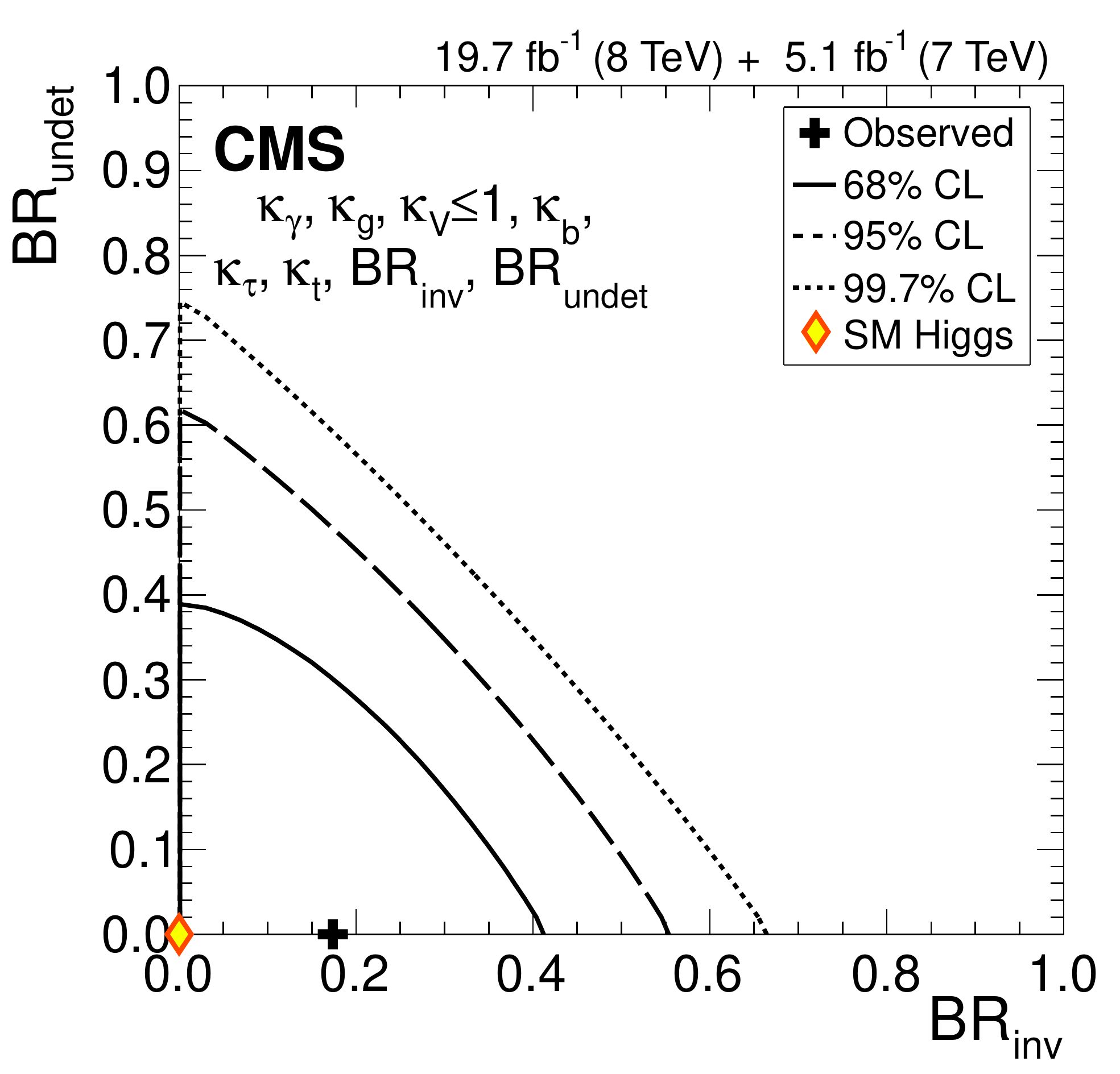}\hfill
\includegraphics[width=0.49\textwidth]{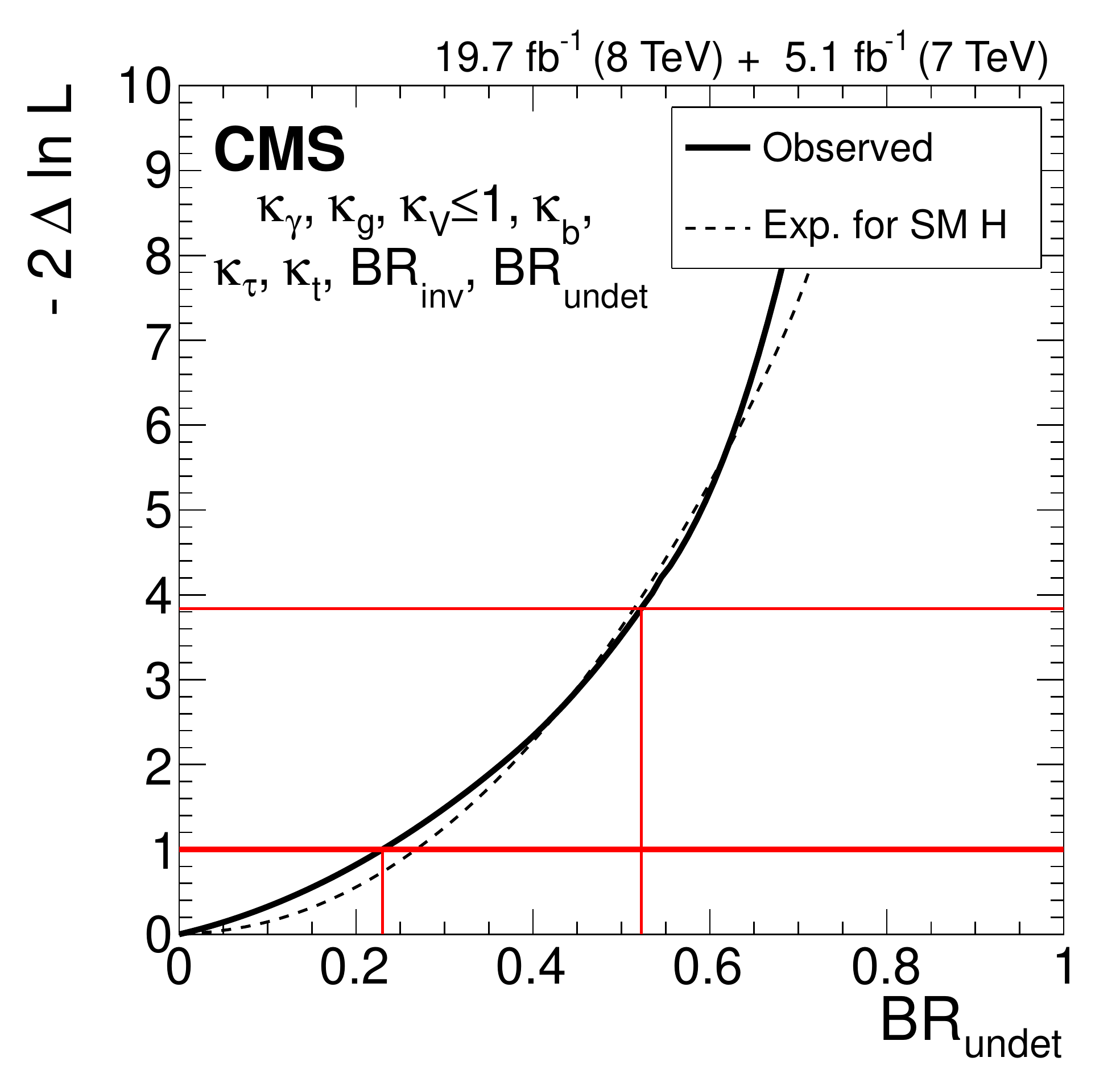}
\caption{
(Left)
The 2D likelihood scan for the \BRinv and \BRundet parameters for a combined
analysis of the \hinv search data and visible decay channels.
The cross indicates the best-fit values.
The solid, dashed, and dotted contours show the 68\%, 95\%, and 99.7\%~CL
confidence regions, respectively.
The diamond represents the SM expectation, $(\BRinv, \BRundet)=(0,0)$.
(Right)
The likelihood scan versus \BRundet.
The solid curve represents the observation in data and the dashed curve
indicates the expected median result in the presence of the SM Higgs boson.
\BRinv is constrained by the data from the \hinv searches and modifiers for both
the tree-level and loop-induced couplings are profiled, but the couplings to
the electroweak bosons are assumed to be bounded by the SM expectation ($\kV \le 1$).}
\label{fig:BSM4}
\end{figure*}

{\tolerance=500
Finally, instead of simply assuming $\BRundet=0$, a simultaneous
fit for \BRinv and \BRundet is performed.
In this case, the data from the \hinv searches constrains \BRinv, while the
visible decays constrain $\BRBSM=\BRinv+\BRundet$.
The 2D likelihood scan for $(\BRinv,\BRundet)$ is shown in
Fig.~\ref{fig:BSM4}~(left), while Fig.~\ref{fig:BSM4}~(right) shows the
likelihood scan for \BRundet when profiling all other parameters, \BRinv
included.
The 95\%~CL confidence interval for \BRundet in data
is \BRundetHinvCEight, while the expected interval for the SM hypothesis is
\BRundetHinvCEightEXPTwoSig.
\par}

\subsection{Summary of tests of the compatibility of the data with the
SM Higgs boson couplings}

\label{sec:summary}

\begin{figure}[bpht]
\centering
\includegraphics[width=0.49\textwidth]{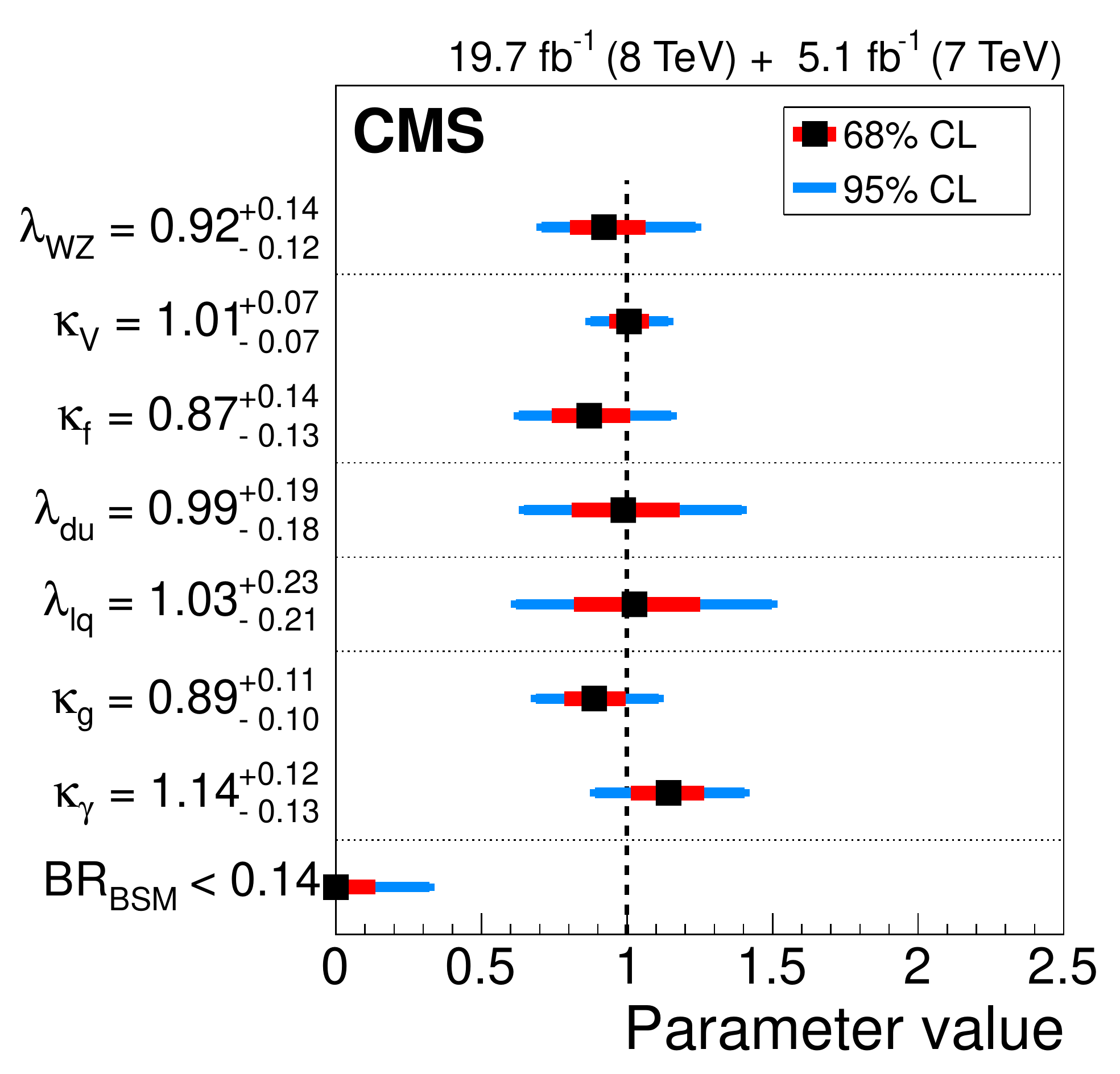}
\caption{Summary plot of likelihood scan results for the different parameters of
interest in benchmark models from Ref.~\cite{LHCHXSWG3} separated by dotted
lines. The \BRBSM value at the bottom is obtained for the model with three
parameters $(\kglu,\kgam,\BRBSM)$. The inner bars represent the 68\%~CL confidence
intervals while the outer bars represent the 95\%~CL confidence intervals.
}
\label{fig:xswg}
\end{figure}

Figure~\ref{fig:xswg} summarizes the results for the benchmark scenarios of
Ref.~\cite{LHCHXSWG3} with fewest parameters and shows that, in those
benchmarks, all results are compatible with the SM expectations.

A much more comprehensive overview of the searches performed for deviations from
the SM Higgs boson expectation is provided in Table~\ref{tab:CouplingTests},
where all results obtained in this section are summarized.

No statistically significant deviations are observed with respect to the SM
Higgs boson expectation.

\begin{table*}[tp]
\centering
\topcaption{
Tests of the compatibility of the data with the SM Higgs boson couplings.
The best-fit values and 68\% and 95\%~CL confidence intervals
are given for the evaluated scaling factors $\kappa_{i}$ or ratios
$\lambda_{ij}=\kappa_{i}/\kappa_{j}$.
The different compatibility tests discussed in the text are separated by horizontal lines.
When one of the parameters in a group is evaluated, others are treated as
nuisance parameters.
}
\label{tab:CouplingTests}
\renewcommand{\arraystretch}{1.2}
\resizebox{\textwidth}{!}{%
\begin{tabular}{r>{\footnotesize}cccc>{\small}l}

\hline

\multirow{2}{*}{Model parameters} & \multicolumn{1}{c|}{\multirow{2}{*}{\begin{tabular}[c]{@{}c@{}}Table in\\Ref.~\cite{LHCHXSWG3}\end{tabular}}} & \multicolumn{1}{c}{\multirow{2}{*}{Parameter}} &\multicolumn{2}{c}{Best-fit result} & \multicolumn{1}{|l}{\multirow{2}{*}{Comment}} \\
 & \multicolumn{1}{c|}{} &  & 68\%~CL & \multicolumn{1}{c|}{95\%~CL} &  \\

\hline
\hline

\kZ, \lWZ (\kf=1) & --- & \lWZ & \lwzONEOneSig & \lwzONE & \begin{tabular}[c]{@{}l@{}}$\lWZ=\kW/\kZ$ from \zz and\\0/\njet[1] \ww channels.\end{tabular} \\ \hline
\kZ, \lWZ, \kf & \begin{tabular}[c]{@{}c@{}}44\\(top)\end{tabular} & \lWZ & \lwzTWOOneSig & \lwzTWO & \begin{tabular}[c]{@{}l@{}}$\lWZ=\kW/\kZ$ from\\full combination.\end{tabular} \\

\hline
\hline

\multirow{2}{*}{\kV, \kf} & \multirow{2}{*}{\begin{tabular}[c]{@{}c@{}}43\\(top)\end{tabular}} & \kV & \CTWOkVOneSig & \CTWOkV & \begin{tabular}[c]{@{}l@{}}$\kV$ scales couplings\\to $\PW$ and $\cPZ$ bosons.\end{tabular} \\
 &  & \kf & \CTWOkFOneSig & \CTWOkF & \begin{tabular}[c]{@{}l@{}}$\kf$ scales couplings\\to all fermions.\end{tabular} \\

\hline
\hline

\kV, \ldu, \ku & \begin{tabular}[c]{@{}c@{}}46\\(top)\end{tabular} & \ldu & \lduOneSig & \lduTwoSig & \begin{tabular}[c]{@{}l@{}}$\ldu=\ku/\kd$, relates\\up-type and down-type\\ fermions.\end{tabular} \\ \hline
\kV, \llq, \kq & \begin{tabular}[c]{@{}c@{}}47\\(top)\end{tabular} & \llq & \llqOneSig & \llqTwoSig & \begin{tabular}[c]{@{}l@{}}$\llq=\kl/\kq$, relates\\leptons and quarks.\end{tabular} \\

\hline
\hline

\multirow{6}{*}{\begin{tabular}[c]{@{}r@{}}\kW, \kZ, \ktop,\\ \\\kb, \ktau, \kmu\end{tabular}} & \multirow{6}{*}{\begin{tabular}[c]{@{}c@{}}Extends\\51\end{tabular}} & \kW & \CFIVEkWOneSig & \CFIVEkW &  \\
 &  & \kZ & \CFIVEkZOneSig & \CFIVEkZ &  \\
 &  & \ktop & \CFIVEktopOneSig & \CFIVEktop & \begin{tabular}[c]{@{}l@{}}Up-type quarks (via \cPqt).\end{tabular} \\
 &  & \kb & \CFIVEkbOneSig & \CFIVEkb & \begin{tabular}[c]{@{}l@{}}Down-type quarks (via \cPqb).\end{tabular} \\
 &  & \ktau & \CFIVEktauOneSig & \CFIVEktau & \begin{tabular}[c]{@{}l@{}}\ktau scales the coupling to tau leptons.\end{tabular} \\
 &  & \kmu & \CFIVEkmuOneSig & \CFIVEkmu & \begin{tabular}[c]{@{}l@{}}\kmu scales the coupling to muons.\end{tabular} \\ \hline

\multirow{2}{*}{$M$, $\epsilon$} & \multirow{2}{*}{Ref.~\cite{EllisYou2013}} & $M$ (\GeV) & \mepsMOneSig & \mepsMTwoSig & \multirow{2}{*}{\begin{tabular}[c]{@{}l@{}} $\kf = v \frac{m_\mathrm{f}^{\epsilon}}{M^{1+\epsilon}}$ and $\kV = v \frac{m_\mathrm{V}^{2\epsilon}}{M^{1+2\epsilon}}$\\(Section~\ref{sec:mepsc5})\end{tabular}} \\
 &  & $\epsilon$ & \mepsEOneSig & \mepsETwoSig &  \\

\hline
\hline

\multirow{2}{*}{\kglu, \kgam} & \multirow{2}{*}{\begin{tabular}[c]{@{}c@{}}48\\(top)\end{tabular}} & \kglu & \CTWOkgluOneSig & \CTWOkglu & \multirow{2}{*}{\begin{tabular}[c]{@{}l@{}}Effective couplings to\\gluons (\Pg) and photons (\PGg).\end{tabular}} \\
 &  & \kgam & \CTWOkgamOneSig & \CTWOkgam &  \\ \cline{2-6}

\kglu, \kgam, \BRBSM & \begin{tabular}[c]{@{}c@{}}48 (middle)\end{tabular} & \BRBSM & \BRBSMOneSig & \BRBSMTwoSig & Allows for BSM decays.\\ \cline{2-5}

\begin{tabular}[c]{@{}r@{}} with \hinv searches\end{tabular} & --- & \BRinv & \BRBSMHinvOneSig & \BRBSMHinvTwoSig & \hinv use implies \BRundet=0.\\ \hline

with \hinv and $\kappa_i=1$ & --- & \BRinv & \BRBSMHinvKSMOneSig & \BRBSMHinvKSMTwoSig & Assumes $\kappa_i=1$ and uses \hinv. \\

\hline
\hline

\multirow{7}{*}{\begin{tabular}[c]{@{}r@{}}\kgluZ, \\ \\\lWZ, \lZglu, \lbZ, \\ \\\lgamZ, \ltauZ, \ltopglu\end{tabular}} & \multirow{7}{*}{\begin{tabular}[c]{@{}c@{}}50\\(bottom)\end{tabular}} &
      \kgluZ   & $0.98~^{+0.14}_{-0.13}$ & $[0.73,1.27]$ & \begin{tabular}[c]{@{}l@{}}$\kgluZ=\kglu\kZ/\kH$, \ie floating \kH.\end{tabular} \\
 &  & \lWZ     & $0.87~^{+0.15}_{-0.13}$ & $[0.63,1.19]$ & \begin{tabular}[c]{@{}l@{}}$\lWZ=\kW/\kZ$.\end{tabular} \\
 &  & \lZglu   & $1.39~^{+0.36}_{-0.28}$ & $[0.87,2.18]$ & \begin{tabular}[c]{@{}l@{}}$\lZglu=\kZ/\kglu$.\end{tabular} \\
 &  & \lbZ     & $0.59~^{+0.22}_{-0.23}$ & $\leq1.07$ & \begin{tabular}[c]{@{}l@{}}$\lbZ=\kb/\kZ$.\end{tabular} \\
 &  & \lgamZ   & $0.93~^{+0.17}_{-0.14}$ & $[0.67,1.31]$ & \begin{tabular}[c]{@{}l@{}}$\lgamZ=\kgam/\kZ$.\end{tabular} \\
 &  & \ltauZ   & $0.79~^{+0.19}_{-0.17}$ & $[0.47,1.20]$ & \begin{tabular}[c]{@{}l@{}}$\ltauZ=\ktau/\kZ$.\end{tabular} \\
 &  & \ltopglu & $2.18~^{+0.54}_{-0.46}$ & $[1.30,3.35]$ & \begin{tabular}[c]{@{}l@{}}$\ltopglu=\ktop/\kglu$.\end{tabular} \\

\hline
\hline

\multirow{6}{*}{\begin{tabular}[c]{@{}r@{}}\kV, \kb, \ktau, \\ \\\ktop, \kglu, \kgam\end{tabular}} & \multirow{6}{*}{\begin{tabular}[c]{@{}c@{}}Similar to\\50 (top)\end{tabular}} & \kV & \CSIXkVOneSig & \CSIXkV &  \\
 &  & \kb & \CSIXkbOneSig & \CSIXkb & \begin{tabular}[c]{@{}l@{}}Down-type quarks (via \cPqb).\end{tabular} \\
 &  & \ktau & \CSIXktauOneSig & \CSIXktau & \begin{tabular}[c]{@{}l@{}}Charged leptons (via \PGt).\end{tabular} \\
 &  & \ktop & \CSIXktopOneSig & \CSIXktop & \begin{tabular}[c]{@{}l@{}}Up-type quarks (via \cPqt).\end{tabular} \\
 &  & \kglu & \CSIXkgluOneSig & \CSIXkglu &  \\
 &  & \kgam & \CSIXkgamOneSig & \CSIXkgam &  \\
\cline{2-5}
with $\kV\leq1$ and \BRBSM & --- & \BRBSM & \BRBSMCSevenOneSig & \BRBSMCSeven & Allows for BSM decays.\\

\cline{2-6}

with $\kV\leq1$ and \hinv & --- & \BRinv & \BRinvHinvCSevenOneSig & \BRinvHinvCSeven & \hinv use implies $\BRundet=0$. \\ \cline{2-6}

\multirow{2}{*}{\begin{tabular}[c]{@{}r@{}}with $\kV\leq1$, \hinv,\\\BRinv, and \BRundet \end{tabular}} & --- & \BRinv & \BRinvHinvCEightOneSig & \BRinvHinvCEight & \multirow{2}{*}{\begin{tabular}[c]{@{}c@{}}Separates \BRinv from \BRundet,\\$\BRBSM=\BRinv+\BRundet$.\end{tabular}} \\
& --- & \BRundet & \BRundetHinvCEightOneSig & \BRundetHinvCEight &  \\

\hline
\hline

\end{tabular}

}
\end{table*}

\section{Summary}
\label{sec:conclusion}

Properties of the Higgs boson with mass near 125\GeV are measured
in proton-proton collisions with the CMS experiment at the LHC.
Comprehensive sets of production and decay measurements are combined.
The decay channels include \gg, \zz, \ww, \tt, \bb, and \mumu pairs.
The data samples were collected in 2011 and 2012 and correspond to
integrated luminosities of up to 5.1\fbinv at 7\TeV and up to 19.7\fbinv
at 8\TeV.
From the high-resolution \gg and \zz channels, the mass of the Higgs boson is
determined to be \MASSHdetail.
For this mass value, the event yields obtained in the different analyses
tagging specific decay channels and production mechanisms are consistent with
those expected for the standard model Higgs boson.
The combined best-fit signal relative to the standard model
expectation is
\MUHATdetail
at the measured mass.
The couplings of the
Higgs boson are probed for deviations in magnitude from the standard model
predictions in multiple ways, including searches for invisible and undetected
decays.
No significant deviations are found.

\begin{acknowledgments}
\hyphenation{Bundes-ministerium Forschungs-gemeinschaft Forschungs-zentren}
We congratulate our colleagues in the CERN accelerator departments for the
excellent performance of the LHC and thank the technical and administrative
staffs at CERN and at other CMS institutes for their contributions to the
success of the CMS effort. In addition, we gratefully acknowledge the computing
centres and personnel of the Worldwide LHC Computing Grid for delivering so
effectively the computing infrastructure essential to our analyses. Finally, we
acknowledge the enduring support for the construction and operation of the LHC
and the CMS detector provided by the following funding agencies: the Austrian
Federal Ministry of Science, Research and Economy and the Austrian Science
Fund; the Belgian Fonds de la Recherche Scientifique, and Fonds voor
Wetenschappelijk Onderzoek; the Brazilian Funding Agencies (CNPq, CAPES,
FAPERJ, and FAPESP); the Bulgarian Ministry of Education and Science; CERN; the
Chinese Academy of Sciences, Ministry of Science and Technology, and National
Natural Science Foundation of China; the Colombian Funding Agency
(COLCIENCIAS); the Croatian Ministry of Science, Education and Sport, and the
Croatian Science Foundation; the Research Promotion Foundation, Cyprus; the
Ministry of Education and Research, Estonian Research Council via IUT23-4 and
IUT23-6 and European Regional Development Fund, Estonia; the Academy of
Finland, Finnish Ministry of Education and Culture, and Helsinki Institute of
Physics; the Institut National de Physique Nucl\'eaire et de Physique des
Particules~/~CNRS, and Commissariat \`a l'\'Energie Atomique et aux \'Energies
Alternatives~/~CEA, France; the Bundesministerium f\"ur Bildung und Forschung,
Deutsche Forschungsgemeinschaft, and Helmholtz-Gemeinschaft Deutscher
Forschungszentren, Germany; the General Secretariat for Research and
Technology, Greece; the National Scientific Research Foundation, and National
Innovation Office, Hungary; the Department of Atomic Energy and the Department
of Science and Technology, India; the Institute for Studies in Theoretical
Physics and Mathematics, Iran; the Science Foundation, Ireland; the Istituto
Nazionale di Fisica Nucleare, Italy; the Ministry of Science, ICT and Future
Planning, and National Research Foundation (NRF), Republic of Korea; the
Lithuanian Academy of Sciences; the Ministry of Education, and University of
Malaya (Malaysia); the Mexican Funding Agencies (CINVESTAV, CONACYT, SEP, and
UASLP-FAI); the Ministry of Business, Innovation and Employment, New Zealand;
the Pakistan Atomic Energy Commission; the Ministry of Science and Higher
Education and the National Science Centre, Poland; the Funda\c{c}\~ao para a
Ci\^encia e a Tecnologia, Portugal; JINR, Dubna; the Ministry of Education and
Science of the Russian Federation, the Federal Agency of Atomic Energy of the
Russian Federation, Russian Academy of Sciences, and the Russian Foundation for
Basic Research; the Ministry of Education, Science and Technological
Development of Serbia; the Secretar\'{\i}a de Estado de Investigaci\'on,
Desarrollo e Innovaci\'on and Programa Consolider-Ingenio 2010, Spain; the
Swiss Funding Agencies (ETH Board, ETH Zurich, PSI, SNF, UniZH, Canton Zurich,
and SER); the Ministry of Science and Technology, Taipei; the Thailand Center
of Excellence in Physics, the Institute for the Promotion of Teaching Science
and Technology of Thailand, Special Task Force for Activating Research and the
National Science and Technology Development Agency of Thailand; the Scientific
and Technical Research Council of Turkey, and Turkish Atomic Energy Authority;
the National Academy of Sciences of Ukraine, and State Fund for Fundamental
Researches, Ukraine; the Science and Technology Facilities Council, UK; the US
Department of Energy, and the US National Science Foundation.

Individuals have received support from the Marie-Curie programme and the
European Research Council and EPLANET (European Union); the Leventis
Foundation; the A. P. Sloan Foundation; the Alexander von Humboldt Foundation;
the Belgian Federal Science Policy Office; the Fonds pour la Formation \`a la
Recherche dans l'Industrie et dans l'Agriculture (FRIA-Belgium); the Agentschap
voor Innovatie door Wetenschap en Technologie (IWT-Belgium); the Ministry of
Education, Youth and Sports (MEYS) of the Czech Republic; the Council of
Science and Industrial Research, India; the HOMING PLUS programme of Foundation
for Polish Science, cofinanced from European Union, Regional Development Fund;
the Compagnia di San Paolo (Torino); the Consorzio per la Fisica (Trieste);
MIUR project 20108T4XTM (Italy); the Thalis and Aristeia programmes cofinanced
by EU-ESF and the Greek NSRF; and the National Priorities Research Program by
Qatar National Research Fund.
\end{acknowledgments}
\bibliography{auto_generated}   

\cleardoublepage \appendix\section{The CMS Collaboration \label{app:collab}}\begin{sloppypar}\hyphenpenalty=5000\widowpenalty=500\clubpenalty=5000\textbf{Yerevan Physics Institute,  Yerevan,  Armenia}\\*[0pt]
V.~Khachatryan, A.M.~Sirunyan, A.~Tumasyan
\vskip\cmsinstskip
\textbf{Institut f\"{u}r Hochenergiephysik der OeAW,  Wien,  Austria}\\*[0pt]
W.~Adam, T.~Bergauer, M.~Dragicevic, J.~Er\"{o}, M.~Friedl, R.~Fr\"{u}hwirth\cmsAuthorMark{1}, V.M.~Ghete, C.~Hartl, N.~H\"{o}rmann, J.~Hrubec, M.~Jeitler\cmsAuthorMark{1}, W.~Kiesenhofer, V.~Kn\"{u}nz, M.~Krammer\cmsAuthorMark{1}, I.~Kr\"{a}tschmer, D.~Liko, I.~Mikulec, D.~Rabady\cmsAuthorMark{2}, B.~Rahbaran, H.~Rohringer, R.~Sch\"{o}fbeck, J.~Strauss, W.~Treberer-Treberspurg, W.~Waltenberger, C.-E.~Wulz\cmsAuthorMark{1}
% ...
\vskip\cmsinstskip
% ...
1:~~Also at Vienna University of Technology, Vienna, Austria\\
2:~~Also at CERN, European Organization for Nuclear Research, Geneva, Switzerland\\
% ...

\end{sloppypar}
\end{document}